\documentclass{article}

\newcommand{\ie}{\mbox{i.\,e.\,\ }}
\newcommand{\iec}{\mbox{i.\,e.\,}}

\newcommand{\eg}{\mbox{e.\,g.\,\ }}
\newcommand{\egc}{\mbox{e.\,g.\,}}
\newcommand{\etc}{etc.\,\ }

\newcommand{\dr}[1]{\ensuremath{\mathrm{d} #1\,}}
\newcommand{\mc}[1]{\ensuremath{\mathcal{#1}}}

\newcommand{\edf}{\ensuremath{=_{_{df}}}}

\newcommand{\ket}[1]{\ensuremath{\left|  #1 \right\rangle}}
\newcommand{\bra}[1]{\ensuremath{\left\langle #1 \right|}}
\newcommand{\bk}[2]{\ensuremath{\left\langle #1 | #2 \right\rangle}}
\newcommand{\proj}[2]{\ensuremath{\ket{#1} \bra{#2}}}
\newcommand{\tpk}[2]{\ensuremath{\ket{#1}\!\otimes\!\ket{#2}}}

\newcommand{\matel}[3]{\ensuremath{\bra{#1} #2 \ket{#3}}}
 
\newcommand{\op}[1]{\ensuremath{\widehat{\textsf{\ensuremath{#1}}}}}
\newcommand{\opad}[1]{\ensuremath{\op{#1}^{\dagger}}}
\newcommand{\id}{\op{\mathsf{1}}}
\newcommand{\denop}{\ensuremath{\rho}}

\newcommand{\tr}{\textsf{Tr}}

\newcommand{\be}{\begin{equation}}
\newcommand{\ee}{\end{equation}}
\usepackage{chicago}
\usepackage{latexsym}

\newcommand{\val}{\mathrm{Val}}
\newcommand{\crossout}{\!\!\!\!\!\! /}
\newcommand{\ev}{\item[Everettian:]}
\newcommand{\sk}{\item[Sceptic:]}

\begin{document}

\newtheorem{theorem}{Theorem}[section]

\newtheorem{theoremapp}{Theorem A.}
\newtheorem{stage}{Stage}
\newtheorem{stageII}{Stage}
\newtheorem{savstep}{Savage's Step}

\newtheorem{lemma}{Lemma}
\newtheorem{lemma'}{Lemma}
\newtheorem{lemma''}{Lemma}

\title{Quantum Probability and Decision Theory, Revisited}
\author{David Wallace \\ Magdalen College, Oxford \\ (Email: \texttt{david.wallace@magd.ox.ac.uk})}
\date{18th November 2002}
\maketitle

\begin{abstract}
An extended analysis is given of the program, originally suggested by
Deutsch, of solving the probability problem in the Everett
interpretation by means of decision theory.  Deutsch's own proof is
discussed, and alternatives are presented which are based upon different decision theories
and upon Gleason's Theorem.  It is argued that decision theory gives 
Everettians most or all of what they need from `probability'. 
Contact is made with Lewis's Principal Principle
linking subjective credence with objective chance: an Everettian
Principal Principle is formulated, and shown to be at least as
defensible as the usual Principle.  Some
consequences of (Everettian) quantum mechanics for decision theory itself are
also discussed.
\end{abstract}

\section{Introduction}\label{intro}

In recent work on the Everett (Many-Worlds) interpretation of quantum
mechanics, it has increasingly been recognised that any version of the
interpretation worth defending will be one in which the basic formalism
of quantum mechanics is left unchanged.  Properties such as the
interpretation of the wave-function as describing a multiverse of
branching worlds, or the ascription of probabilities to the branching
events, must be emergent from the unitary quantum mechanics rather than
added explicitly to the mathematics.  Only in this way is it possible to
save the main virtue of Everett's approach: having an account of quantum
mechanics consistent with the last seventy years of physics, not one in
which the edifice of particle physics must be constructed afresh
\cite[p.\,44]{saundersmetaphysics}.\footnote{This is by no means
\emph{universally} recognised.  Everett-type interpretations can perhaps
be divided into three types: 
\begin{description}
\item[(i)] Old-style ``Many-Worlds''
interpretations in which worlds are added explicitly to the quantum
formalism (see, \egc, \citeN{dewitt} and \citeN{deutsch85}, although
Deutsch has since abandoned this approach; in fact, it is hard to find any remaining
defendants of type (i) approaches). 
\item[(ii)] ``Many-Minds'' approaches in which some intrinsic property of the mind 
is essential to understanding how to reconcile indeterminateness and probability with 
unitary quantum mechanics (see,
\egc, \citeN{albertloewer}, Lockwood \citeyear{lockwood,lockwoodbjps1}, \citeN{donald}, and
\citeN{sudbery}).
\item[(iii)] Decoherence-based approaches, such as those defended by myself 
(Wallace \citeyearNP{wallacestructure,wallaceworlds}), Saunders
\citeyear{saundersdecoherence,saundersmetaphysics,saundersprobability},
Deutsch \citeyear{deutschlockwood,deutschstructure}, Vaidman
\citeyear{vaidman,vaidmanencyclopaedia} and \citeN{zurekprobability}.
\end{description}
For the rest of this paper, whenever I refer to ``the Everett
interpretation'', I shall mean specifically the type (iii) approaches.
This is simply for brevity, and certainly isn't meant to imply anything
about what was intended in Everett's original \citeyear{everett} paper.}

Of the two main problems generally raised with Everett-type
interpretations, the preferred-basis problem looks eminently solvable
without changing the formalism.  The main technical tool towards
achieving this has of course been decoherence theory, which has provided powerful
(albeit perhaps not conclusive) evidence  that the quantum state has a
\textit{de facto} preferred basis and that this basis allows us to
describe the universe in terms of a branching structure of approximately
classical, approximately non-interacting worlds. I have argued elsewhere
(Wallace \citeyearNP{wallacestructure,wallaceworlds}) that there are no purely
conceptual problems with using decoherence to solve the preferred-basis
problem, and that the inexactness of the process should give us no cause
to reject it as insufficient.

The other main problem with the Everett interpretation concerns the concept of
probability: given that the Everettian description of measurement is a 
deterministic, branching process, how are we to reconcile that with the stochastic
description of measurement used in practical applications of quantum
mechanics?  It has been this problem, as much as the preferred basis
problem, which has led many workers on the Everett interpretation to
introduce explicit extra structure into the mathematics of quantum theory
so as to make sense of the probability of a world as (for instance) a measure over
continuously many identical worlds.  Even some proponents of the 
Many-Minds variant on Everett (notably \citeNP{albertloewer} and Lockwood 
\citeyearNP{lockwood,lockwoodbjps1}), 
who arguably have no difficulty with the preferred-basis problem,
have felt forced to modify quantum mechanics in this way.

It is useful to identify two aspects of the problem.  The first might be
called the \emph{incoherence problem}: how, when every outcome actually
occurs, can it even make sense to view a measurement as indeterministic?
Even were this solved, there would then remain a \emph{quantitative
problem}: why is that indeterminism quantified according to the quantum probability rule
(\iec, the Born rule),
and not (for instance) some other assignment of probabilities to branches?

In my view, the incoherence problem has been essentially resolved by Saunders, building
on Parfit's reductionist approach to personal identity.
(Saunders' approach is summarised in section \ref{dectoQM}).
This then leaves the quantitative problem as the major conceptual
obstacle to a satisfactory formulation of the Everett interpretation.

Saunders himself has claimed \citeyear{saundersprobability} that the
quantitative problem is a non-problem: that once we have shown the
coherence of ascribing probability to quantum splitting, we can simply
postulate that the quantum weights are to be interpreted as
probabilities:
\begin{quote}
Neither is it routinely required, of a physical theory, that a proof be
given that we are entitled to interpret it in a particular way; it is
anyway unclear as to what could count as a proof.  Normally it is enough
that the theory can be subjected to empirical test and confirmation;
quantum mechanics can certainly be \emph{applied}, on the understanding
that relations in the Hilbert-space norm count as probability \ldots It
is not as though the experimenter will need to understand something
more, a philosophical \emph{argument}, for example, before doing an
experiment. [\citeNP{saundersprobability}, p.\,384; emphasis his.]
\end{quote}
Whether one accepts such a claim depends upon one's attitude to the
philosophy of probability in general.  Many have claimed (\egc,
\citeNP{mellor}) that objective probability is simply another theoretical
posit, like charge or mass, in which case presumably all that is
required to introduce probability into a theory is a mathematical
structure satisfying the Kolmogorov axioms together with the statement
that the structure is to be interpreted as probability.
 If this is acceptable in classical
physics, then it seems no less so in quantum mechanics.

But there is a more demanding view of probability, eloquently defended
by Lewis \citeyear{lewisprobability,lewis94} and more recently argued
for in the quantum context by \citeN{papineau}: whatever (objective) probability
\emph{is}, our empirical access to it is via considerations of
rationality and behaviour: it must be the case that it is rational to
use probability as a guide to action.  This seems to call for just the
`proof' which Saunders rejects: some argument linking rational action to
whatever entities in our theory are called `probabilities'.
Granted, there is no really convincing such account (to my knowledge) in
classical philosophy of probability, but there is at least some prospect
that either straight frequentism \cite[pp.\,344--347]{howsonurbach} or Lewis's
Best-Systems Analysis variant on frequentism can be made to suffice But
clearly neither are applicable to the Everett interpretation, which
seems to leave it at a significant disadvantage.

In this context it is extremely interesting that David Deutsch has
claimed \cite{deutschprobability} to derive the quantum probability rule from decision theory:
that is, from considerations of pure rationality. It is rather
surprising how little attention his work has received in the
foundational community, though one reason may be that it is very unclear
from his paper that the Everett interpretation is assumed from the
start.\footnote{Nonetheless it \emph{is} assumed: 
\begin{quote}
However, in other respects he [the rational agent] will not behave as if
he believed that stochastic processes occur.  For instance if
\emph{asked} whether they occur he will certainly reply `no', because
the non-probabilistic axioms of quantum theory require the state to
evolve in a continuous and deterministic way.
[\citeNP[pp.\,13]{deutschprobability}; emphasis his.]
\end{quote}
}
If it is tacitly assumed that his work refers instead to some
more orthodox collapse theory, then it is easy to see that the proof is
suspect; this is the basis of the criticisms levelled at Deutsch by
Barnum \textit{et al}, \citeyear{BCFFS}.  Their attack on Deutsch's paper seems to have been
influential in the community; however, it is at best questionable
whether or not it is valid when Everettian assumptions are made
explicit.

The purpose of this paper, then, is threefold: first, to give as clear
an exegesis as possible of Deutsch's argument, making explicit its tacit
assumptions and hidden steps, and to assess its validity; second, to
place his argument in the context of general decision theory and show
how we can thereby improve upon it, making its assumptions more plausible
and its argument more transparent; thirdly, to assess the implications
of Deutsch's proof (and my variations on it) both for the Everett
interpretation and for decision theory itself.

Since decision theory is probably unfamiliar to many people working in
the foundations of quantum mechanics, I begin (section \ref{classdec})
by expounding its general principles and goals.  In section
\ref{dectoQM} I explain how decision theory can be applied to 
quantum-mechanical contexts, and especially to the quantum games which
feature so heavily in Deutsch's proof.

Sections \ref{quantgames}--\ref{gleason} are the core of the paper.
Section \ref{quantgames} introduces Deutsch's games, and shows that
with very few assumptions beyond the mathematical formalism of quantum
mechanics it is possible to prove some strong results about a rational
agent's preferences between such games.  These results are then central in
the proofs of sections \ref{deutschproof}--\ref{gleason}, which give
a reconstruction of Deutsch's proof as well as three variants on it: one
closely related to Deutsch's proof, one based upon axioms inspired by
Savage's \citeyear{savage} axiomatization of decision theory, and one based
upon Gleason's theorem.  The last of these is only tentatively proved,
and shows that the usefulness of Gleason's theorem in a realistic
reconstruction of probability is more questionable than has sometimes
(\eg Barnum \textit{et al} \citeyearNP{BCFFS}) been suggested.  In sections \ref{quantprobdiscussion}--\ref{conclusion}
I discuss the implications of Deutsch's proof and its variants: section
\ref{quantprobdiscussion} is an analysis of the extent to which the
proof allows quantum probability to satisfy the rationality requirements
of Lewis and Papineau, whilst section \ref{conclusion}, a dialogue
between an Everettian and a Sceptic, discusses the possible problems and
weak points of this approach and concludes with a summary of its
implications.

\section{Classical decision theory}\label{classdec}

In this section, I will give a brief exposition of classical (\iec, standard) decision
theory: both its aims and its technical details.  The latter will be
relevant because, as we will see, the quantum-mechanical derivations of
probability which form the core of the paper are in many ways rather
closely modelled on the classical ones.

\subsection{The decision problem}\label{decisionintro}

Decision theory is concerned with the preferences of rational agents ---
where ``rational'' is construed in a rather narrow sense.  If someone
were to choose to jump into an alligator pit we would be inclined to
call their choice irrational, but from a decision-theoretic viewpoint it
would simply be unusual.  If an agent, however, were to say that they
preferred alligator pits to snake pits, snake pits to scorpion pits, and
scorpion pits to alligator pits, \emph{then} decision theory would deem
their preferences irrational.  Decision theory, then, is concerned with
the logical and structural  \emph{constraints} which rationality places on an agent's structure
of preferences, but is not intended to come anywhere near determining those preferences wholly.

It is also concerned, in large part, with decision-making under
uncertainty.  If for any action which an agent takes they are certain
what the outcome will be, then the constraint alluded to above --- that,
if A is preferred to B and B to C, then A is preferred to C --- is
really all that decision theory has to say about the agent's
preferences.  But when the agent has to choose between a number of acts
none of which have a perfectly predictable outcome --- betting
on horses, for instance, or choosing whether or not to cross the road 
--- then considerations of pure rationality can place strong constraints
on that agent's preferences.

To understand this further, let us define a certain sort of
\emph{decision problem}.  This problem is somewhat stylised but
nonetheless can be used to describe a wide class of real-world decision
problems; it is a mildly enlarged version of the decision problem
considered by \citeN{savage}.

The decision problem is specified by the following sets:
\begin{itemize} 
\item A set \mc{C} of \emph{consequences}, to be regarded as the 
``atomic holistic entities that have value to the individual''
\cite{fishburn}.  Typical consequences might be receiving a 
thousand-euro cheque, or being hit by a bus. 
\item A set \mc{M} of \emph{chance setups}, to be regarded as
situations in which a number of possible events might occur, and where
it is in general impossible for the agent to predict with certainty \emph{which} will
occur.  Examples might be a rolled die (in which case there is
uncertainty as to which face will be uppermost) or a section of road
over a five-minute period (in which case there is uncertainty as to
whether or not a bus will come along).
\item A set \mc{S} of \emph{states}, to be regarded as 
possible, in general coarse-grained, 
states of the world at some time; typical states might be that state in
which lottery ticket 31942 is drawn, or in which there is a bus coming
down the road.  For each $M \in \mc{M}$ we can define a subset $\mc{S}_M
\subseteq \mc{S}$ of states which might occur as a consequence of the
chance setup. ($\mc{S}_M$ is taken as consisting of mutually exclusive,
jointly exhaustive states.)
(We will define an \emph{event} for $M$ as some subset of $\mc{S}_M$.
The \emph{event space} for $M$ is the set $\mc{E}_M$ of events, \ie the power set of 
$\mc{S}_M$.)
\item A set \mc{A} of \emph{acts}, to be regarded as possible choices of
action for the individual (usually in the face of uncertainty as to which state
is to arise).  For our purposes each act can be understood as a pair $f=(M,\mc{P})$ where 
$M$ is a chance setup and \mc{P} is a function from the set $\mc{S}_M$ of states consistent
with $M$, to the
set \mc{C} of consequences. (Thus, performing an act is a two-part process:
the chance setup $M$ must be allowed to run, and \mc{P} fixes the
consequences for the individual of the various results of $M$.)
 In a sense \mc{P} represents some sort of
``bet'' placed on the outcome of $M$ (such as on which die-face is
uppermost), although it need not be a formal wager: in the case of the
road, for instance, one payoff scheme might refer to crossing the road,
so that \mc{P}(bus in road)=being hit by bus, \mc{P}(no bus)=getting
safely to other side; another might refer to choosing not to
cross.\footnote{It is important for the development of the theory that states do not
\emph{per se} have value to agents; an agent values a state's obtaining
only insofar as it is associated via an act to a good consequence.
\citeN{joyce} has criticised this assumption in the general case: it is
hard to see how the state `the Earth is destroyed' can be rendered innocuous by appropriate
choice of consequence! However, in this paper the states will by and large be readout states
of quantum measurement processes, for which the assumption is far more reasonable.} 
\item A preference order $\succ$ on the set of acts, so that $f\succ g$
if and only if the agent would prefer it that $f$ is performed than that $g$
is.  If $f$ and $g$ are acts which it is within the agent's power to
bring about, this implies that the agent would choose $f$ rather than
$g$; more generally it is a hypothetical preference: \emph{if} the agent
were able to choose between $f$ and $g$, they would choose $f$.
\end{itemize}

(In developing the axiomatics of decision theory we can simply treat
acts, states and consequences as primitives; to apply the theory to the
actual world, of course, there are many subtleties as to exactly how to
carve up the world into states, how to distinguish between states and
consequences, how to analyse an act as a pair $(M,\mc{P})$ etc.  In the quantum 
decision problems we analyze,
however, it will be reasonably clear what the correct analysis is.)

The decision problem, put informally, is then: how do considerations of
rationality constrain an agent's preferences between elements of \mc{A}?

\subsection{Expected utility}\label{exputilsect}

The standard answer to the decision problem is as follows: act $f$ is
preferred to act $g$ iff $EU(f)>EU(g)$, where for any act $f=(M,\mc{P})$
the \emph{expected utility} $EU(f)$ of $f$ is defined by
\be \label{exputil} 
EU(f) = \sum_{x \in \mc{S}_M} \Pr (x | M) \mc{V}[\mc{P}(x)],\ee
and where
\begin{itemize}
\item $\Pr(x|M)$, a real number between zero and one, is the conditional probability
that $x$ will obtain given that $M$ occurs;
\item $\mc{P}(x)$ is the consequence associated by $f$ with state $x\in \mc{S}_M$;
\item $\mc{V}(c)$, a real number, is the value to the individual of
consequence $c$.
\end{itemize}
(In fact, if the set $\mc{S}_M$ is infinite the notion of probability must be applied
to (a certain Boolean sub-algebra of) events of $M$ rather than states \textit{simpliciter}.)

If the notions of `probability' and `value' are treated as primitive,
then this \emph{expected-utility rule} gives a complete answer to the
decision problem, and very strongly constrains the agent's choices
between events with uncertain outcomes: given the probabilities and the
values of consequences, then the preference order amongst acts is fixed.
(The question would remain: how is the rule itself justified?)

However, it is rather unclear what these primitive notions of
probability and value actually refer to.  Understanding them
\emph{qualitatively} is not difficult: for the value function, to say
that $\mc{V}(c)>\mc{V}(d)$ is to say that the agent would prefer to
receive $c$ than to receive $d$.  As for probability, if `more likely than' can be
treated as primitive then we can understand `$\Pr (x|M)>\Pr (y|M)$' as
saying that $x$ is more likely than $y$ to occur.  If we do not want to
treat it as primitive then we can understand it in terms of preferences between bets:
to say that one state is more likely than another is to say that we
would rather bet on the occurrence of the first state than of the
second (assuming we don't care \textit{per se} which state occurs).

But to define the expected-utility rule we need to understand the
notions quantitatively.  If the above qualitative understanding is all that constrains them, then
we could replace $\Pr$ and \mc{V} by arbitrarily monotonically
increasing functions of $\Pr$ and \mc{V}, for as yet we have given no meaning to
ideas like $c$ is \emph{twice} as valuable as $d$, or $x$ is
\emph{half} as likely as $y$.  

We might hope to find reasons outside decision theory to make
quantitative sense of \mc{V} and $\Pr$.  Maybe probabilities are
objective chances (see section \ref{objectivechance}); maybe \mc{V}
measures some ``moral worth'' of a consequence to us (as originally 
suggested by Bernoulli;
see \citeN[pp.\,91--104]{savage} for a historical discussion).  

But this is not the approach taken by decision theory.  The aim,
instead, is to derive some quantitative aspects of $\Pr$ and \mc{V} --- and,
in particular, the EUT --- from
the agent's preferences between acts (not just consequences).  (In doing
so, of course, we abandon any hope that the expected-utility rule always
tells us what to do in situations of uncertainty, and fall back on the
idea of decision theory as \emph{constraining} our preferences rather
than determining them completely.)

It is fairly easy to see how, \emph{given} a quantitative notion of
value and a preference ordering on acts, we could use it to define 
quantitative probabilities.  Suppose
$c$, $d$ and $e$ are consequences such that our agent is indifferent between
an act where he receives $c$ just if some state in event A occurs and $d$ otherwise, 
and another where
he receives $e$ with certainty; then we \emph{define} the probability of
$A$ by
\be \Pr(A)\edf(\mc{V}(e)-\mc{V}(d))/(\mc{V}(c)-\mc{V}(d)).\ee
In effect, this defines probability by betting: the probability of an
event is the shortest odds at which we would be prepared to bet on it.\footnote{To see this
more clearly, specialise to the case where $\mc{V}(d)=0.$}  (This
notion of probability was first suggested by \citeN{ramsey}, and has 
been explored \textit{in extenso} since then; see, \egc, \citeNP{definetti} or \citeNP{mellor}.)

Conversely, if we already have a quantitative notion of
probability then we can use it to define quantitative value of
consequences (this was first advocated by Pareto; again, see \citeN[pp.\,91--104]{savage} for a brief history).  
If an agent is indifferent between receiving $e$ with
certainty, and receiving $c$ if state $A$ obtains and $d$ otherwise, then 
\be (\mc{V}(e)-\mc{V}(d))/(\mc{V}(c)-\mc{V}(d))\edf \Pr(A).\ee
This fixes all ratios of differences of values, and hence fixes values up to a
multiplicative and an additive constant.

Plainly, in both cases there would be a need for additional assumptions
to ensure that these definitions were self-consistent and to derive the
expected-utility rule.  In any case, though, they are not completely
satisfactory as each takes one of $\Pr$ and $\mc{V}$ as primitive.  A
more satisfactory approach would generate \emph{both} from purely
qualitative axioms about preference.  Only one of the two would need to be
thus generated: the
other could then be defined in terms of the first, as above.

This is, in fact, possible.  The structure of such an approach is as
follows: we introduce, by means of some decision-theoretic postulates,
enough structure to the qualitative orderings of either \mc{C} or $\mc{E}_M$, that 
it is possible to prove some representation theorem guaranteeing,
as appropriate, either an effectively
unique value function, or a unique
probability function.\footnote{Value functions are in general only unique up to overall
multiplicative and additive constants; this can be seen from the expected-utility rule, from which it
is clear that multiplicative and/or additive changes to \mc{V} do not affect 
whether $EU(f)>EU(g)$. We should also stress that uniqueness here is \emph{only relative to a given observer's act preferences}. 
There is nothing at all to prevent an agent preferring death to ice
cream or having preferences which constrain the probability of the sun
rising tomorrow to be zero.}   Then we apply the methods above to generate
either \mc{V} from $\Pr$, or vice versa.

In either case, the ultimate goal is the same: from some basic axioms of
rational preference, prove the following representation theorem:
\begin{quote}
\textbf{Expected Utility Theorem (EUT):} Any agent's preferences amongst
acts determine a unique probability measure on events and a value
function, unique up to multiplicative and additive constants, such that
the preferences are completely represented by the expected-utility rule.
\end{quote}
This result will be partly descriptive,
partly normative: only by knowing the agent's preferences amongst some
substantial subset of acts can $\Pr$ and $\mc{V}$ be determined, but
once they are determined the remaining preferences are fixed.  As such
we will have a  strong constraint on the actions of rational agents.

In the next few sections, we will show how the EUT can be derived in two ways: one beginning with
a derivation of \mc{V}, one with $\Pr$.  As we will see, the approach which first proves a
representation theorem for probability and then derives utility is
technically rather more complex than its converse, but foundationally is
far more satisfactory. For clarity, I will begin with the simpler
utility-based approach
(in section \ref{add}); the probability-based approach is in sections
\ref{vnmsect}--\ref{savage}. First, though, I shall make some general
remarks about decision-theoretic axioms.

\subsection{The nature of decision-theoretic axioms}\label{decnature}

To prove our representation theorems, we have to introduce a number of
axioms of decision.  In specifying these, we will need to make use of the notion of a \emph{weak
ordering}.  Recall\footnote{Or take as a definition; the nomenclature in the literature
is somewhat variable.} that a weak ordering is a relation (which we will
always denote $\succeq$) which is:
\begin{itemize}
\item transitive: if $x \succeq y$ and $y \succeq z$, then $x \succeq
z$;
\item total: either $x \succeq y$, or $y \succeq x$.
\end{itemize}
We will write $x \simeq y$ whenever $x \succeq y$ and $y \succeq x$, and 
$x \succ y$ whenever $x \succeq y$ but  $x \simeq \crossout
y$; the relation can equally be specified in terms of $\succ$, defining
$x\simeq y$ whenever $x \succ \crossout y$ and $y \succ \crossout x$
(though the axiomatization is mildly more complicated).

We will also make use of the idea of a \emph{null event}: a null event
\mc{N}
is one to which a rational agent is completely indifferent.  Formally
this means that the agent is indifferent between any two payoff schemes
which differ only on \mc{N}; informally, it means that \mc{N} has
probability zero, though of course this cannot be taken as a definition.

The axioms themselves will be seen to break into two categories,
which can be described as follows (I follow Joyce's\citeyear{joyce} terminology 
[need to check ref to Suppes in Joyce]):
\begin{enumerate}
\item Axioms of pure rationality.  These are intended to be immediately
self-evident principles of decision-making: the rule about transitivity
of preferences mentioned in section \ref{decisionintro} is one of them.  
\item Structure axioms.  These are mathematically-inspired axioms, not
nearly so self-evident as the axioms of pure rationality: their purpose
is to rule out possibilities such as infinitesimal probability or
infinitely valuable consequences.
\end{enumerate}
In all versions of classical decision theory explored to date (not just
in those presented here) it is necessary to assume some fairly strong
structure axioms to get a unique representation theory.  A partial
justification for them as ``axioms of coherent extendability'' has 
been advocated by \citeN{kaplanstructure} and others (see \citeN{joyce} for some comments on 
this approach).
Nonetheless it is generally accepted that structure axioms are less satisfactory than
axioms of pure rationality, and that their use should be
minimized.

Whilst it is in general fairly straightforward to distinguish between
the two sorts of axiom, I should mention one controversy. We will be
assuming, as an axiom of pure rationality, that the preference order
between acts is a weak ordering: that is, for any two acts an agent either
prefers one to another or is indifferent between them.  This is
questionable: it is defensible to argue that an agent may simply have no
opinion as to which act is better.  Whether this is ultimately a
sustainable viewpoint depends in part on one's theory of desires and
preferences: Lewis, for instance, regards preferences as wholly
determined by dispositions to act \cite{lewisprobability}, 
in which case preference ordering is necessarily total; \citeN{joyce}
criticises this view.  Further investigation of this controversy lies
beyond this paper.

\subsection{Defining value functions through additivity}\label{add}

If we wish to treat quantitative value as prior to quantitative
probability (section \ref{exputilsect}'s first approach to the expected-utility rule) then we
will need some non-probabilistic way of understanding the idea that one
consequence is twice as valuable as another.  The only one of which I am
aware is composition: $c$ is twice as valuable as $d$ if I would be
indifferent between receiving $c$, and receiving $d$ twice.  

More formally, let us assume that there is some (associative, commutative)
operation $+$ of \emph{composition} 
on the set of consequences, so
that the consequence $c+d$ is to represent receiving both $c$ and
$d$. (We will as usual use `$n c$' to abbreviate `$c+c+ \ldots c$' ($n$ times)'). 
This strongly restricts the elements of \mc{C}, of course: many
consequences, such as, ``becoming president of the EU'', cannot be
received indefinitely many times.

We will define a weak ordering $\succ$ on any set \mc{C} with a
composition operation as \emph{additive} if it satisfies:
\begin{description}
\item[A1] There exists a ``zero consequence'' $0$ such that $c+0=c$ for
all $c$;\footnote{Note that the zero consequence is unrelated to the `null event' of
section \ref{decnature}: the latter is an event which (informally) I am certain will
not occur, whereas the former is a consequence which I may be certain I will receive,
but am indifferent about receiving.}
\item[A2] $c \succ d$ if and only if $c+e \succ d + e$ for
any $e$;
\item[A3] Whenever $c \succ d \succ 0$, there exists some integer $n$
such that $n d \succ c$;
\item[A4] Whenever $c \prec 0$ and $d \succ 0$, there exists some integer
$n$ such that $c+ n d \succ 0$.
\item[A5] For any $c,d$ where $c \succ d$, and for any $e$, there
exist integers $m,n$ such that $n c \succ m e \succ n d$.
\end{description}
It is easy to prove that $A1-A5$ are equivalent to the existence of a
value function \mc{V}, unique up to a multiplicative constant, such that
(a) $\mc{V}(c)>\mc{V}(d)$ iff $c \succ d$ and (b)
$\mc{V}(c+d)=\mc{V}(c)+\mc{V}(d)$.  (The proof that such a value function
implies $A1--A5$ is trivial. The essence of the proof of the converse
is:  pick an
arbitrary consequence $c \succ 0$, assign it value 1, and put
$\mc{V}(d)= n$ whenever $d \simeq n c$.  For negative values, put 
$\mc{V}(d)=-n$ whenever $d+n c \simeq 0$; for rational values, 
put $\mc{V}(d)=m/n$ whenever $n d \simeq m c.$  Postulates 3 and 4
ensure that all consequences have finite value; postulate 5 handles
irrational value.  The details of this proof are in the appendix.)

A2--A5, when we understand them as applying to consequences (or acts)
are all structure axioms, and most are pretty innocuous: A3 and A4 rule
out infinitely valuable consequences, and A5 rules out infinitesimally
different values.  A2, however, is a very substantive assumption which
isn't true for most of us: 
let $c$ be the consequence of becoming a billionaire; let
$d$ be the consequence of getting a wonderful, hassle-free cruise in
the South Pacific; let $e$ be the consequence of receiving $100,000$
euros.  For most of us (\ie., assuming that the reader isn't already a
billionaire), even a wonderful holiday is over-priced at $100,000$
euros, so $e \succ d$; however, a billionaire might very
well pay $100,000$ euros to guarantee a good time, so $e + c
\prec d+c.$

As such, the assumption that a preference ordering is additive is at best
an approximation, applicable to certain special circumstances (gambling with small
sums is an obvious example).  If we \emph{do} assume additivity, though,
we can state a set of axioms which imply the expected utility rule:
\begin{description}
\item[U0: Act availability]  The set \mc{A} of acts 
consists of all acts $(M,\mc{P})$, for some fixed $M\in\mc{M}$ and for
arbitrary functions
$\mc{P}:\mc{S}_M\rightarrow \mc{C}$.  (Since $M$ is fixed, when writing acts we
will drop the $M$ and identify acts with their payoff functions:
$(M,\mc{P})\equiv \mc{P}.$)
\item[U1: Preference ordering]  There exists a weak ordering $\succ$ on
\mc{A}  (this in turn defines a weak ordering on \mc{C}, by restricting 
$\succ$ to
constant $\mc{P}$ and identifying such $\mc{P}$ with their constant value.) 
\item[U2: Dominance] If $\mc{P}_1(s)\succeq \mc{P}_2(s)$ for all $s \in
\mc{S}_M$, then $\mc{P}_1 \succeq \mc{P}_2$.
\item[U3: Composition]  There is an operation $+$ of composition on
\mc{A} such that 
\be (M,\mc{P}_1)+(M,\mc{P}_2)=(M,\mc{P}_1+\mc{P}_2)\ee
(where addition on consequences is defined similarly to the weak
ordering on consequences, by restriction of $+$ to constant acts).
\item[U4: Act additivity]  The weak ordering $\succ$ on \mc{A} is
additive (that is, satisfies A1--A5) with respect to $+$.

\end{description}
From these axioms we can deduce that:
\begin{enumerate}
\item The order $\succ$ on the set \mc{C} of consequences defines a
value function \mc{V}, unique up to multiplication (this is just the
result proved above, of course).
\item To each event $s \in \mc{S}_M$ there exists some unique real number 
$\Pr (s)$
(between 0 and 1) such that $\mc{P}_1 \succ \mc{P}_2$ iff $\sum_s
\Pr(s) \mc{V}\cdot \mc{P}_1 (s) > \sum_s
\Pr(s) \mc{V}\cdot \mc{P}_2 (s)$. This is of course the expected-utility theorem.  
\end{enumerate}

Of the axioms:
\begin{itemize}
\item U0 is a structure axiom, concerning those acts which we can
consider.  Some such structure axiom will be used throughout the paper,
and its validity will not be discussed further.
\item U1 and U2 are axioms of pure rationality.  U1 is familiar; U2 is
the assertion that, if one act is guaranteed to give  consequences
as good as another act, then we should not prefer the second act to the first.
\item U3 and U4 encode additivity, which has been extended from
consequences to acts in order to ensure the well-definedness of the
probabilities.  To understand the justification of additivity, think of
acts as the placing of bets on all the possible outcomes of 
some fixed chance event $M$: additivity
says that our preference on possible bets doesn't depend on what bets we
have already placed, which again is implausible in general but may be a
reasonable approximation in some circumstances.
\end{itemize}

\subsection{Defining utility from probability: the von Neumann-
Morgenstern approach}\label{vnmsect}

Though the derivation of EUT from additivity gives some insight into
decision theory, additivity is an unreasonably strong assumption.  We
therefore consider the alternative strategy for deriving EUT sketched in section
\ref{exputilsect}, which begins by proving a representation theorem for
probability and then goes on to define the value function.  In the next section
we discuss that representation theorem; in this
section we address the question of exactly what is needed to derive
quantitative value from quantitative probability.

This question was originally answered by \citeN{VN-morgenstern}.  They
proved a result which, in modern terminology (I follow \citeNP{fishburn})
may be expressed as follows:
\begin{quote}
Let $c_1, \ldots, c_n$ be a set of possible consequences (intended to be
regarded as jointly exhaustive).  Define a
\emph{gamble} $F$ as an act  which will lead to consequence $c_i$ with
probability $\Pr_F(i)$, with $\sum_i \Pr_F(i)=1$.  Define convex sums of 
gambles as 
follows: if $F$ and $G$ are
gambles and $0 \leq\lambda \leq 1$, then $\lambda F + (1-\lambda)G$ is
the gamble which assigns probability $\lambda \Pr_F(i) + (1-
\lambda)\Pr_G(i)$ to consequence $c_i$.
Then assume:
\begin{description}
\item[VNM0: Gamble availability] \mc{A} is some set of gambles, which contains 
all constant gambles (that is, gambles $F$ for which $\Pr_F(i)=1$ for some $i$) 
and is closed under convex sums.
\item[VNM1: Transitive preferences] There exists a weak order $\succ$ on the 
set \mc{A}.
\item[VNM2: Sure-thing principle] If $F \succ G$ and $0<\lambda <1$ then for 
any gamble $H$, $\lambda F + (1-\lambda)H
\succ \lambda G + (1-\lambda)H$.
\item[VNM3: Gamble structure axiom] If $F \succ G$ and $G \succ H$ then there 
exist some $\alpha, \beta \in (0,1)$ such that $\alpha F + (1-\alpha)H \succ G$ 
and  $G \succ \beta F + (1-\beta)H$.
\end{description}

Then (von Neumann and Morgenstern showed) there exists a real function $\mc{V}$
on the set of consequences, uniquely given up to affine transformations and 
such that 
\begin{itemize}
\item $c_i \succ c_j$ iff $\mc{V}(c_i)> \mc{V}(c_j).$ (Here constant
gambles are identified with their values.)
\item If $F$ and $G$ are gambles, then $F\succ G$ iff $EU(F) \succ EU(G)$,
where
\be EU(F)=\sum_i \Pr_F(i) \mc{V}(c_i).\ee
\end{itemize}

\end{quote}

Of the assumptions above, VNM0 is a structure axiom analogous to U0
and VNM1 is the usual transitive-preferences assumption.
VNM2 is essentially a statement about
the meaning of probability, an example of a class of principles called
`Sure Thing Principles' by \citeN{savage};
in words, it could be stated as 

\begin{quote}``either $x$ or $y$ will occur.  Of the
two bets I could choose to make, I'm indifferent between them if $x$
occurs, and I'd prefer to have made Bet 1 if $y$ occurs;
therefore, I should go for Bet 1, as it's a sure thing that I'll either 
prefer that to Bet 2, or at least not mind which bet I made.''\end{quote}

Sure thing principles (and dominance assumptions, come to that) are more 
controversial than  at first they appear;
see \citeN{gardenfors} for a range of criticisms of them.  Nonetheless we will make
use of them without further analysis.\footnote{One reason for this is that
the most convincing objections to the Sure Thing principle concern
situations where we feel we have more knowledge about one situation than
about another; as we shall see, in analysing quantum probabilities we
shall assume throughout that we have perfect knowledge.}

The other assumption, VNM3, is another structure axiom: in
effect, it says that if one bet is preferred to another, then there will
exist some change to the probabilities involved in the bets which is so small that 
the preference order is unchanged.  This rules out the
possibility of infinitesimal value differences between consequences.

What is shown by the von Neumann-Morgenstern approach?  That if we
assume the probabilities known, assume a merely qualitative preference
order on gambles (with no need for compositions of acts or for a zero
consequence), and make some extremely reasonable-seeming
assumptions about how probabilities affect our preferences, then we can
deduce the EUT.  We now turn to the problem of deducing the
probabilities themselves.

\subsection{Defining probability: the Savage axioms}\label{savage}

The project of deriving quantitative probabilities from qualitative
preferences was carried out with great clarity by \citeN{savage}, and we
will sketch his analysis here.  He first defines a notion of qualitative
probability by the method given in section \ref{exputilsect}: that is,
event $A$ is more probable than event $B$ ($A \succ B$) just if, for any
consequences $c$ and $d$ with $c \succ d$, the agent would prefer ($c$
if $A$ obtains and $d$ otherwise) to ($c$ if $B$ obtains and $d$
otherwise). Then he gives an axiom system which allows him to turn this
qualitative notion of `more probable than' into a quantitative
probability measure.
  
Savage's axiom system is extremely powerful, and hence inevitably rather 
complex:
it consists of the following (taken directly from \citeNP{savage}, with the 
tacit axiom
S0 made explicit, following \citeNP{joyce}.)

\begin{description}
\item[S0: Act availability]   The set \mc{A} of acts 
consists of all acts $(M,\mc{P})$, for some fixed $M\in\mc{M}$ and for
arbitrary functions
$\mc{P}:\mc{S}_M\rightarrow \mc{C}$.  (As such, when writing acts we
will drop the $M$ and identify acts with their payoff functions:
$(M,\mc{P})\equiv \mc{P}.$)
\item[\textbf{S1: Preference ordering}] There exists a weak ordering
$\succ$ on the set of acts.  (The ordering defines, by an obvious restriction, 
an ordering on consequences: define $\mc{P}_c$ and $\mc{P}_d$ as \emph{constant
acts}, with $\mc{P}_c(x)=c$ and $\mc{P}_d(x)=d$ for all $x$; then define
$c \succ d$ iff $\mc{P}_c \succ \mc{P}_d$. This also allows us to compare acts
and consequences: act $F_1$ is preferred to consequence $c$ iff it is preferred
to the constant act $\mc{P}_c$. )
\item[\textbf{S2: Non-triviality}] Not all consequences are equally
valuable: that is, there are at least 2, $c$ and $d$ say, such that
$c \succ d$.
\item[\textbf{S3: Sure-Thing Principle}]  If two acts $F_1$ and $F_2$ agree with each
other on some event $C \subseteq \mc{S}$, then whether or not $F_1 \succ F_2$ 
is independent of the actual form of $F_1$ (and thus $F_2$) on $C$.
\item[\textbf{S4: Dominance}]  
If $\{C_i\}$ is some partition of $\mc{S}_M$ into finitely
many events, if $F_1$ and $F_2$ are acts with $F_1=c_{1,i}$ on restriction to $C_i$ and
$F_2=c_{2,i}$ on restriction to $C_i$, and if $c_{1,i} \succeq c_{2,i}$ for all $i$,
then $F_1 \succeq f_2$.  If in addition there is at least one $i$ with $C_i$ non-null  
such that $c_{1,i}\succ c_{2,i}$, then $F_1 \succ F_2$.
\item[\textbf{S5: Probability}]  Given any two events $C$ and $D$, then
either $C$ is more probable than $D$, or it is less probable, or the two are equiprobable.
\item[\textbf{S6: Savage's Structure Axiom}] If $F_1$ and $F_2$ are acts with $F_1 
\succ f_2$, and $c$ is an arbitrary consequence, then there exists a partition of $\mc{S}_M$ into finitely many
events $\{C_i\}$ such that if either or both of $F_1$ and $F_2$ are modified to be 
equal to $c$ on an arbitrary $C_i$, the preference order is unchanged.
\item[\textbf{S7: Boundedness}] If $F_1$ and $F_2$ are acts such that
$F_1 \succ F_2(s)$ for every $s$, then $F_1 \succ F_2$ (and vice versa).
\end{description}
Most of these axioms are axioms of pure rationality: S1, S3 and S4 we have 
essentially met before, S2 guarantees that agents have \emph{some} preferences 
(and hence that their beliefs about probability can be manifested in their choice 
of bets), and S7 says that if one act is preferred to all the consequences of 
another act, it is preferred to that other act \textit{simpliciter}.
S5 is rather more substantive: it says that the probability of a given
event is independent of the act performed, $\Pr(A|\mc{P})=\Pr(A)$.  (This
is why we have to restrict, in S0, to acts within a fixed chance setup
$M$.)
S0 is a structure axiom equivalent to U0 or VNM0.  S6 is a new and very
substantive structure axiom: it effectively says that:
\begin{enumerate}
\item events are continuously divisible, so that any event can be broken
into $N$ sub-events each of which are equally probable.
\item No event has infinitesimal probability.
\item No consequences are infinitely valuable.
\item No consequences differ in value only infinitesimally.
\end{enumerate}

Savage's derivation of the EUT then proceeds as follows.  Firstly it is
shown (by virtue of S0-S5) that `more probable than' is a weak ordering
on the set of events.  S6 is then used to establish that there is one
and only one probability measure on the events which is compatible with
this weak ordering: the method, in essence, is to break $\mc{S}_M$ into
a very large group $N$ of equiprobable events $B_1, \ldots, B_N$, each of which 
must be assigned probability $1/N$, and approximate each event by some finite 
union of these $B_N$.  If an event contains $M$ of the $B_M$ it must
have probability greater than $M/N$; if it is contained within the union
of $M'$ of them then it must have probability less than $M'/N$.
Iterating this process for successively larger $N$ gives, in the limit,
unique probabilities for each event.

It is perhaps worth recalling the fact that this `unique' probability
measure is unique \emph{only} given an agent's actual preferences
between acts. Different agents might well assign different probabilities
to an event; indeed, quite arbitrary assignments of probabilities are
compatible with the Savage axioms.

With probability defined, all Savage needs to do is establish that 
VNM0--3 apply; this done he can apply the von Neumann-Morgenstern result
to define values and prove the EUT.  This is straightforward: VNM0, VNM1 and VNM2
are easy consequences of S0, S1 and S3 respectively (though in general
S6 is needed to prove this) and the structure axiom VNM3 follows
directly from S6.
 
S7 has a rather special place in Savage's scheme: it is relevant only
if we wish to consider acts with countably infinitely many consequences
(all the development so far, including the VNM axioms, has been in terms of finite-consequence
games).  From S7 it can be proved that \mc{V} is bounded, and as a
consequence of this that expected utility still represents preferences
in the countably-infinite case.

\subsection{Objective chance}\label{objectivechance}

Technically speaking, the program outlined above seems in reasonably
good health: it is possible to quibble about the structure axioms or some
of the principles of rationality, but in general the strategy of
deriving probabilities from preferences seems workable.  However, it is
less clear that the purely subjective notion of probability which
emerges is a satisfactory analysis of probability.  In particular, it
seems incompatible with the highly objective status played by
probability in science in general, and physics in particular.  
Whilst it is coherent to advocate the abandonment of objective
probabilities, it seems implausible: it commits one to believing, for
instance, that the predicted decay rate of radioisotopes is purely a matter of
belief.  

In the face of this problem, it has often been argued
\cite{lewisprobability,mellor} that subjective probabilities can coexist with some
other probabilities --- `objective chances', to use Lewis's term.  
But how are these objective chances to be understood?  \citeN{papineau}
has identified two ways in which objective chances are linked to 
non-probabilistic facts: an \emph{inferential link} whereby we use
observed frequencies to estimate objective chances, and a 
\emph{decision-theoretical link} whereby we regard it as rational to
base our choices on the objective chances.  In the presence of a
subjective notion of probability (such as that derivable from one of the
decision-theoretic strategies above) we can formalise this second link
via Lewis's \emph{Principal principle}, which says --- effectively ---
that the rational subjective probability of a chance event occurring
should be set 
equal to its objective chance.  From this principle we can derive the
inferential link also, at least in simple cases (see \cite[pp.\,106--108]{lewisprobability}).

But whether we have to justify Lewis's Principle, or Papineau's two
links directly, we owe some account of what sort of things objective
chances are, and why they should influence our behaviour the way that
they do.  The traditional approach has been to \emph{identify}
objective chances with frequencies, but this generally leads to circularity
(we wish it to be the case that it is \emph{very probable} that the
frequency is close to the objective chance).  Lewis has a more sophisticated
account \cite{lewis94} in which the objective chances are those given
by the laws which best fit the particular facts about the world (hence
both frequency considerations, and those based on symmetry or
simplicity, get to contribute), though he acknowledges that he can see
only ``dimly'' how these facts can constrain rational action.

What is not acceptable (to Lewis, nor I believe to anyone) is simply to introduce 
chance as a primitive concept and just stipulate that it is linked to rational 
action in the required way.   Lewis again:
\begin{quote}
Don't call any alleged feature of reality ``chance'' unless you've
already shown that you have something, knowledge of which could
constrain rational credence \ldots I don't begin to see, for instance,
how knowledge that two universals stand in a certain special relation
$N^*$ could constrain rational credence about the future
coinstantiation of those universals.
Unless, of course, you can can convince me first that this special
relation is a chancemaking relation: that the fact that $N^*(J,K)$ makes
it so, for instance, that each $J$ has a $50\%$ chance of being $K$.
But you can't just tell me so.  You have to show me.  Only if I already
see --- dimly will do! --- how knowing the fact that $N^*(J,K)$ should
make me believe to degree $50\%$ that the next $J$ will be a $K$ will I
agree that the $N^*$ relation deserves the name of chancemaker that you
have given it. \cite[pp.\,484--485]{lewis94}
\end{quote}

If Lewis were wrong, of course --- if it were legitimate just to posit
that certain quantities were objective chances --- then there would be
\emph{no} probability problem --- and, in particular, as I will argue in section \ref{dectoQM}, 
there is no probability problem in the Everett interpretation.  Given the
legitimacy (as argued for by Saunders; see section \ref{SIviewpoint}) 
of regarding quantum
branching as subjectively indeterministic, we could just stipulate that
some Quantum Principal Principle held, linking rational belief
\textit{ex hypothesi} to mod-squared amplitude.  

Papineau, in fact, turns this argument around: he argues that although
it is unsatisfactory merely to stipulate that mod-squared amplitude is
objective chance, no other account of objective chance does better!  So
it is changing the goal-posts to claim that there is a probability
problem in Everett but not in classical physics: the two are on equal
footing.

However, there is a more positive way to see the situation: \emph{if} it
could be argued that quantum probabilities really should be treated as
probabilities --- if, that is, some EUT could be derived for quantum
mechanics, with mod-squared amplitude in place of subjective probability --- then
not only would the problem of probability in the Everett interpretation
be resolved, but progress would have been made on a wider philosophical problem 
with the very notion of objective chance.  This will be one goal of the
remainder of the paper.

\section{Applying decision theory to quantum events}\label{dectoQM}

To achieve this goal, we will need to transfer much of the machinery of
classical decision theory to a quantum context.  Before this can be
done, though, we need to understand how a theory designed to deal with
uncertainty between possible outcomes deals with quantum events where
all the outcomes are realised: this will be the task of this section.

\subsection{Quantum branching events}

Our assumptions about the quantum universe can be summarised as follows:
\begin{itemize}
\item The Universe can be adequately modelled, at least for the purposes of analysing 
rational decision-making, by assuming it to be a closed 
system, described by a pure Hilbert-space state which evolves at all times
according to the unitary dynamics of quantum theory.
\item In the Hilbert space of the theory, a \textit{de facto} preferred
basis is defined by decoherence theory (see \citeN{zurek} for 
introduction to this topic, \citeN{zurekreview} for a technical review, and
Wallace \citeyear{wallacestructure,wallaceworlds} for a more philosophical
analysis.)  This basis is approximately diagonalized in both position
and momentum, and allows the Universe to be interpreted, to a very good
approximation, as a branching system of approximately classical regions.
\item Certain subsystems of the Universe can be interpreted as
observers, who exist within the approximately classical regions defined
by the preferred basis.  
\end{itemize}

How are we to understand the branching events in such a theory?  I have
argued elsewhere \cite{wallacestructure} that they can be understood,
literally, as replacement of one classical world with several --- so
that in the Schr\"{o}dinger Cat experiment, for instance, after the
splitting there is a part of the quantum state which should be
understood as describing a world in which the cat is alive, and another
which describes a world in which it is dead.  This multiplication comes
about not as a consequence of adding extra, world-defining elements to
the quantum formalism, but as a consequence of an ontology of
macroscopic objects (suggested by \citeNP{realpatterns}) according to 
which they are treated as patterns in the underlying microphysics.

This account applies to human observers as much as to cats: such
an observer, upon measuring an indeterminate event, branches into
multiple observers with each observer seeing a different outcome.  Each
future observer is (initially) virtually a copy of the original
observer, bearing just those causal and structural relations to the
original that future selves bear to past selves in a non-branching
theory.  Since (arguably; see \citeN{parfit} for an extended defence)
the existence of such relations is all that there is to personal
identity, the post-branching observers can legitimately be understood as
future selves of the original observer, and he should care about them
just as he would his unique future self in the absence of branching.

\subsection{The objective-deterministic viewpoint on quantum decisions}

How should such an observer, if he is also a rational agent, choose
between various actions which cause branching?  We will suppose for
simplicity that the agent has effectively perfect knowledge of the
quantum state and the dynamics, so that for each choice of action he might make,
he can predict with certainty the resultant post-branching quantum state.  This might
suggest that we need only that part of decision theory dealing with
decision-making in the absence of uncertainty; call this the
\emph{objective-deterministic} (OD) viewpoint.

The decision theory available from the OD viewpoint is sparse indeed:
all we need is a transitive preference order over possible quantum states, 
then we choose from the available acts that one which leads --- with
certainty, remember --- to the best available quantum state.  If we
adopt the Savage framework for decision-making, for instance, the only
axiom which survives is S1, which requires the preference ordering to be
transitive: all the others deal with situations involving uncertainty.

Clearly such a structure would not be rich enough to prove any
interesting results at all --- let alone establish the quantum
probability rule.  However, even if we cannot use the other Savage axioms 
we might hope to use close analogues of them.  Dominance, for instance
(S4) suggests the following quantum analogue:

\begin{description}
\item[OD version of Dominance:]
Suppose an agent is about to be split into copies $C_1, \ldots, C_N$,
and then rewarded according to one of two possible reward-schemes
$\mc{P}_1$, $\mc{P}_2$ (so that either for each $i$ the $i$th copy gets reward
$\mc{P}_1(C_i)$ for each $i$, or it gets $\mc{P}_2(C_i)$.)  
If for no $i$ will future copy $C_i$  prefer $\mc{P}_2(C_i)$ to $\mc{P}_1(C_i)$, 
then the agent should not prefer $\mc{P}_2$ to $\mc{P}_1$; if in addition there
is at least one $i$ for which $\mc{P}_1(C_i)$ is preferred by copy $C_i$
to $\mc{P}_2(C_i)$, then the agent should prefer $\mc{P}_1$ to $\mc{P}_2$.
\end{description}

By no stretch of the imagination can this be construed as a statement of
classical decision theory --- no such theory contains axioms dealing
specifically with agents who undergo fission!  This is not to say that 
such a statement cannot be defended as rational, but it would have to be
treated as rational in the same way that avoiding alligator pits is rational, 
as opposed to the way in which avoiding intransitive preferences is
rational. 

In this case, a rational justification  might go something like
this:
\begin{quote}
I have the same structural and causal relation to my future copies as I
have to my future self in the absence of branching; hence insofar as it
is rational to care about my unique future self's interest, it is
rational to care about the interests of my future copies.  If I choose
$\mc{P}_1$ over $\mc{P}_2$, no future copy will be worse off; hence
$\mc{P}_1$ is at least as good as $\mc{P}_2$.  If, further, at least one
of my future copies is better off under $\mc{P}_1$ than $\mc{P}_2$, then
I should choose $\mc{P}_1$.
\end{quote}

The problem with such justifications is that there are worryingly similar justifications
available for other assumptions which we must at all cost \emph{avoid}
making.  The obvious example would be the assumption that I should care
equally about all of my future selves irrespective of the amplitudes of
their respective branches
(after all, they are all equally \emph{me}, and none are cognizant of
the amplitudes of their, or other, branches).  Obviously this assumption would be
disastrous for any recovery of the quantum rules, but it is arguably
just as intuitive as the quantum analogue of
Dominance.\footnote{Anticipating the discussion of Section \ref{papdec},
I should note that there \emph{may} be a more principled defence available for the
application of the decision-theoretic axioms, even from the perspective
of the OD viewpoint: \emph{following} a measurement the results of which
I do not know, my lack of knowledge of the result is just ordinary
ignorance and I can apply decision theory with impunity; then, invoking
van Fraassen's Reflection Principle, I can move those probabilities
backwards in time to the moment \emph{before} the measurement. However,
I do not think this approach (which emerged in conversation with Hilary
Greaves, to whom I am indebted here) can deal adequately with the issues
to be raised in Sections \ref{abandonOD} and \ref{justMN}; as such, I
shall not discuss it further.}

\subsection{Abandoning the OD viewpoint}\label{abandonOD}

The problem above is arguably not the worst for the OD viewpoint.
There is an important sense in which it asks the wrong question.
It asks, in effect, how we should act given the known truth of
the Everett interpretation --- akin to asking how we should act if a
matter-transporter were to be invented tomorrow.  But the real question
should be: should we believe the Everett interpretation in the first
place?

In this case, it might be suggested that rationality considerations are
just irrelevant: if the Everett interpretation is the best available
explanation of our empirical situation, adopt it; if not, don't.  This
is too quick, though, as we can see by analysing the notion of 
explanation in this context.

The sort of evidence that the Everett
interpretation is being asked to explain is along the lines of, ``running
such-and-such experiment $N$ times led to $M$ clicks''; what sort of
explanation could be given here?
The simplest sort of explanation to analyse would be one couched in terms of 
a deterministic (and non-branching) theory and a known initial state: if the initial state
is known with certainty and it predicts the experimental results which
in fact occur, then (\textit{ceteris paribus}, \ie modulo considerations of
simplicity and so forth) it is explanatory of
them.  This is the sense in which Newtonian gravity explains planetary
motion, for instance.

But the \textit{prima facie} randomness of quantum-mechanical empirical
data suggests (again, if we ignore the possibility of branching) that 
either the initial state, or the dynamical evolution,
cannot be known with certainty: either we have some probability
distribution over initial states, or the dynamics is intrinsically
indeterministic.  In either case, the theory is explanatory if it
predicts that the experimental results are typical of the sort of
results which would occur with high probability.\footnote{It would be conceptually
simpler to regard a theory as explanatory, \textit{ceteris paribus}, iff it predicts 
that the experimental results \emph{themselves} have high probability.
But this is inadequate: any given sequence of quantum events (or indeed
coin-tosses) will usually have very low probability.  It is classes of
such sequences (such as the class of all those with frequencies in a
given range) which are assigned high probability by explanatory
theories.  While the question of how to analyse the notion of a ``typical''
result  is rather subtle, the Everett interpretation does not introduce
any new problems to the analysis; I shall therefore not discuss it
further.}\label{footnote}

But if we wish to make a decision-theoretic analysis of the concept of
probability, then statements about probability must be understandable in terms of 
rational action (either directly, or through some Lewis-style Principal Principle).
The connection, qualitatively speaking, is that it is rational to act as
though events of very high probability will actually occur, and this
suggests an account of explanation directly in terms of rationality:

\begin{quote}
A theory \mc{T} is explanatory of empirical data \mc{D}  (\textit{ceteris
paribus}) if, had I believed \mc{T} before collecting the data, it would
have been rational for me to expect \mc{D} (or other data of which
\mc{D} is typical).
\end{quote}

This account has the advantage of applying both to deterministic and
indeterministic theories, and both to situations where we have perfect
knowledge of the initial state and situations where we do not.   Can it
be applied to the Everett interpretation, though?  There seems no reason
why not --- provided that we can make sense of the notion of rationality
independently of the particular theories under consideration.  If we
allow a theory to set its own standards of rationality, then the account
collapses into circularity: consider, for instance, the theory that the
world is going to end tomorrow \emph{and that it is rational to believe
that the world is going to end tomorrow}.

This view of rationality does not commit us to the view that
rationality itself is not a legitimate subject of study and, where
necessary, revision.  Certainly we are physical systems, only
imperfectly rational \cite[pp.\,94--99]{dennettintentional}; certainly 
we can consider and discuss what ideally rational behaviour is (as shown
by the lively debates
over (for instance) the sure-thing principle and Newcomb's problem
 in the literature of decision theory).  As we have
learned from \citeN{wordandobject}, every part of our conceptual scheme is in principle
open to revision; however, this has to be done in a piecemeal way,
keeping most of the scheme fixed whilst certain parts are varied.
(Recall Quine's frequent reference to Neurath's metaphor of science as a
boat, and his extension of the metaphor to philosophy: we must rebuild
the boat, but we have to remain afloat in it whilst we do so; we can
rebuild it plank by plank, but at any stage most of the planks must be left
alone.)  In the case of theory change in physics, to vary both our
theory and the rational standards by which theories are judged would be
to go too far.

So, to analyse the Everett interpretation's ability to recover the
probability rule --- and so explain our empirical data --- we need a
viewpoint on rationality which is not itself radically altered by the
conceptual shift from single-world to branching-world physics.  To
achieve this viewpoint we must abandon the OD viewpoint and seek a new
one.

\subsection{The subjective-indeterministic viewpoint}\label{SIviewpoint}

To find our new perspective, we return to the situation of an agent
undergoing splitting.  From our God's-eye view we can regard the
splitting as a deterministic multiplication of the number of observers,
but how should the agent himself view it: if he is awaiting splitting,
what should he \emph{expect}?

\citeN{saundersprobability} has argued persuasively that the agent should treat the
splitting as subjectively indeterministic: he should expect to become
one future copy or another but not both, and he should be uncertain as
to which he will become.  His argument proceeds by analogy with classical splitting,
such as that which would result from a Star Trek matter transporter or an operation
in which my brain is split in two.  It may be summarised as follows: in
ordinary, non-branching situations, the fact that I expect to become my
future self supervenes on the fact that my future self has the right
causal and structural relations to my current self so as to \emph{count}
as my future self.  What, then, should I expect when I have two or more
such future selves?  There are only three logical possibilities:
\begin{enumerate}
\item I should expect to become both future selves.
\item I should expect to become one or the other future self.
\item I should expect nothing: oblivion.
\end{enumerate}

Of these, (3) seems absurd: the existence of either future self would
guarantee my future existence, so how can the existence of \emph{more}
such selves be treated as death?  (1) is at least coherent --- we could
imagine some telepathic link between the two selves --- but on any
remotely materialist account of the mind this link will have to
supervene on some physical interaction between the two copies which is
not in fact present.  This leaves (2) as the only option, and in the
absence of some strong criterion as to which copy to regard as
``really'' me, I will have to treat it (subjectively) as
indeterministic.

(In understanding Saunders' argument, it is important to realise that
there are no further physical facts to discover about expectations which could 
decide between (1-3): on the contrary, \textit{ex hypothesi} all the physical facts
are known.  Rather, we are regarding expectation as a higher-level
concept supervenient on the physical facts --- closely related 
to our intuitive idea of the passage of time --- and
asking how that concept applies to a novel but physically possible
situation). 

Of course (argues Saunders) there is nothing particularly important
about the fact that the splitting is classical; hence the argument
extends \textit{mutatis mutandis} to quantum branching, and implies that
agents should treat their own branching as a subjectively
indeterministic event.  We will call this the \emph{subjective-indeterministic} 
(or SI) viewpoint, in contrast with the OD viewpoint
which we have rejected for decision-theoretic purposes.\footnote{This
dichotomy of viewpoints --- the ``God's-eye'' and ``personal'' views ---
resembles that presented by \citeN{sudbery}; his motivation and
philosophy of mind, however, seem rather different, as is his resolution
of the probability problem.}

But if branching is subjectively indeterministic, the agent can apply
ordinary, classical decision theory without modification!  The whole
point of such decision theory is to analyse decision-making under
uncertainty, and from the SI viewpoint --- that is, from the viewpoint
 of the agent himself --- that is
exactly the exercise in which he is involved when he is choosing between
quantum-mechanical acts.  

The SI viewpoint, then, is exactly what we need to judge the explanatory 
adequacy of the Everett interpretation: it allows us to transfer the
axioms of decision theory directly across to the quantum-mechanical
case.  In the next three sections we will show how this process is
sufficient to establish the quantum probability rule.

\section{Quantum games and measurement neutrality}\label{quantgames}

We have seen how decision theory offers two distinct routes to the
Expected-Utility representation theorem: through additivity of value
(section \ref{add}), and through a representation theorem for
probabilities (section \ref{savage}).  We will shortly see that both
methods can be adapted straightforwardly to quantum mechanics (with
Deutsch's proof effectively a form of the first method).

However, before this can be done we need to make sense of what,
precisely, are the quantum acts which we are considering.  In this
section, then, we will define a certain large sub-class of quantum-mechanical acts, 
called ``quantum games'', and consider some properties of that class.  This will be
common ground for the various derivations of expected utility presented
in later sections, and provides the framework whereby the subjective
probabilities given by decision theory are constrained to equal the
probabilities predicted by quantum mechanics.

\subsection{Quantum measurements}\label{QMdef}

In the Everett framework, a measurement is simply one physical process amongst
many, and will be modelled as follows: let $\mc{H}_s$ be the Hilbert space of some subsystem
of the Universe, and $\mc{H}_e$ be the Hilbert space of the measurement device; 
let \op{X} be a self-adjoint operator on $\mc{H}_s$, with discrete spectrum.  

Then a \emph{non-branching measurement} of \op{X} consists of:
\begin{enumerate}
\item Some state \ket{\mc{M}_0} of $\mc{H}_e$, to be interpreted as its
initial (pre-measurement) state; this state must be an element of the
preferred basis picked out by decoherence.
\item Some basis \ket{\lambda_a} of eigenstates of \op{X}, where
$\op{X}\ket{\lambda_a}=x_a \ket{\lambda_a}.$ (Since we allow for the
possibility of degeneracy, we may have $x_a =x_b$ even though $a \neq
b$.)
\item Some (orthogonal) set $\{\ket{\mc{M};x_a}\}$ of ``readout states'' of $\mc{H}_s\otimes \mc{H}_e$,
also elements of the decoherence basis, one for each state $\ket{\lambda_a}$. 
The states must physically display $x_a$, in
some way measurable by our observer (\egc, by the position of a needle).
\item Some dynamical process, triggered when the device is activated,
and defined by the rule
\be \tpk{\lambda_a}{\mc{M}_0} \longrightarrow \ket{\mc{M};x_a}\ee
(Since all dynamical processes in unitary quantum
mechanics have to be linear, this rule uniquely determines the process.)
\end{enumerate}

What justifies calling this a `measurement'?  The short answer is that
it is the standard definition; a more principled answer is that the point
of a measurement of \op{X} is to find the value of \op{X}, and that
whenever the value of \op{X} is definite, the measurement process will
successfully return that value.  (Of course, if the value of \op{X} is
not definite then the measurement process will lead to branching of the
device and the observer; but this is inevitable given linearity.)

The ``non-branching'' qualifier in the definition above refers to the
assumption that the measurement device does not undergo quantum branching when
\op{H} is prepared in one of the states \ket{\lambda_a}.  This is a
highly restrictive assumption, which we will need to lift.  We 
do so in the next definition. a \emph{general measurement} of \op{X}
consists of:
\begin{enumerate}
\item Some state \ket{\mc{M}_0} of $\mc{H}_e$, to be interpreted as its
initial (pre-measurement) state; this state must be an element of the
preferred basis picked out by decoherence.
\item Some basis \ket{\lambda_a} of eigenstates of \op{X}, where
$\op{X}\ket{\lambda_a}=x_a \ket{\lambda_a}.$
\item Some set $\{\ket{\mc{M};x_a; \alpha}\}$ of ``readout states'' of 
$\mc{H}_s\otimes \mc{H}_e$,
also elements of the decoherence basis, \emph{at least} one for each state \ket{\lambda_a} 
(the auxiliary label $\alpha$ serves to distinguish states associated with the same $\ket{\lambda_a}$).  
The states must physically display $x_a$, in
some way measurable by our observer (\egc, by the position of a needle).
\item Some dynamical process, triggered when the device is activated,
and defined by the rule
\be \tpk{\lambda_a}{\mc{M}_0} \longrightarrow \sum_\alpha \mu(\lambda_a;\alpha) 
\ket{\mc{M};x_a;\alpha}\ee
where the $\mu(\lambda_a;\alpha)$ are complex numbers
satisfying $\sum_\alpha |\mu(\lambda_a;\alpha)|^2=1$.
\end{enumerate}

In a general measurement, the measurement device (and thus the observer)
undergoes branching even when $\op{X}$ has definite value; however, the
observer can predict that, in such a case, all his/her future copies will correctly learn 
the value of \op{X}.  In practice, of course, most physical measurements
are unlikely to be non-branching.

We end with three comments on the definition of measurement:
\begin{enumerate}
\item We are not restricting our attention to so-called 
``non-disturbing'' measurements, in which
$\ket{\mc{M};x_a}=\tpk{\lambda_a}{\mc{M}';x_a}.$  In general
measurements will destroy or at least disrupt the system being measured,
and we allow for this possibility here.
\item In practice, the
Hilbert space $\mc{H}_e$ would probably have to be expanded to include
an indefinitely large portion of the surrounding environment, since the
latter will inevitably become entangled with the device.  We have also
not allowed for the possibility of the device having
a large number of possible initialisation states. But neither of these
idealisations appear to have any practical consequences for Deutsch's argument,
nor for the rest of this paper.
\item Since a readout state's labelling is a matter not only of physical
facts about that state but also of the labelling conventions used by the
observer, there is no physical difference between a measurement of
\op{X} and one of $f(\op{X})$, where $f$ is an arbitrary one-to-one
function on the spectrum of \op{X}: a measurement of $f(\op{X})$ may be
interpreted simply as a measurement of $\op{X}$, using a different
labelling convention.  More accurately, there \emph{is} a physical
difference, but it resides in the brain state of the observer (which
presumably encodes the labelling convention in some way) and not in the
measurement device.
\end{enumerate}

To save on repetition, let us now define some general conventions for
measurement: we will generally use \op{X} for the operator being
measured, and denote its eigenstates by \ket{\lambda_a}; the eigenvalue
of \ket{\lambda_a} will be $x_a$.  (Recall that we allow for the possibility of
degenerate \op{X}, so that may have $x_a=x_b$ even though $a \neq b$.)

\subsection{Defining quantum games}\label{defininggames}

What form does the decision problem take for a quantum agent?
Our (mildly stylised) description of the problem in classical decision
theory involved an agent who was confronted with some chance setup and
placed bets on the outcome.  This suggests an obvious quantum version:
our agent measures some quantum state, and receives a reward which
depends on the result of the measurement.
For simplicity, let us suppose that the agent has perfect knowledge of the
physical state  that the Universe will be in, post-branching.  Hence, he
knows what experiences all of his future copies will have, and what the
amplitudes are for each such experience.  Nonetheless (as discussed in 
section \ref{SIviewpoint}) the process is
\emph{subjectively} indeterministic for him: he expects to become one of the
possible future copies but does not know which.

Let us develop the details of this.  Suppose that there exists some
large class of quantum systems, for which we will need to assume the
following:
\begin{description}
\item[Q1] For each system in the class there exists at least
one device capable of measuring some discrete-spectrum observable of that system.
\item[Q2] For each (positive integral) $n$, there exists some system with
an $n$-dimensional Hilbert space and some way of measuring a 
\emph{non-degenerate} observable of that system. 
\item[Q3] For any system in the class, and any measurable observable \op{X} on that system,
it is possible to perform any unitary transformation \op{U} which
permutes the eigensubspaces of \op{X} (that is, any \op{U} such that
$\op{U}\op{X}\opad{U}=f(\op{X})\equiv \sum_a f(x_a) \proj{\lambda_a}{\lambda_a}$ 
for some real function $f$).   
\item[Q4] On any \emph{pair} of systems in the class, on which the
operators $\op{X}$ and $\op{Y}$ are respectively measurable, it is
possible to perform any joint unitary transformation which permutes the
eigensubspaces of $\op{X}\otimes\op{Y}$.
\item[Q5] Any system in the class can be prepared in any pure state. 
\end{description}

Thus for any system in the class, we can prepare it in an arbitrary
state, operate on it with a certain set of unitary operators, and
measure it in some basis.  Operations of this form will be the
\emph{chance setups} of the quantum decision problem, analogous to the
set \mc{M} of classical chance setups: we will denote the set of such
operations by $\mc{M}_Q$.

(Incidentally, the reader who finds the conditions Q1--Q5 too pedantic
for what is in any case supposed to be a set of operations which an agent can
contemplate performing and not necessarily a set of actually performable
operations, is welcome to replace them with
\begin{description}
\item[Q$\infty$] The class contains systems of every finite dimension, 
and any system in the class can be prepared in any state
and measured with respect to any observable, and arbitrary unitary
operations can be performed on systems and pairs of systems in the
class.)
\end{description}

Quantum acts (`games') then involve preparing a system, measuring it, and then
receiving some reward which is dependent on the result of the
measurement. It can be specified  by the following rather cumbersome
notation: an ordered quartet $\langle
\ket{\psi}, \op{X}, \mc{P}, \omega \rangle$, where:
\begin{itemize}
\item \ket{\psi} is a pure state in the Hilbert space \mc{H} of some system (for
notational simplicity the Hilbert space is not exhibited in the notation
for a game);
\item \op{X} is a measureable observable on \mc{H} with pure discrete spectrum;
\item \mc{P}, the \emph{payoff function}, is a function from the
spectrum of \op{X} to some set \mc{C} of consequences;
\item $\omega$ is the complete specification of a physical process by
which 
\begin{enumerate}
\item \mc{H} is prepared in state \ket{\psi};
\item a measurement of \op{X} is made on \mc{H};
\item In each branch where the measurement device shows $x_a$, 
the consequence $\mc{P}(x_a)$ is given to the observer (in that branch).
\end{enumerate}
\end{itemize}

We will suppose (the notion will be formalised later) that an agent's
preferences define a weak ordering on the space of games, so that an
agent prefers to play game $\mc{G}_1$ to game $\mc{G}_2$ iff $\mc{G}_1
\succ \mc{G}_2$. As usual, we will write $\mc{G}_1 \simeq \mc{G}_2$ just
in case neither game is preferred to the other. (Two such games will be
referred to as `value-equivalent', pending the introduction of a
quantitative value function.)

It will be completely crucial to later results that a game is merely
specified by such a quartet, not identified with it: games are certain
sorts of \emph{physical processes}, not mathematical objects, and it is
at this point open whether a given game can be specified by more than
one quartet. (This will in fact turn out to be the case, with important
consequences). Nonetheless, by construction each quartet identifies a
unique game, so we can consider preferences as holding between quartets
without ambiguity.

We will require that it is physically possible to realise games with 
arbitrary payoff function \mc{P}.  This in turn requires the set of
consequences to be specified quite abstractly, with much of the physical
details of how they are realised contained within the $\omega$ (this is,
however, equally true for classical decision theory).

Note that whether or not a given physical process $\omega$ realises a
given measurement involves counterfactuals: to say that a device is a
measuring device for \op{X} is to say that it would fulfil our
requirements for such a device, \emph{whatever} state was inserted into
it.  There is no counterfactual element to realising a payoff, however:
payoffs need be given only in those branches which are actually in the
superposition, so if with certainty eigenvalue $x$ will not be recorded,
then the value of $\mc{P}(x)$ may be changed completely arbitrarily
without affecting the physical situation.

Deutsch's own notation (among other simplifications) made no
reference to the physical contingencies specified in $\omega$, tacitly
assuming them to be irrelevant.  We can make this assumption explicit as:

\begin{quote}
\textbf{Measurement neutrality:} Rational agents are
indifferent to 
the physical details involved in specifying a quantum game; that is,
$\langle\ket{\psi},\op{X},\mc{P},\omega_1\rangle \simeq 
\langle\ket{\psi},\op{X},\mc{P},\omega_2 \rangle$
for any $\omega_1$, $\omega_2$.
\end{quote}

If measurement neutrality holds, we can omit $\omega$ from the
specification of a game .  Measurement
neutrality is nowhere stated explicitly by Deutsch, but it is tacit in his
notation and central to his proof, as will be seen.

The set of quantum games, $\mc{A}_Q$, can now be defined:
\begin{quote} $\mc{A}_Q$ is the set of all physical processes labelled
by ordered triples $\langle \ket{\psi}, \op{X}, \mc{P}\rangle$ where
\ket{\psi} is a state of a preparable system, \op{X} is a measurable
observable of that system, and \mc{P} is an arbitrary payoff function on
the spectrum of \op{X}.
\end{quote}

In understanding this definition it is important to note that $\mc{A}_Q$
is a set of physical processes, not a set of ordered triples.  Just as
one triple may be realised in many ways, so one and the same physical
process may realise many different triples.  This fact is absolutely
crucial to Deutsch's proof, as  section \ref{ETsect} will show.

\subsection{Justifying measurement neutrality}\label{justMN}

At first sight there is scarcely any need to justify an assumption as
obvious as measurement neutrality: who cares exactly \emph{how} a
measurement device works, provided that it works?  What justifies
this instinctive response is presumably something like this:
let  A and B be possible
measurement devices for some observable $\op{X}$, and for each
eigenvalue $x_a$ of \op{X} let the agent be indifferent between the 
$x_a$-readout states of A and those of B.  Then if the agent is currently planning to use
device A, he can reason, ``Suppose I get an arbitrary result $x_a$.  Had I used device B I 
would still have got result $x_a$, and would not care about the difference caused in
the readout state by changing devices; therefore, I should be
indifferent about swapping to device B.''

The only problem with this account is that it assumes that this sort of
counterfactual reasoning is legitimate in the face of (subjective)
indeterminism, and this is at best questionable (see, \egc, \citeN{redhead} for
a discussion, albeit not in the context of the Everett interpretation).

For a defence secure against this objection, consider how the
traditional Dirac-von Neumann description of quantum mechanics treats
measurement.  In that account, a measurement device essentially does two
things.  When confronted with an eigenstate of the observable being
measured, it reliably evolves into a state which displays the associated
eigenvalue.  In addition, though, when confronted with a superposition
of eigenstates it causes wave-function collapse onto one of the
eigenstates (after which the device can be seen as reliably evolving
into a readout state, as above).  

In the Dirac-von Neumann description, it is rather mysterious why a
measurement device induces collapse of the wave-function.  One has the
impression that some mysterious power of the device, over and above its
properties as a reliable detector of eigenstates, induces the collapse,
and hence it is \textit{prima facie} possible that this power might
affect the probabilities of collapse (and thus that they might vary from
device to device) --- this would, of course, violate measurement neutrality.  
That this is not the case, and that the probabilities 
associated with the collapse are dependent only
upon the state which collapses (and indeed are equal to those stipulated
by the Born rule) is true by fiat in the Dirac-von Neumann description.

It is a strength of the Everett interpretation (at least as seen from
the subjective-indeterministic viewpoint) that it recovers the subjective validity of the
Dirac-von Neumann description: once decoherence (and thus branching) occurs, subjectively
there has been wave-function collapse.   Furthermore there is no
``mysterious power'' of the measurement device involved: measurement devices
by their nature amplify the superposition of eigenstates in the state to
be measured up to macroscopic levels, causing decoherence, and this in
turn leads to subjective collapse.

But this being the case, there is no rational justification for denying
measurement neutrality.  For the property of magnifying superpositions
to macroscopic scales is one which all measurement devices possess
equally, by definition --- so if this is the only property of the
devices relevant to collapse (after which the system is subjectively deterministic, and so 
differences between measurement devices are irrelevant) then no other
properties can be relevant to a rational allocation of probabilities.
The only relevant properties must be the state being measured, and the
particular superposition which is magnified to macroscopic scales ---
that is, the state being measured, and the observable being measured on it.


\subsection{Physical equivalence of different games}\label{ETsect}

The following two axioms will be common to all the quantum decision
theories which we will present:
\begin{description}
\item[X0: Act availability] The set of acts (\iec, games) is $\mc{A}_Q$,
as defined at the end of section \ref{defininggames}.
\item[X1: Transitive preferences] There exists a weak ordering $\succ$ on $\mc{A}_Q$ 
which satisfies measurement neutrality (that is, $F \simeq G$ whenever
acts $F$ and $G$ are described by the same triple $\langle
\ket{\psi},\op{X}, \mc{P}\rangle$).
\end{description}

Even with this minimal amount of decision theory,
it is already possible to derive one important set of results used in Deutsch's proof.
These results involve the realisation
that certain quantum games, described by different labels in our notation (that
is, with different choices of \ket{\psi}, \op{X} and \mc{P}) are in
reality the same game. 

The essential idea used in all of these proofs is this:
measurement neutrality asserts that the correct description of a game
\textit{qua} game is given by the state, the measured operator and the
payoff, and that the remaining physical details are irrelevant.  But
this is not the right way to carve up the space of games \textit{qua}
physical systems: one game can be realised in many ways, but also one
and the same physical process may be understood as realising many
different games.  By going back and forth between the two sorts of
indifference implied by measurement neutrality and by existence of multiple labels for the
same game (implied by the physics of
playing the game), we are able to prove our equivalences.

We begin with the following result:

\begin{quote} 
\textbf{Payoff Equivalence Theorem (PET):} 
Let \mc{P} be a payoff scheme for \op{X}, and let $f$ be any function from the spectrum of \op{X} 
to the reals satisfying
\be \label{eqn8} f(x_a)=f(x_b) \longrightarrow \mc{P}(x_a)=\mc{P}(x_b).\ee 
(Hence the function $\mc{P}\cdot f^{-1}$ is well-defined even though $f$
may be non-invertible.)
Then the games $\langle \ket{\psi}, \op{X}, \mc{P} \rangle$ and $\langle \ket{\psi},
f(\op{X}), \mc{P} \cdot f^{-1} \rangle$ can be realised by the same
physical process; they therefore have the same value.
\end{quote}

\noindent \textbf{Proof:} Recall that our definition of a measurement process
involves a set of states \ket{\mc{M};x_a} of the decoherence-preferred basis, 
which are understood as readout states --- and that the rule associating
an eigenvalue $x_a$ with a readout state \ket{\mc{M};x_a} is just a
matter of convention.  Change this convention, then: regard
\ket{\mc{M};x_a} as displaying $f(x_a)$ --- but also change the payoff
scheme: replace a payoff $\mc{P}(f(x_a))$ upon getting result $f(x_a)$
with a payoff $(\mc{P}\cdot f^{-1})(f(x_a))\equiv \mc{P}(x_a)$.  These
two changes replace the game $\langle \ket{\psi}, \op{X}, \mc{P} \rangle$ with $\langle \ket{\psi},
f(\op{X}), \mc{P} \cdot f^{-1} \rangle$ with $f$ satisfying (\ref{eqn8}) --- 
but no physical change at all has occurred, just a change of labelling convention.  

But the value function does not assign values to triples $\langle\ket{\psi},
\op{X}, \mc{P}\rangle$ --- it assigns them to acts, construed as
physical sequences of events.  If two ``different'' games can be
realised by the same sequence of events, then, they are really the same
game, and should be assigned the same value.  Measurement neutrality
then tells us that all other realisations of the same game --- that is,
all other quartets $\langle \ket{\psi}, \op{X}, \mc{P}, \omega \rangle$
and $\langle \ket{\psi}, f(\op{X}), \mc{P}\cdot f^{-1}, \omega'
\rangle$ --- are also value-equivalent. $\Box$

A similar physical equivalence holds between transformations of the
state and of the operator to be measured. 

\begin{quote}
\textbf{Measurement Equivalence Theorem (MET):} 
\begin{enumerate}
\item Let \op{U} be any unitary operator which permutes, possibly trivially, the
eigensubspaces of \op{X}: \ie
$\op{X}\op{U}\ket{\lambda_a}=\pi(x_a)\op{U}\ket{\lambda_a}$,
where $\pi$ is some permutation of the spectrum of \op{X}.
Then the games $\langle \ket{\psi},\op{X},\mc{P} \rangle$
and $\langle \op{U} \ket{\psi}, \op{U}\op{X}\opad{U},\mc{P} \rangle
\equiv \langle \ket{\psi}, \pi^{-1}(\op{X}), \mc{P} \rangle$
have the same value.
\item In particular, suppose \op{X} is nondegenerate and let $f$ be a
permutation of its spectrum: $f (x_a) \equiv x_{\pi(a)}.$  Define
$\op{U}_f$ by $\op{U}_f \ket{\lambda_a}=\ket{\lambda_{\pi(a)}}.$ Then
the games 
$\langle \ket{\psi}, \op{X}, \mc{P} \rangle$ and $\langle \op{U}_f\ket{\psi},
f^{-1}(\op{X}),\mc{P}\rangle$ have the same value.
\end{enumerate}
\end{quote}

\noindent \textbf{Proof:} (2) is an immediate corollary of (1), which we prove as follows.  (Note
that the realisability of the unitary transformations in (1) and (2)
follows from assumption Q3.)
Let the distinct eigenvalues of \op{X} be $x_1, \ldots x_M$; let $d(i)$
be the dimension of the $x_i$-eigensubspace of \op{X}.  We change our labelling for
eigenvectors, denoting them by $\ket{x_i; j}$, where $j$ is a label
ranging from $1$ to $d(x_i)$.\footnote{We implicitly
restrict to finite-dimensional Hilbert spaces, but the generalisation is
trivial provided that \op{X} has pure discrete spectrum: just replace the finite set 
of indices used to label degenerate 
eigenvectors with an infinite set.}  \op{U} carries this basis to
another eigenbasis of \op{X}, whose elements are similarly denoted
$\ket{*;x_i;j}$: $\op{U}\ket{x_i;j}=\ket{*;\pi(x_i);j}$.  (Note that the
unitarity of \op{U} forces $d(x_i)=d(\pi(x_i))$, so this is 
well-defined.)

Let us realise the game $\langle \op{U}\ket{\psi},
\op{U}\op{X}\opad{U},\mc{P} \rangle$ by the following process:
\begin{enumerate}
\item[1] Prepare the system in state $\ket{\psi}$, so that the overall
quantum state is 
\be\ket{\psi}\otimes\ket{\mc{M}_0}\equiv\left(\sum_i 
\sum_{j=1}^{d(x_i)}\alpha_{i,j} \ket{x_i;j}\right)\otimes 
\ket{\mc{M}_0}\ee
where \ket{\mc{M}_0} is the initial state of the measurement device.
\item[2a] Operate on the state to be measured with the operator $\op{U}$, changing the 
overall state into state 
\be \left(\sum_i\sum_{j=1}^{d(x_i)} \alpha_{i,j} \ket{*;\pi(x_i);j}\right) 
\otimes \ket{\mc{M}_0}.\ee
\item[2b] Measure $\pi^{-1}(\op{X})$ using the following dynamics:
\be \tpk{*;x_i;j}{\mc{M}_0}\longrightarrow\ket{\mc{M};\pi^{-1}(x_i);j}\ee
where for each $x_i$, the states $\ket{\mc{M};x_i;j}$ are a set of
readout states giving readout $x_i$.
(This fits our definition of a non-branching measurement process; the
generalisation to a branching process is trivial.)
\item[3] The final state is now 
\be \sum_i \sum_{j=1}^{d(x_i)} \alpha_{i,j} \ket{\mc{M};x_i;j}.\ee  
In the branches in which result
$x_i$ is recorded, give a reward of value $\mc{P}(x_i)$.
\end{enumerate}

This is indeed a realisation of the game: in steps $1$ and $2a$ the
state $\op{U} \ket{\psi}$ is prepared, in step $2b$ the operator $\pi^{-1}(\op{X})$ is  
measured, and in step $3$ the payoff is made.
But now suppose that we avert our eyes from the dynamical details of
steps $2a$ and $2b$, considering them to be a black box.  Then the
effect of this black box is simply to carry out the transformation
\be \left(\sum_i \sum_{j=1}^{d(x_i)} \alpha_{i,j} \ket{x_i;j}\right)\otimes 
\ket{\mc{M}_0} \longrightarrow \sum_i \sum_{j=1}^{d(x_i)} \alpha_{i,j} \ket{\mc{M};x_i;j}.\ee
This transformation, though, fits the definition for a measurement of
\op{X}, and we have an alternative description of the same physical
events: in step $1$ the state \ket{\psi} is prepared, in steps $2_a$ and
$2_b$ the operator \op{X} is measured, and in step 3 payoff is made.
This is a realisation of the game $\langle \ket{\psi}, \op{X},\mc{P} \rangle$ 
--- so again we have two different games realised by the same physical
process and thus being assigned the same value. $\Box$

Two important (and immediate) corollaries of the MET concern the role of symmetry.

\begin{quote}
\textbf{Operator Symmetry Principle:}
Suppose that $\op{U}$ is defined as for part (1) of the MET, and that
$\op{U}\op{X}\opad{U}=\op{X}$ (equivalently, suppose that the
permutation $\pi$ is trivial).  Then 
$\langle\op{U}\ket{\psi},\op{X},\mc{P}\rangle$ and 
$\langle\ket{\psi},\op{X},\mc{P}\rangle$ have the same value.
\end{quote}
\begin{quote}
\textbf{State Symmetry Principle:}
Suppose that \op{X} is non-degenerate, that $\op{U}_f$ is defined as for 
part (2) of the MET, and that 
\ket{\psi} is invariant under the action of $\op{U}_f$; that is, that $\op{U}_f
\ket{\psi}=\ket{\psi}$.  Then  
$\langle\ket{\psi},\op{X},\mc{P}\rangle$ and
$\langle\ket{\psi},f(\op{X}),\mc{P}\rangle$ have the same value.
\end{quote}

In other words, the symmetries of a state being measured imply
relationships between the values of measuring different observables upon
that state, and vice versa.

The next equivalence theorem we prove is a corollary of the PET.

\begin{quote} 
\textbf{Operator equivalence theorem (OET):} Let $\op{X}$ and $\op{X}'$ have
the same spectrum,and suppose that they have a certain set of eigenstates in common. 
Let $\op{X}$ and $\op{X}'$ agree on the subspace \mc{S} spanned by those eigenstates, 
and let $\ket{\psi}\in \mc{S}.$  Define \mc{P} and $\mc{P}'$ to be payoff functions 
on the spectra of \op{X} and $\op{X}'$ respectively, which agree on the
spectrum of $\op{X}|_\mc{S}.$

Then $\langle\ket{\psi},\op{X},\mc{P}\rangle\simeq\langle\ket{\psi},\op{X}',\mc{P}'\rangle.$
\end{quote}

\noindent \textbf{Proof:} Without loss of generality, assume that 0 is in the spectrum neither of
$\op{X}$ nor $\op{X}'$.
We define:
\begin{itemize}
\item $\op{X}_0$ is the operator equal to $\op{X}$ on \mc{S} and equal
to zero otherwise (clearly $\op{X}_0=\op{X}'_0$).
\item $f$ is that function on the spectrum of $\op{X}$ defined by
$f(x)=x$ for $x$ in the spectrum of $\op{X}_0$, and $f(x)=0$
otherwise.
\item $\mc{N}$ is some arbitrary consequence.
\item $\mc{P}_1$ is a payoff function for $\op{X}$, such that
$\mc{P}_1(x)=\mc{P}(x)$ whenever $x$ is in the spectrum of $\op{X}|_\mc{S}$
and $\mc{P}_1(x)=\mc{N}$ otherwise.  
\item $\mc{P}_0$ is a payoff function for $\op{X}_0$, such that
$\mc{P}_0(x)$=$\mc{P}_1(x)$ for $x\neq 0$ and $\mc{P}_0(0)=\mc{N}$.
\end{itemize}

As was explained after the definition of a game, payoffs are not
specified counterfactually, and hence the value of a payoff function is
arbitrary on any $x_a$ which with certainty will not occur.  Hence
without changing the game as a physical process, we can replace the
payoff function \mc{P} by $\mc{P}_1$ (since the
observer will, with certainty, get one of the results in the spectrum of
$\op{X}|_\mc{S}$).  

Now we apply the PET:
\be
\mc{V}(\ket{\psi},\op{X},\op{P}_1)=\mc{V}(\ket{\psi},f(\op{X}),\mc{P}_1 \cdot f^{-1}).\ee
But $f(\op{X})=\op{X}_0$ and $\mc{P}_0 = \mc{P}_1 \cdot f^{-1}$, so in
fact we have
\be \langle\ket{\psi},\op{X}, \op{P}\rangle \simeq \langle\ket{\psi},\op{X}_0,
\op{P}_0\rangle.\ee
An identical argument tells us that
\be \langle\ket{\psi},\op{X}', \op{P}'\rangle \simeq\langle\ket{\psi},\op{X}_0,
\op{P}_0\rangle,\ee and the theorem follows. $\Box$

Something should be said about the role of the OET in Deutsch's proof.
It appears to be necessary to the proof unless we are to make quite
strong spectral assumptions about \op{X}, but is not explicitly used.
It might be that Deutsch avoids using either by defining measurements
(tacitly) in a state-dependent way: a measurement of \op{X} on the state $\sum_a\alpha_a
\ket{\lambda_a}$ could have been defined as any transformation with final state of form
\be \sum_a \alpha_a\ket{\mc{M};x_a}.\ee

However, the notion of measurement defined in section \ref{QMdef} was
intentionally state-independent (and thus counterfactual): that is, whether something does or does
not count as a measurement device does not depend on which microstate triggers 
it.  This seems intuitively reasonable: after all, a
device which always emits the result ``spin up'' would not count as a
legitimate spin-measurement device even if the state it was measuring
happened to have spin up!  

The next equivalence theorem we prove shows that it is essentially the amplitudes
of results and not the details of the state which matter.

\begin{quote} 
\textbf{State equivalence theorem (OET):} Let $\op{X}$ and $\op{X}'$ be self-adjoint
operators with discrete spectrum such that $\op{X}\ket{\lambda_a}=x_a\ket{\lambda_a}$
and $\op{X}'\ket{\lambda'_a}=x_a\ket{\lambda'_a}$; let $\ket{\psi}=\sum_{a=1}^N\ket{\lambda_a}$ and
$\ket{\psi'}=\sum_{a=1}^N\ket{\lambda'_a}$;   Let \mc{P} and $\mc{P}'$ be payoffs for
$\op{X}$ and $\op{X}'$ respectively, which agree on the set $\{x_1, \ldots, x_N\}$.

Then $\langle \ket{\psi}, \op{X}, \mc{P}\rangle 
\simeq \langle \ket{\psi'}, \op{X}', \mc{P}'\rangle.$
\end{quote}

Note that $\ket{\psi}$ and $\ket{\psi'}$ need
not be in the same Hilbert space, and that $\ket{\lambda_1}, \ldots, \ket{\lambda_N}$ 
($\ket{\lambda'_1}, \ldots, \ket{\lambda'_N}$) may not be
the full set of eigenvectors for \op{X} ($\op{X}'$).

\noindent \textbf{Proof:} Assume without loss of generality that 
$\mathrm{Dim}(\mc{H})\leq\mathrm{Dim}(\mc{H}')$. 
From the OET, we may replace $\op{X},$
with an operator $\op{Y}$, which agrees with $\op{X}$ on the span of the
\ket{\lambda_a} and which is non-degenerate elsewhere, with eigenvectors
\ket{\mu_i} and distinct eigenvalues $y_i$, with $y_i \neq x_a$ for all $i,a$: thus we have
\be \op{Y}=\sum_a x_a \proj{\lambda_a}{\lambda_a} + \sum_i y_i
\proj{\mu_i}{\mu_i}.\ee
Similarly, we may replace $\op{X}'$ with 
\be \op{Y}'= \sum_a x_a \proj{\lambda'_a}{\lambda'_a} + \sum_i y_i
\proj{\mu'_i}{\mu'_i}+\sum_j z_j \proj{\nu_j}{\nu_j}.\ee
(The third sum in the definition of $\op{Y}'$ occurs because the
dimension of $\mc{H}'$ may exceed the dimension of $\mc{H}$; for
convenience we will require that the $z_j$ are all distinct from one
another and from the $x_a$ and $y_i$.)

Let $\mc{P}_1$ be a payout scheme which coincides with $\mc{P}$ (and
thus $\mc{P}'$ on the set $\{x_1, \ldots, x_N\}$).  Let $\mc{P}'_1$ be a payout
scheme for $\op{Y}'$ which coincides with
$\mc{P}_1$ on the spectrum of $\op{Y}$.

Now consider the following operation (which is performable given Q4 and Q5):
\begin{enumerate}
\item Prepare \mc{H} in an \emph{arbitrary} state $\ket{\psi}$, and $\mc{H}'$ in some
fixed state $\ket{0'}$.
\item Perform a joint operation on $\mc{H}\otimes\mc{H}'$,
defined by:
\[\tpk{\lambda_a}{0'}\longrightarrow \tpk{0}{\lambda'_a};\]
\[\op{U}\tpk{\mu_i}{0'}\longrightarrow \tpk{0}{\mu'_i}\]
where \ket{0} is some fixed state of $\mc{H}$.  
At the end of this process, the joint state is $\tpk{0}{\psi'}$
for some state $\ket{\psi'}$; discard the fixed state \ket{0}.
(If $\mc{H}=\mc{H}'$,
replace this operation with the simpler one
$\ket{\lambda_a}\rightarrow\ket{\lambda'_a}$,
$\ket{\mu_i}\rightarrow\ket{\mu'_i}$, which just leaves \mc{H} in state
\ket{\psi'}.)
\item Measure $\op{Y}'$ by some process 
\[ \tpk{\lambda'_a}{\mc{M}_0}\longrightarrow \ket{\mc{M};x_a};\]
\[ \tpk{\mu'_i}{\mc{M}_0}\longrightarrow \ket{\mc{M};y_i};\]
\[ \tpk{\nu'_j}{\mc{M}_0}\longrightarrow \ket{\mc{M};z_j}.\]
\item In the branch where the result is $x$, provide a payout
$\mc{P}'_1(x)$.
\end{enumerate}

As usual in these proofs, this scheme admits of two descriptions.  If we
regard step 1 as the preparation of state $\ket{\psi}\in \mc{H}$ and
steps 2--3 as a measurement of $\op{Y}$ for that state, the process
instantiates the game $\langle \ket{\psi},\op{Y},\mc{P}_1 \rangle$.  If
however we regard 1--2 as the preparation of the state
$\ket{\psi'}\in \mc{H}'$, and 3 as a measurement of $\op{Y}'$ on that
state, then the process instantiates $\langle
\ket{\psi'},\op{Y}',\mc{P}'_1 \rangle$.  These games are thus of equal
value; hence when \ket{\psi} is as in the statement of the SET (so that we can
apply the OET) so are 
$\langle \ket{\psi}, \op{X}, \mc{P}\rangle$ and $\langle \ket{\psi'}, \op{X}', \mc{P}'\rangle.$
$\Box$

\subsection{The Grand Equivalence Theorem}

The results of the previous section, jointly, have powerful consequences
for decision-making, and all are utilized (tacitly) at various points in
Deutsch's proof.  However, it is possible to take them further than
Deutsch does: together, they imply a very powerful result of which
they in turn are immediate consequences.  To state this result, recall
that the \emph{weight} of a branch is simply the squared modulus of the
amplitude of that branch (relative to the pre-branching amplitude, of
course); thus if the state of a measuring device following measurement
is
\be\sum_a \alpha_a \ket{\mc{M};x_a},\ee
then the weight of the branch in which result $x_a$ occurs is
$|\alpha|^2$.  In a game, the weight of a consequence will be defined as
the sum of the weights of all branches in which that consequence occurs.

\begin{quote}
\textbf{Grand Equivalence Theorem}
For decision purposes, a game is completely specified by giving all the
distinct possible consequences of that game, together with their weights.
\end{quote}

\noindent \textbf{Proof:} Suppose a game 
$\langle \ket{\psi}, \op{X}, \mc{P}\rangle$
has $N$ possible
consequences $c_1, \ldots, c_N$ with weights $w_1, \ldots w_N$,  
then we will show that it is equivalent to what I will call a
`canonical'
game, $\langle \ket{\psi_0},\op{X}_0, \mc{P}_0\rangle$, where
\begin{itemize}
\item \ket{\psi_0} is a state in an $N$-dimensional Hilbert space
$\mc{H}_0$.  (That such games exist follows from Q2).
\item $\op{X}_0=\sum_{n=1}^N n \proj{n}{n}.$
\item $\ket{\psi_0}=\sum_{n=1}^N \sqrt{w_n} \ket{n}.$
\item $\mc{P}_0(n)=c_n.$
\end{itemize}
If each game with the same set of consequences and weights is equivalent
to the same canonical game, then they are all equivalent to each other
and the theorem will follow.

We proceed as follows.  For each $c_n$, let $M_n$ be the set of all
eigenvalues of \op{X} for which $\mc{P}(x)=c_n$, and let $M_0$ be the set
of all eigenvalues which (given \ket{\psi}) will not be found to occur in any branch 
post-measurement.  Since payoff functions are not specified
counterfactually, we can replace \mc{P} by any payoff function $\mc{P}'$
which is constant (say, equals $c_0$) on $M_0$.

Let $S_n$ be the subspace 
spanned by all eigenvectors of \op{X} with eigenvalues in $M_n$, and
$S_0$ the subspace spanned by eigenvectors with eigenvalues in $M_0$.
Define the function $f$ on the spectrum of \op{X} by
$f(x)=j$ whenever $x \in M_j$.  
By the PET, we can replace \op{X} with
\be \op{X}'=\sum_{j=0}^N j \mathrm{P}_j,\ee
where $\mathrm{P}_j$ is the projector onto $S_j$, and 
$\mc{P}'$ by $\mc{P}''$, where $\mc{P}''(j)=c_j$.

For each $n\geq 1$, let \ket{\psi_n} be an arbitrary eigenstate of \op{X} with
eigenvalue in $M_n$.  
The vector $\mathrm{P}_j\ket{\psi}$ has amplitude $\sqrt{w_j}$ (up to phase), 
and any
unitary transformation which leaves each $S_j$ fixed will also leave
\op{X} fixed.
Thus by the Operator Symmetry Principle we can
replace $\ket{\psi}$ by 
\be \ket{\psi'}=\sum_{n=1}^N \sqrt{w_n} \ket{\psi_n}.\ee

Thus we have shown that our game is equivalent to $\langle
\ket{\psi'},\op{X}',\mc{P}''\rangle.$  We now apply the SET to conclude
that this  is in turn equivalent to our canonical game.

\subsection{Composite games}

At various points in Deutsch's paper, it is necessary to make use of the
idea of `composite' games: that is, games where a quantum state is
measured by where, instead of giving a payoff dependent on the result of
the measurement, another game is played, where the \emph{game} is
dependent on the result of the measurement: effectively, the payoff
function \mc{P} takes values not in \mc{C} but in $\mc{A}_Q$.  Composite
games, of course, can themselves be composed, and so forth: the set of
all such composite games will be denoted by $\mc{A}_{CQ}$.

However, given measurement neutrality the decision problem is not really
extended by the move from $\mc{A}_Q$ to $\mc{A}_{CQ}$.  For any
composite game can be replaced by one huge simple game, with all the
states which might be needed in sub-games prepared in advance and one
(very complex) measurement performed on all of them at once.  It then
follows from measurement neutrality (most directly from the GET) that
this replacement does not change the value of the game.

\section{Deutsch's Proof}\label{deutschproof}

Deutsch's proof of the expected-utility rule is best understood as a
``quantization'' of the additivity proof of the EUT given in section
\ref{add}: it makes essential use of the additivity of consequences.
(Deutsch's own discussion of additivity \cite[p.\,5]{deutschprobability} seems to
imply that additivity is a mere convention, but the discussion of
section \ref{add} should make it clear that additivity places strong
constraints on preferences.)  In this section I will give a
reconstruction of Deutsch's own proof; I will then reformulate Deutsch's
axioms slightly in order to present an alternative and more direct additivity proof.
I conclude the section with a brief discussion of how Deutsch's proof fails in the presence
of `hidden variables'.

\subsection{Deutsch's postulates}

Deutsch develops his theory entirely within the framework of an additive
value function on consequences; he does not, however, explicitly make the assumption 
that the value of sequential \emph{acts} is additive.  In fact, in his system the 
values of acts are derivable from the values of the consequences, given a qualitative
preference order on acts: any constant act (where the act, with
certainty, leads to some consequence $c$) is assigned value
$\mc{V}(c)$,\footnote{Technically, we should prove that this assignation
is consistent, given that many different acts might all have constant
consequences: this is a trivial consequence of MET and PET, though.} 
and another act $A$ is assigned value \mc{V}(c) iff the agent is indifferent
between performing $A$ and performing the constant act.  This requires
there to exist quite a large class of consequences, one for each real number
between the largest- and smallest-value consequence.

Deutsch's class of acts is the composite set $\mc{A}_{CQ}$; in his 
system, the value of composite games follows from the value of simple ones.
This follows from his Substitutivity assumption (p.\,5 of his paper): if 
$\mc{G}=\langle \ket{\psi}, \op{X}, \mc{P}\rangle$, and if $\mc{G}'$ is
the composite game played by measuring $\ket{\psi}$ and then playing
game $\mc{G}(x)$ if we get result $x$, then a sufficient (but not
necessary) condition for $\mc{G}\simeq\mc{G}'$ is that
$\mc{V}[\mc{P}(x_a)]=\mc{V}[\mc{G}(x_a)]$ for all $x_a$ in the spectrum
of \op{X}.  In words, this means that an agent is indifferent between a
game where on some outcome he receives reward $c$, and another where on
that same outcome he plays a game which is worth $c$.  Fairly clearly,
this is another form of the Sure-Thing Principle.

Substitutivity allows us to simplify the notation for games: let
$\mc{P},\mc{P}'$
be any payoff functions for the same \op{X} such that
$\mc{P}(x_a)\simeq\mc{P}'(x_a)$.  Then substitutivity tells us
that irrespective of the state,
\be \langle \ket{\psi},\op{X},\mc{P}\rangle\simeq\langle
\ket{\psi},\op{X},\mc{P}'\rangle.\ee
This means we can ignore the details of what the consequences are, and
just use their numerical values: henceforth, then, \mc{P} will be taken to be a
function from the spectrum of \op{X} into the reals.

The main use  which Deutsch makes of Additivity is to prove the
\begin{quote}\textbf{Additivity Lemma:}
\be \mc{V}(\ket{\psi},\op{X},\mc{P}+k)
=\mc{V}(\ket{\psi},\op{X},\mc{P})+k.\ee
\end{quote}
He does so by considering the physical process of playing some quantum game
$\mc{G}=\langle \ket{\psi},\op{X},\mc{P}\rangle$ 
and then receiving a reward of value $k$ with certainty.  This is
physically identical to measuring $\ket{\psi}$ and then receiving two
rewards upon getting result $x_a$: one of value $\mc{P}(x_a)$ and one of
value $k$; by additivity this is equivalent to receiving a single reward
of value $\mc{P}(x_a)+k$.  

To complete this proof Deutsch needs a second use of additivity, to show
that the value of playing $\mc{G}$ and then receiving $k$ must be $\mc{V}(\mc{G})+k$; 
however, this requires him to assume value to be additive across acts
and not just consequences.  This would be a fairly innocuous extra
assumption for him to make; for completeness, though, note that his result can
be also be derived from additivity of consequences and substitutivity:
assume that the fixed reward $k$ is received \emph{before} playing
\mc{G}, then note by substitutivity that receiving $k$ and then playing
$\mc{G}$ must have the same value as receiving $k$ and then receiving
a reward worth $\mc{V}(\mc{G})$, and that by additivity this latter
process is worth $k+\mc{V}(\mc{G}).$  The result then follows.  

Deutsch requires two further assumptions.  One is a dominance principle
(tacitly introduced on page 12): if $\mc{P}(x_a)\geq \mc{P}'(x_a)$ for
all $x_a$, then a game using \mc{P} as payoff scheme is preferred or equivalent to one
using $\mc{P}'$.  (A similar assumption, recall, was used in the
classical discussion of additive value; note that the equivalence of two
games with equal-valued consequences for each given state, and hence the replacement
of $\mc{P}$ with a real-valued function, follows as easily from Dominance as
from Substitutivity.)

The last of Deutsch's assumptions, the Zero-Sum Rule, is on the face of
it more contentious.  This rule states that for any game \mc{G}, if the
payoff $\mc{P}$ is replaced by $-\mc{P}$ then the value of the game is
replaced by $-\mc{V}(\mc{G}).$  This is a trivial consequence of
consequence additivity for constant games but it is unclear what
Deutsch's 
motivation for it is when considering general games; in section \ref{altproof} we shall derive it from
act additivity. Saunders has argued for the Zero-Sum Rule directly by
considering the situation from the viewpoint of the banker:
\begin{quote}
\ldots banking too is a form of gambling. The only difference between
acting as one who bets, and as banker who accepts the bet, is that
whereas the gambler pays a stake in order to play, and receives payoffs
according to the outcomes, the banker receives the stake in order to act
as banker, and pays out the payoffs according to the outcomes. The 
zero-sum rule is the statement that the most one will pay in the hope of
gaining a utility is the least that one will accept for fear of losing
it. \cite{saundersdecision}
\end{quote}
Deutsch's postulates can then be axiomatized as follows.  
\begin{description}
\item[D0: Act availability] The set of acts is $\mc{A}_{CQ}$.
\item[D1: Transitive preferences (acts)] There is a weak order $\succ$
on $\mc{A}_{CQ}$ which satisfies measurement neutrality (and which
defines a weak order on \mc{C} via the constant acts).
\item[D2: Additive preferences (consequences)] There is a composition
operation $+$ on \mc{C} such that $\succ$ is additive with respect to
$+$ (that is, satisfies A1--A5).  (Hence there exists some value
function \mc{V} on consequences.)
\item[D3: Consequence availability] Each act in $\mc{A}_{CQ}$ is
value-equivalent to some constant act.
(Hence \mc{V} can be extended to acts.)
\item[D4: Substitutivity] Forming a compound game from any game, by
substituting for its consequences games of equal value to those
consequences, does not change the value of that game.
\item[D5: Zero-sum Rule] $\mc{V}(\ket{\psi},\op{X},\mc{P})=
-\mc{V}(\ket{\psi},\op{X},-\mc{P}).$
\item[D6: Dominance] If $\mc{P} \geq \mc{P}'$ then
\[\mc{V}(\ket{\psi},\op{X},\mc{P})\geq
\mc{V}(\ket{\psi},\op{X},\mc{P}').\]
\end{description}

\subsection{Deutsch's Proof}

In this section, we put together the results so far to achieve Deutsch's goal: to prove the 
quantum probability rule.  The proof given below follows Deutsch's own proof rather
closely, although some minor changes have been made for clarity or to
conform to my notation and terminology.  As such, it uses the various
equivalence theorems of section \ref{ETsect}, rather than the single
Grand Equivalence Theorem.

As usual, \ket{\lambda_a} will always denote an eigenstate of \op{X}
with some eigenvalue $x_a$.  By default, the operator measured will be $\op{X}$ and
the payoff function will be $f(x)=x$, restricted to the spectrum of $\op{X}$; thus,
$\langle\ket{\psi} \rangle \equiv
\langle\ket{\psi},\op{X},\mc{P}\rangle.$

\begin{stage} \label{stage1} Let $\ket{\psi}=(1/\sqrt{2})(\ket{\lambda_1}+\ket{\lambda_2})$.  Then
$\mc{V}(\ket{\psi})=1/2(x_1+x_2)$.
\end{stage}

We know (from the Additivity Lemma) that Deutsch can show that
\be\mc{V}(\ket{\psi},\op{X},\mc{P}+k)=\mc{V}(\ket{\psi},\op{X},\mc{P})+k.\ee
The PET simplifies this to
\be\mc{V}(\ket{\psi},\op{X}+k)=\mc{V}(\ket{\psi},\op{X})+k.\ee
Similarly,  the Zero-Sum Rule together with another use of the PET gives us
\be\mc{V}(\ket{\psi},-\op{X})=-\mc{V}(\ket{\psi},\op{X}),\ee
and combining these gives
\be\label{eq1}\mc{V}(\ket{\psi},-\op{X}+k)=-\mc{V}(\ket{\psi},\op{X})+k.\ee

Now, let $f$ be the function of reflection about the point $1/2
(x_1+x_2)$.  Then
$f(x)=-x+x_1+x_2$.  Provided that \op{X} is non-degenerate and that 
the spectrum of $\op{X}$ is invariant
under the action of $f$, we can define the operator $\op{U}_f$ as in
section \ref{ETsect}.  Since \ket{\psi} is a symmetry of \ket{\psi},
the State Symmetry Principle tells us that
\be\mc{V}(\ket{\psi},-\op{X}+x_1+x_2)=\mc{V}(\ket{\psi},\op{X}).\ee
Combining this with (\ref{eq1}), we have
\be \mc{V}(\ket{\psi},\op{X})=-\mc{V}(\ket{\psi},\op{X})+x_1+x_2,\ee
which solves to give $\mc{V}(\ket{\psi},\op{X})=1/2(x_1+x_2)$, as
required.

In the general case where $\op{X}$ is degenerate, or has a spectrum which is not invariant
under the action of $f$, we use the OET to replace \op{X} with
$\op{X}'$, which agrees with \op{X} on the span of
$\{\ket{\lambda_1},\ket{\lambda_2}\}$ and equals $1/2(x_1+x_2)$ times the identity
otherwise.  The result then follows, except in the case where
$x_1=x_2$; in this case the result is trivial.

Deutsch refers to this result, with some justice, as `pivotal': it is
the first point in the proof where a value has been calculated for a
superposition of different-value states, and the first time in our discussion
of decision theory that we have forced the probabilities to take specific values,
independent of the subjective views of our agent.

It is crucial to understand the importance in the proof of the symmetry
of \ket{\psi} under reflection, which in turn depends on the equality of
the amplitudes in the superposition; the proof would fail for
$\ket{\psi}=\alpha\ket{\lambda_1}+\beta\ket{\lambda_2},$ unless $\alpha
=\beta$.  

\begin{stage}\label{stage2} If $N=2^n$ for some positive integer $n$, and if
$\ket{\psi}=(1/\sqrt{N})(\ket{\lambda_1}+\cdots+\ket{\lambda_N})$, then
\be \mc{V}(\ket{\psi})=(1/N)(x_1+ \cdots +x_N).\ee
\end{stage}

The proof is recursive on $n$, and I will give only the first step (the
generalisation is obvious).  
Define:
\begin{itemize}
\item $\ket{\psi}=(1/2)(\ket{\lambda_1}+\ket{\lambda_2}+\ket{\lambda_3}+\ket{\lambda_4})$ 
\item $\ket{A}=(1/\sqrt{N})(\ket{\lambda_1}+\ket{\lambda_2})$;
$\ket{B}=(1/\sqrt{N})(\ket{\lambda_3}+\ket{\lambda_4})$
\item $y_A=(1/2)(x_1+x_2)$; $y_B=(1/2)(x_3+x_4)$.
\item 
$\op{Y}=y_A \proj{A}{A} + y_B\proj{B}{B}.$
\end{itemize}

Now, the game $\mc{G}=\langle \ket{\psi},\op{Y} \rangle$ has value
$1/4(x_1+x_2+x_3+x_4)$, by Stage \ref{stage1}.
In the $y_A$ branch, a reward of value $1/2(x_1+x_2)$ is given; by
Substitutivity the observer is indifferent between receiving that reward
and playing the game $\mc{G}_A=\langle \ket{\psi},\op{X}\rangle$, since the
latter game has the same value.  A similar observation applies in the $y_B$
branch.

So the value to the observer of measuring \op{Y} on \ket{\psi} and then playing either
$G_A$ or $G_B$ according to the result of the measurement is
$1/4(x_1+x_2+x_3+x_4)$.  But the physical process which instantiates
this sequence of games is just
\be   \left(\sum_{i=1}^4\frac{1}{2}\ket{\lambda_i}\right)\otimes
\ket{\mc{M}_0} \rightarrow \sum_{i=1}^4\frac{1}{2}\ket{\mc{M};x_i},\ee
which is also an instantiation of the game $\langle
\ket{\psi},\op{X}\rangle$; hence, the result follows.

\begin{stage}\label{stage3}
Let $N=2^n$ as before, and let $a_1, a_2$ be positive integers
such that $a_1+a_2=N$.  Define \ket{\psi} by 
$\ket{\psi}=\frac{1}{\sqrt{N}}(\sqrt{a_1}
\ket{\lambda_1}+\sqrt{a_2}\ket{\lambda_2})$.
Then 
\be\mc{V}(\ket{\psi})=\frac{1}{N}(a_1 x_1+a_2 x_2).\ee
\end{stage}

One way to measure \op{X} on a general superposition 
$\alpha\ket{\lambda_1}+\beta \ket{\lambda_2}$ of $\ket{\lambda_1}$ and $\ket{\lambda_2}$
is to use an $N$-dimensional auxiliary Hilbert space $\mc{H}_a$ spanned by states \ket{\mu_i}, 
and prepare it in state
\be \ket{1}=\frac{1}{\sqrt{a_1}}\sum_{i=1}^{a_1}\ket{\mu_i} \,\,\,\,\mathrm{or}\,\,\,\, 
\ket{2} = \frac{1}{\sqrt{a_2}}\sum_{i=a_1+1}^{N}\ket{\mu_i}\ee
according to whether \op{X} takes value $x_1$ or $x_2$; this process can
be combined with the erasure and destruction of the initial state.  If
we then define a non-degenerate operator $\op{Y}=\sum_i y_i
\proj{\mu_i}{\mu_i}$ on $\mc{H}_A$ and measure it, the overall dynamical
process will be
\be \alpha \ket{\lambda_1}+\beta \ket{\lambda_2} \longrightarrow
\alpha \ket{1}+\beta \ket{2} \longrightarrow 
\frac{\alpha}{\sqrt{N}}\sum_{i=1}^{a_1}\ket{\mc{M};y_i}
+ \frac{\beta}{\sqrt{N}}\sum_{i=a_1+1}^{N}\ket{\mc{M};y_i}.\ee
If a payoff of $x_1$ is provided whenever the measurement readout is
$y_i$ for $i \leq a_1$, and $x_2$ otherwise, then the overall process
realises the game $\langle \alpha\ket{\lambda_1}+\beta \ket{\lambda_2},
\op{X}\rangle$.

However, the selfsame process can also be regarded as a realisation of the
measurement of the \emph{degenerate} observable $f(\op{Y})$ (where
$f(y_i)=x_1$ for $i \leq a_1 $ and $f(y_i)=x_2$ otherwise) on the state
\be \ket{\phi}=\alpha \ket{1}+\beta \ket{2}.\ee  In the particular case where 
$\alpha =\sqrt{a_1/N}$ and $\beta
= \sqrt{a_2/N}$, then $\ket{\phi}=(1/N)(\ket{\mu_1}+\cdots
+\ket{\mu_N}$, and this second game has value $(1/N)(a_1 x_1 + a_2
x_2)$, by Stage \ref{stage2}; hence the result is proved.

Deutsch then goes on to prove the result for arbitrary $N$ (\iec, not 
just $N=2^n$); however, that step can be skipped from the proof without
consequence.

\begin{stage}\label{stage4}
Let $a$ be a positive real number less than 1, and let
$\ket{\psi}=\sqrt{a}\ket{\lambda_1}+\sqrt{1-a}\ket{\lambda_2}$.
Then $\mc{V}(\ket{\psi})=a x_1 + (1-a)x_2$.
\end{stage}

Suppose, without loss of generality, that $x_1 \leq x_2,$ and make the
following definitions:
\begin{itemize}
\item $\mc{G}=\langle \ket{\psi} \rangle$. 
\item $\{a_n\}$ is a decreasing sequence of numbers of form
$a_n=A_n/2^n$, where $A_n$ is an integer, and such that
$\lim_{n\rightarrow \infty}a_n = a$.  (This will always be possible, as numbers of this form
are dense in the reals.)
\item $\ket{\psi_n}=\sqrt{a_n}\ket{\lambda_1}+\sqrt{1-
a_n}\ket{\lambda_2}$.
\item $\ket{\phi_n}=(1/\sqrt{a_n})(\sqrt{a}\ket{\lambda_1}+\sqrt{a_n-a}\ket{\lambda_2}.$ 
\item $\mc{G}_n = \langle\ket{\psi_n}\rangle$.
\item $\mc{G}_n'= \langle\ket{\phi_n} \rangle$.
\end{itemize}

Now, from Stage \ref{stage3} we know that $\mc{V}(\mc{G}_n)=a_n x_1 + (1-a_n) x_2.$
We don't know the value of $\mc{G}_n'$, but by the postulate of Dominance we know that it is 
at least $x_1$.  Then, by Substitutivity, the value to the observer of measuring \ket{\psi_n}, 
then receiving $x_2$ euros if the result is $x_2$ and playing
$\mc{G}_n'$ if the result is $x_1$, is at least as great as the
$\mc{V}(\mc{G}_n).$

But this sequence of games is, by strong measurement neutrality, just a
realisation of \mc{G}, for its end state is one in which a reward of
$x_1$ euros is given with amplitude ${a}$ and a reward of $x_2$ euros with
amplitude $\sqrt{1-a}$.  It follows that $\mc{V}(\mc{G})\geq
\mc{V}(\mc{G}_n)$ for all $n$, and hence that $\mc{V}(\mc{G})\geq a x_1
+ (1-a) x_2.$   

A similar argument with an increasing sequence establishes that
$\mc{V}(\mc{G})\leq a x_1+ (1-a) x_2$, and the result is proved.


\begin{stage}
Let $\alpha_1, \alpha_2$ be complex numbers such that
$|\alpha_1|^2+|\alpha_2|^2=1$,
and let $\ket{\psi}=\alpha_1\ket{\lambda_1}+\alpha_2\ket{\lambda_2}$.
Then $\mc{V}(\ket{\psi})=|\alpha_1|^2 x_1 + |\alpha_2|^2 x_2$.
\end{stage}

This is an immediate consequence of the Operator Symmetry Principle, as
the operator $\op{U}=\sum_a \exp(i \theta_a)
\proj{\lambda_a}{\lambda_a}$ leaves \op{X} invariant.

\begin{stage}
The quantum probability rule is the correct strategy to determine preference: that is,
if $\ket{\psi}=\sum_i\alpha_i\ket{\lambda_i}$, then
$\mc{V}(\ket{\psi})=\sum_i|\alpha_i|^2 x_i$.
\end{stage}

This last stage of the proof is simple and will not be spelled out in
detail.  It proceeds in exactly the same way as the proof of Stage
\ref{stage2}: any $n$-term measurement can be assembled by successive 
2-term measurements, using Substitutivity and weak measurement
neutrality.

\subsection{Alternate form of Deutsch's proof}\label{altproof}

In this section I shall give an alternative proof of Deutsch's
result.  It differs from Deutsch's own proof in two ways: the Grand
Equivalence Theorem is used directly, and additivity of consequences is
replaced by additivity of acts.  The latter is a mild strengthening of
Deutsch's axioms, but seems fairly innocuous --- and, in any case, seems
substantially more plausible than the Zero-sum rule even though strictly
it implies it.  It allows us to streamline the axiomatization, removing
reference to compound games and making the axioms virtually identical to
the classical structure U0--U4.

In the classical case, we used act additivity to combine bets to construct an 
arbitrary bet.  We can do so in the quantum case also, if we allow that games may 
include not just the holistic process of preparing, betting on and measuring a quantum
state, but also that of betting on a quantum state which has already
been prepared, and which is to be measured by another party.  If so,
then act additivity implies that the value of placing such a bet is
unaffected by which bets have already been placed: this means  that
\be
\mc{V}(\ket{\psi},\op{X},\mc{P}_1)+\mc{V}(\ket{\psi},\op{X},\mc{P}_2)=
\mc{V}(\ket{\psi},\op{X},\mc{P}_1+\mc{P}_2).\ee

Our new axiom scheme, then, will be:

\begin{description}
\item[D0$'$: Act availability] The set of acts is $\mc{A}_Q$.
\item[D1$'$: Transitive preferences] There is a weak order $\succ$
on $\mc{A}_Q$ which satisfies measurement neutrality (and which
defines a weak order on \mc{C} via the constant acts).
\item[D2$'$: Dominance] If $\mc{P}(x_a)\succeq\mc{P}'(x_a)$ for all $x_a$ then
\be\langle\ket{\psi},\op{X},\mc{P}\rangle \succeq
\langle\ket{\psi},\op{X},\mc{P}'\rangle.\ee
\item[D3$'$: Composition] There is an operation $+$ of composition on
\mc{C}, and another such operation $+$ on $\mc{A}_Q$ such that 
\be (M,\mc{P}_1)+(M,\mc{P}_2)=(M,\mc{P}_1+\mc{P}_2).\ee
\item[D4$'$: Act additivity]  The weak ordering $\succ$ on $\mc{A}_Q$
is additive (that is, satisfies A1--A5) with respect to composition.
\end{description}

This list is very similar to U0--U4: the only real differences are the use
of the quantum acts $\mc{A}_Q$ and the assumption of measurement
neutrality. 

Our new proof is  as follows. U4 implies the existence of an additive value function \mc{V} on
acts, and hence (via constant acts) on consequences.  We define the \emph{expected utility} of a
game by $EU(\mc{G})=\sum_i w_i V_i$, where the sum ranges over the
distinct numerical values of the consequences with non-zero weight (\ie
the consequences which actually occur in some branch) and $w_i$ is the
weight of the consequences with value $V_i$, \ie the sum of the squared
moduli of all branches in which payoffs of value $V_i$ are made.

As with Deutsch's own proof, we suppose $\mc{P}(x)=x$ by default, and
hold fixed the observable \op{X} to be measured: this allows us to write
$\langle \ket{\psi} \rangle$ for $\langle \ket{\psi}, \op{X},
\mc{P}\rangle$.  In this case, we also write $EU(\ket{\psi})$ for $EU(\mc{G})$.

\begin{stageII}\label{stage1'}
Any game is characterised by the distinct numerical values of the
payoffs given in its branches, and their weights.
\end{stageII}

The GET tells us that any game is characterised by the distinct
consequences and their weights.  From Dominance, it follows
that two games which differ only by substituting some of the payoffs for
equal-valued payoffs are equivalent, and two payoffs are equivalent iff
they have the same numerical value.

\begin{stageII}\label{stage2'}
If \mc{G} is an equally-weighted superposition of eigenstates of \op{X},
$\mc{V}(\ket{\psi})=EU(\psi)$.
\end{stageII}

Without loss of generality, suppose
$\ket{\psi}=(1/N)(\ket{\lambda_1}+ \cdots + \ket{\lambda_N}).$

Let $\pi$ be an arbitrary permutation of $1, \ldots, N$, and define
$\mc{P}_\pi$ by $\mc{P}_\pi(x_i)=x_{\pi(i)}$.  Then by act additivity,
\be \sum_\pi
\mc{V}(\ket{\psi},\op{X},\mc{P}_\pi)=\mc{V}(\ket{\psi},\op{X},\sum_\pi
\mc{P}_\pi)=(n-1)! \sum_i x_i\ee
since $\sum_\pi \mc{P}_\pi$ is just the constant payoff function that
gives a payoff of $(n-1)!(x_1+ \cdots +x_N)$ irrespective of the result
of the measurement.

But each of the $n!$ games $\langle \ket{\psi},\op{X},\mc{P}_\pi\rangle$ 
is a game in which each consequence $x_i$ occurs with weight $1/N$.  
Hence, by the GET, all have equal value, and that value is just 
$\mc{V}(\ket{\psi})$.  Thus, $n! \mc{V}(\ket{\psi})=(n-1)!(x_1 + \cdots
+x_N)$, and the result follows.

\begin{stageII}\label{stage3'}
If $\ket{\psi}=\sum_i a_i \ket{\lambda_i}$, where the $a_i$ are all
rational, then $\mc{V}(\ket{\psi})=EU(\ket{\psi})$.
\end{stageII}

Any such state may be written 
\be \ket{\psi}=(1/\sqrt{N})\sum_i \sqrt{m_i}\ket{\lambda_i},\ee
where the $m_i$ are integers satisfying $\sum_i m_i=n$.  Such a game
associates a weight $m_i/N$ to payoff $x_i$.

But now consider an equally-weighted superposition \ket{\psi'} of $n$ eigenstates of
$\op{X}$ where a payoff of $x_1$ is given for any of the first $m_1$
eigenstates, $x_2$ for the next $m_2$, and so forth.  Such a game is
known (from stage \ref{stage2'}) to have value $(1/N)(m_1 x_1 + \cdots +
m_N x_n) \equiv EU(\ket{\psi})$.  But such a game also associates a weight
$m_i/N$ to payoffs of value $x_i$, so by stage \ref{stage1'} we have
$\langle \ket{\psi} \rangle \equiv \langle \ket{\psi'}\rangle$ and the
result follows.

\begin{stageII}
For all states \ket{\psi} which are superpositions of finitely many eigenstates of \op{X},
$\mc{V}(\ket{\psi})=EU(\ket{\psi})$. 
\end{stageII}

By the GET, it is sufficient to consider only states 
\be \ket{\psi}=\sum_i \alpha_i \ket{\lambda_i}\ee
with positive real
$\alpha_1$.  Let $\ket{\mu_i}$, $(1 \leq i \leq N)$, be a further set of
eigenstates of \op{X}, orthogonal to each other and to the
\ket{\lambda_i} and with eigenstates $y_i$ distinct from each other and
all strictly less than 
all of the $x_i$ (that we can always find such a set of states, or
reformulate the problem so that we can, is a consequence of GET, or SET
if preferred).
For each $i$, $1\leq i\leq N$, let $a_i^n$ be an increasing
series of rational numbers converging on $(\alpha_i)^2$, and define
\be \ket{\psi_n}=\sum_i \sqrt{a_i^n}\ket{\lambda_i}+\sum_i \sqrt{a_i-
a^n_i}\ket{\mu_i}.\ee

It follows from stage \ref{stage3'} that $\mc{V}(\ket{\psi_n})=EU(\ket{\psi_n})$, and
from Dominance that for all $n$, $\mc{V}(\ket{\psi}) \geq \mc{V}(\ket{\psi_n})$. 
Trivially $lim_{n\rightarrow\infty} EU(\ket{\psi_n})=EU(\ket{\psi})$, so
$\mc{V}(\ket{\psi})\geq EU(\ket{\psi})$.  Repeating the construction
with all the $y_i$ strictly greater than all the $x_i$ gives
$\mc{V}(\ket{\psi})\leq EU(\ket{\psi})$, and the result follows.

\subsection{Hidden variables}

As mentioned in section \ref{intro}, Deutsch's argument (and, as will be seen, my generalisations of 
it) rely essentially on the assumption that the Everett interpretation is correct. 
It may then be instructive to see how the argument fails in one particular set of non-Everett 
interpretations: those involving `hidden variables', such as the de Broglie-Bohm theory 
(\citeNP{bohm,holland}).

Recall that in a hidden-variable theory, the physical state of a system is represented not just
by a Hilbert-space vector \ket{\psi}, but also by some set $\omega$ of hidden variables, so that
the overall state is an ordered pair $\langle \ket{\psi}, \omega \rangle$. (In the de Broglie-Bohm
theory, for instance, $\omega$ is the position of the corpuscles.)

The Deutsch argument and its generalisations rely on the invariance of the state under certain unitary
transformations, and to apply the argument to a hidden-variable theory we need to know how these
transformations act on the hidden variables. This will in general depend upon the hidden-variable theory
in question; however, if the hidden variables are intended to represent spatial positions of particles
(as is generally the case, and in particular is true for de Broglie-Bohm corpuscles) we can specialise
to position measurements and to spatial translations and reflections, whose effects upon corpuscle positions
are clear.

Suppose, in particular, that we consider a measurement of the spatial position of a particle in one dimension, and  
assume that the quantum state is $\ket{\psi}=(1/\sqrt{2})(\ket{x}+\ket{-x})$, where $\ket{x}$ and $\ket{-x}$ 
are eigenvectors of position with eigenvalues $x$ and $-x$ respectively. Deutsch's argument relies on the
fact that this state is invariant under reflections around the origin. If the argument is to generalise to hidden
variables, we will then require that their positions are also invariant under reflection --- which forces them
to be located at the origin. However, since the hidden variable is supposed to represent the \emph{actual} position
of the particle, it must be located either at $+x$ or $-x$, since these are the only possible results of a position
measurement on state \ket{\psi}; it follows that the Deutsch argument cannot apply to such systems.

We could, of course, try to get round this problem by considering a probability distribution over hidden variables
and requiring the \emph{distribution} to be symmetric. Fairly clearly, this forces a distribution assigning probability 0.5 to both $+x$ and $-x$. The Deutsch argument can now be applied, and yields the unedifying conclusion that \emph{if} the particle is at position $+x$ 50 \% of the time, it is rational to bet at even odds
that it will be found there when measured.

\section{Going beyond additivity}\label{QSproof}

We have seen that, in the classical case, additivity is ultimately not
required as an assumption, at least in deriving probabilities.  In this
section we will show that additivity may be dispensed with
in the quantum case also, and replaced by a purely qualitative set of
axioms.

\subsection{Quantum versions of the Savage axioms}

The postulates which we shall need to prove the improved version of
Deutsch's result are modelled closely on Savage's axioms (described in
section \ref{savage}).  They are as follows:

\begin{description}
\item[QS0: Act availability] The set of acts is $\mc{A}_Q$.
\item[QS1: Value Preference: Acts] There is a weak asymmetric order
$\succ$ on the set \mc{A}, which satisfies measurement neutrality.   (As usual
$\succ$ induces a weak order on the set of consequences, via the constant acts.)
\item[QS2 : Non-triviality] There are at least two consequences $c$, $d$
with $c \succ d$.
\item[QS3: Sure-Thing Principle] In two games 
$\mc{G}_1=\langle \ket{\psi},\op{X},\mc{P}_1\rangle$ and 
$\mc{G}_2=\langle \ket{\psi},\op{X},\mc{P}_2\rangle$, if the payoff functions 
$\mc{P}_1$ and $\mc{P}_2$ agree on some subset \mc{R} of the spectrum of
\op{X}, then the preference order between $\mc{G}_1$ and $\mc{G}_2$ does not depend
on what actual value is taken by $\mc{P}_1$ (and thus $\mc{P}_2$) on \mc{R}.
\item[QS4: Dominance] Let $\mc{G}_1$ and $\mc{G}_2$ be as above.  If 
$\mc{P}_1(s) \succeq \mc{P}_2(s) $ for all measurement outcomes $s$ then 
$\mc{G}_1 \succeq \mc{G}_2$; if in
addition $\mc{P}_1(s) \succ \mc{P}_2(s)$ for some $s$ which actually
occurs in some post-measurement branch, then $\mc{G}_1 \succ \mc{G}_2$.
\item[QS5: Structural assumption] See below.
This axiom will play the role of Savage's structural
axiom S5.  It turns out that there are various choices for QS5; these will be discussed later.
\end{description}

Of these postulates:
\begin{itemize}
\item QS0 and QS1 are essentially the same as Deutsch's D0 and D1.
\item QS2--QS4 are almost verbatim copies of Savage's S2--S4.
\item QS5 will play the role of Savage's structural
axiom S5.  It turns out that there are various choices for QS5; these will be discussed later.
\item The only Savage axioms which is not represented here are the
probability axiom S5, which ensures that all events have comparable
probability, and the boundedness axiom $S7$, which allows the theory to
be extended to games with infinitely many distinct consequences.  
We will see that S5 can be derived in the quantum case; whether or not a quantum version
S7 is needed depends on the form of QS5, as discussed below.
\end{itemize}

\subsection{Probabilities for quantum events}\label{QSprob}

Suppose we consider a given measurement $M\in \mc{M}_Q$, specified by the state
\ket{\psi} to be measured and the measured observable \op{X}.  Let
$\mc{S}_M$ be the set of all elements of $\op{X}$'s spectrum which are
possible outcomes of the measurement: that is, all eigenvalues with at least one
associated eigenvector $\ket{\lambda}$ such that $\bk{\psi}{\lambda}\neq 0.$
Then the events for $M$ will be all subsets of
$\mc{S}_M$, and $\tilde C$ will denote $C$'s complement in $\mc{S}_M$.
The weight of an event will be defined in the obvious way, as the sum of the weights of
all branches contributing to that event.

Define a \emph{bet on C} as a game $\mc{G}=\langle \ket{\psi},\op{X},\mc{P}\rangle$, 
such that \mc{P} equals $x$ on $C$ and $y$ on $\tilde C$, where $x$ and
$y$ are consequences and $x \succ y$.  Such a bet will be denoted
$\langle M,C,x,y \rangle$.

We will define a qualitative probability measure on events as follows:
given any two events $C$ and $C'$, 
\begin{itemize}
\item $C$ is more probable than $C'$ (written
$C \succ C'$) if and only if 
$\langle M,C,x,y \rangle \succ \langle M',C',x,y \rangle$
for all $x$, $y$ such that $x \succ y$.   
\item $C$ is equiprobable to $C'$
(written $C \simeq C'$) if and only if
$\langle M,C,x,y \rangle \simeq \langle M',C',x,y \rangle$
for all $x$, $y$ such that $x \succ y$.
\end{itemize}

Note that this definition
allows comparison of events associated with different measurements; note
also that ``$C \simeq C'$'' is not synonymous with ``neither $C \succ C'$
nor $C \prec C'$''.  In fact, we have not yet proved that an arbitrary pair of 
events are comparable in respect of probability at all (recall that this
is an axiom --- S5 --- in the Savage framework).

The probability relations inherit obvious properties from the order on
acts: specifically, $\succ$ is a partial ordering on events, and
$\simeq$ is an equivalence relation.

We will now prove:

\begin{quote}
\textbf{Probability theorem:} 
\begin{enumerate}
\item All events are comparable in probability, with
$C \succ C'$ iff $C$ has strictly greater weight than $C'$. 
\item There exists one and only one function $\Pr$ on the set of all events such that
\begin{enumerate}
\item $\Pr(C) > \Pr(C')$ iff $C \succ C'$;
\item When restricted to the events of a given measurement $M$, $\Pr$ is an additive measure:
that is, for disjoint events $C,D\in \mc{S}_M$, $\Pr (C \cup D)=\Pr(C)+\Pr(D);$
\item For any $M$, $\Pr(\mc{S}_M)=1$.
\end{enumerate}
That function is the weight function: $\Pr(C)$ equals the weight of $C$.
\end{enumerate}
\end{quote}

To begin, note that  any bet $\langle M,C,x,y\rangle$, where $C$ has weight $w$, 
is a game where consequence $x$ occurs with weight $w$ and $y$ occurs with weight $(1-w)$.
By the GET, this is sufficient to specify the game completely for decision-theoretic purposes,
so we can write $\langle w,x,y\rangle$ for any such bet without relevant ambiguity.  It follows 
immediately that two events are equiprobable whenever they have equal weight.

Now suppose $0<w<w'\leq 1$, and set 
$\ket{\psi}=\sqrt{w}\ket{\lambda_1}+\sqrt{w'-w}\ket{\lambda_2}+
\sqrt{1-w'}\ket{\lambda_3}$, 
where $x_1,x_2,x_3$ are all distinct.
Define \mc{P} by 
\be\mc{P}(x_1)=x,\,\mc{P}(x_2)=y,\,\mc{P}(x_3)=y,\ee 
and $\mc{P}'$ by 
\be\mc{P}'(x_1)=x,\,\mc{P}'(x_2)=x,\,\mc{P}'(x_3)=y.\ee
Then by Dominance, if $x \succ y$ then $\langle
\ket{\psi},\op{X},\mc{P}'\rangle\succ \langle
\ket{\psi},\op{X},\mc{P}\rangle$.  But $\langle
\ket{\psi},\op{X},\mc{P}\rangle$ is a realisation of the bet $\langle
w,x,y\rangle$, and $\langle
\ket{\psi},\op{X},\mc{P}'\rangle$ realises $\langle w',x,y\rangle.$  Hence any
event of weight $w'$ is more probable than one of weight $w$; this
completes the proof of part 1.

The proof of part 2 proceeds in a similar way to my alternative proof of
Deutsch's result: we begin with equally weighted superpositions and
proceed successively to rationally weighted superpositions and general
superpositions.

Define a \emph{probability function} as any function on the set of
events satisfying (a)-(c) of part 2 of the probability theorem; let $\Pr$ be an arbitrary probability
function, and let $W$ be the weight function, assigning to each event
its weight.  In view of part 1, we must have $\Pr(C)=\Pr(C')$ iff
$W(C)=W(C').$

\begin{lemma'}\label{problemma1}
$\Pr(C)=W(C)$ is a probability function.  
\end{lemma'}

Given part 1, it is easy to verify that $\Pr(C)=W(C)$ satisfies (a)-(c) of part (2). 

\begin{lemma'}
If $\mc{S}_M={x_1,\ldots x_N}$, all distinct, and if each state in $\mc{S}_M$ has
the same weight, then $\Pr(\{x_i\})=W(x_i)=1/N.$
\end{lemma'}

Since all the states have the same weight, each event $\{x_i\}$ must have
the same probability.  Since $\Pr$ is required to be additive and
$\Pr(\mc{S}_M)=1$, this forces $Pr(\{x_i\})=1/N$.

\begin{lemma'}\label{problemma3}
Any event $C$ with rational weight satisfies $\Pr(C)=W(C)$.
\end{lemma'}
Let $W(C)=K/N$.  Consider any measurement $M$ for which $\mc{S}_M=\{x_1, \ldots, x_N\}$ 
is a set of equally-weighted states; then by additivity of $\Pr$,
$\Pr(\{x_1, \ldots, x_K\})=K/N.$  But $W(\{x_1, \ldots, x_K\})=K/N=W(C)$,
and $\Pr$ is the same for any two events with the same weight.

\begin{lemma'}
$\Pr(C)=W(C)$ is the unique probability function.
\end{lemma'}

In view of lemmas \ref{problemma1}--\ref{problemma3}, all that is left
to prove is that $\Pr(C)=W(C)$ on irrationally-weighted events.  Let $C$
be an arbitrary event, with weight $w$; let $\{w_n\}$ be an increasing
sequence of rational numbers convergent on $w$, and $\{C_n\}$ a sequence of events
with $W(C_n)=w_n$.   Then clearly we must have $\Pr(C)\geq \Pr(C_n)$ for
all $n$, and hence $\Pr(C)\geq W(C)$.  Repeating with a decreasing
sequence shows that $\Pr(C) \leq W(C)$, and the result follows.

It is interesting to compare the Probability Theorem with the analogous
result in Savage's framework.  There too, it is provable that an agent's
preferences determine a unique probability measure on the space of
events, and in fact the method of proof is pretty similar: first ``more
probable than'' is shown to order the events, then the space of events
is carved into arbitrarily many equiprobable (or almost equiprobable)
sub-events and these are used to show that there is a unique probability
function on th events compatible with this ordering.

There are important differences, however:
\begin{enumerate}
\item In the quantum approach, rather fewer axioms of pure rationality
are needed. We can dispense with the assumption that all events are 
comparable in respect of probability (S5), and with the sure-thing 
principle (S3).
\item There is also no need to use Savage's structural assumption (S6), 
which he requires to show that the set of events can be arbitrarily divided up.
In fact, for any given quantum measurement the set of states is \emph{finite}
and so the events certainly cannot be carved arbitrarily finely;
however, measurement neutrality lets us replace any measurement with
an equivalent one which has more events, which is an adequate
substitute.  The rich structure of QM excuses us from a need to
postulate this structure at the decision-theoretic level.
\item Most importantly, all rational agents must agree on their
probabilistic assignments in the quantum case, whereas there is scope in
Savage's system for many different assignments.
\end{enumerate}

\subsection{Choices of structural axiom}

The next step in Savage's proof of the EUT (having obtained a unique
probability measure) is to show that the von Neumann-Morgenstern axioms
VNM0--3 are satisfied (after which the EUT follows directly from von
Neumann's and Morgenstern's result.)  To do so Savage again needs to use
his structural axiom S6 (which we avoided using in proving the
Probability Theorem), and now we too will be forced to use some
analogous axiom.

However, the fact that we did not need S6 for discussions of probability
suggests that it may be too strong for our purposes, and that it might
be possible to weaken it.  Recall (section \ref{savage}) that S6 does
four things for Savage: it guarantees that the space of events is
continuously divisible, it rules out infinitesimal probabilities, and
it requires that no two events  differ infinitely or infinitesimally in
value.  It is only the latter two properties which we need, since we
have probability in hand; this suggests the following variant of S6.

\begin{description}
\item[QS5a: Comparability of acts] Given any three games $\mc{G}_1, 
\mc{G}_2, \mc{G}_3$ with $\mc{G}_1 \succ \mc{G}_2$, there is some
$\mc{G}_1 - \mc{G}_3$ bet which is preferred to $\mc{G}_2$, and some 
$\mc{G}_2 - \mc{G}_3$ bet to which $\mc{G}_1$ is preferred.
\end{description}

(Here a $\mc{G}-\mc{G}'$ bet has the obvious meaning: a bet on some event $A$ such
that we get to play $\mc{G}$ if $A$ obtains, and otherwise play $\mc{G}'$.)
In the presence of measurement neutrality, Q5a obviously entails S6.

There is, however, a completely different strategy available, which
again makes use of the rich structure of quantum mechanics to reduce
structural constraints on decision-making.  This strategy replaces QS5a
with 

\begin{description}
\item[QS5b: Stability] Suppose $\mc{G}_1$ and $\mc{G}_2$ are games with
$\mc{G}_1 \succ \mc{G}_2$; then this preference is stable under
arbitrarily small perturbations of the states being measured in the two
games. Symbolically, this is to say that, if $\langle\ket{\psi},\op{X}_1,\mc{P}_1\rangle
\succ \langle \ket{\phi},\op{X}_2,\mc{P}_2\rangle$ then there exists some $\epsilon >0$ such
that, if $\ket{\psi'}$ and $\ket{\phi'}$ are any states satisfying
$|\bk{\psi}{\psi'}|>1-\epsilon $ and $|\bk{\phi}{\phi'}|>1-\epsilon $,
then $\langle \ket{\psi'}\op{X}_1,\mc{P}_1\rangle \succ \langle 
\ket{\phi'},\op{X}_2,\mc{P}_2\rangle.$ 
\end{description}

Q6b turns out to be just as effective as Q6a in ruling out infinitesimal
or infinite values. It is clearly not an axiom of `pure' decision theory: it makes
essential reference to the quantum mechanics of the decision problem
under consideration.  This being the case, why use it instead of the
`pure' structure axiom QS5a?

Partly, QS5b is preferable because it allows a simple extension of the expected-utility rule
to infinite games, whereas QS5a has to be
supplemented to do so (see section \ref{infinitegames}).

Most importantly, though, QS5b admits of a very reasonable
justification --- more reasonable, perhaps, even than QS5a.  For without
it, it would not be possible for any agent without Godlike powers to act
in accordance with his preferences.  After all, without it then the
agent would have to prepare the state to be measured --- not to mention
the measuring device --- with infinite precision.  Any finite precision,
no matter how good, will fail to tell the agent which act to prefer
unless QS5b holds.  For instance, suppose QS5b fails to hold for two
acts $f$, $g$ (where $f \succ g$) and suppose $f$ involves measuring
the $z-$ component of spin of some spin state $\ket{\psi}$.  Preparing
that state will presumably involve aligning an initial spin with some
magnetic field (or something similar) and any finite error in the
alignment of that field will lead to finite errors in the preparation of
\ket{\psi}.

But given the failure of QS5b, \emph{for any} $\epsilon >0$ --- however
small --- there will exist some $\ket{\psi'}$ with
$|\bk{\psi}{\psi'}|>1-\epsilon$ --- but where measuring 
$\ket{\psi'}$ instead of $\ket{\psi}$ is \emph{not} preferred to $g$.
This means that even if in principle the agent knows that he would
prefer $f$ to $g$, he can never know whether any given act resembling
$f$ is preferred to $g$.

This shows that at the very least an agent with preferences violating 
QS5b would be unable to use decision theory as a practical guide to
action.  Either such an agent is reduced to catatonia, or he must adopt
some secondary theory of decision-making which is practically
implementable, even if (as it must) it conflicts in places with his
bizarre ``real'' preferences; this secondary theory will have to obey
QS5b, and arguably better describes the agent's `real' preferences
between acts than his purely verbally expressed preference for (say)
measurements whose outcomes have rationally-valued weights over those
with irrationally-valued weights.  In fact, if we follow Lewis's advice
and treat preferences as purely determined by dispositions to action,
then violation of QS5b by an agent is not just pathological, but
physically impossible.

\subsection{Expected utilities for quantum events}\label{QMexputil}

Savage's proof of VNM0--3 is straightforward but rather lengthy.  I will
only sketch it here, as well as the minor variations to it required in
the quantum case; see \citeN{savage} for the full details.

\begin{savstep}
Show that two acts are value-equivalent whenever they
assign the same probability as each other to each consequence.  
(In other words, $f$ and $f'$ are value-equivalent whenever there exist
partitions $C_i$, $C'_i$ of the state space such that $P(C_i)=P(C'_i)$
and $f(x)= f'(x')$ whenever $x \in C_i$ and $x'\in C'_i$.)
\end{savstep}
This is Savage's Theorem 5.2.1.  In quantum mechanics, 
the equivalent result is that two acts are value-equivalent whenever
they each assign weights $p_1, \ldots, p_n$ to each of $n$
consequences $c_1, \ldots, c_n$; this follows immediately from
the GET and the Probability Theorem.

\begin{savstep}
Define a gamble as an equivalence class of acts each of 
which assign the same probability as one another to each consequence;
denote a gamble $\mathtt{f}$ which assigns probabilities $p_1, \ldots, p_n$ to
consequences $c_1, \ldots, c_n$ by $\sum_i p_i c_i$.  Define a mixture
$\sum_j \sigma_j \mathtt{f}_j$ of gambles $\mathtt{f}_j=\sum_i p_{ij} c_{ij}$
as the gamble $\sum_{ij,} (\sigma_i p_{ij})c_{ij}$, and show that such
mixtures always exist (this is a consequence of the structural axiom
$S6$).
\end{savstep}
These definitions go through unchanged in the quantum case; to show that
mixtures always exist, let $\ket{\psi}=\sum_{ij}\sqrt{\sigma_i
p_{ij}}\ket{\lambda_{ij}}$
where the $\ket{\lambda_{ij}}$ are eigenstates of $\op{X}$ with
eigenvalues $x_{ij}$, all discrete; let $\mc{P}$ be a payoff scheme
assigning consequence $c_{ij}$ to measurement result $x_{ij}$.  Then
$\langle \ket{\psi},\op{X},\mc{P}\rangle$ instantiates the mixture of
gambles.

\begin{savstep}\label{step522}
Show that, if $\mathtt{f}$, $\mathtt{g}$ and $\mathtt{h}$ are gambles
and $0<\rho \leq 1$, then $\rho \mathtt{f}+(1-\rho)\mathtt{h} \succ 
\mathtt{g}+(1-\rho)\mathtt{h}$ iff $\mathtt{f}\succ \mathtt{g}$.
\end{savstep}

This is Savage's Theorem 5.2.2, and its proof makes essential use of
$S6$.  The quantum proof is identical save for the substitution of either QS5a 
or QS5b for S6; as it is moderately cumbersome it
will not be given here (see \citeNP[p.\,72]{savage}).

\begin{savstep}\label{step523}
Show that, if $\mathtt{f}_1 \prec \mathtt{f}_2$ and $ \mathtt{f}_1
\preceq \mathtt{g} \preceq \mathtt{f}_2$, then there is one and only one
$\rho$ such that $\rho \mathtt{f}_1 +(1-\rho)\mathtt{f}_2 \simeq
\mathtt{g}$.
\end{savstep}
This is Savage's Theorem 5.2.3; it is an easy consequence of step
\ref{step522} and S6.  Though the quantum proof is virtually identical
(again substituting QS5a or QS5b for  S6) I will give it to illustrate the sort of
use that is made in the theory of QS5a/QS5b.

From the quantum version of step \ref{step522} and the principle of
Dedekind cuts, we know that there is one and only one $\rho$ such that
\be \sigma \mathtt{f}_1 + (1-\sigma)\mathtt{f}_2 \prec
\mathtt{g} \,\, \mathrm{if}\,\, \sigma > \rho;\ee
\be \label{stepeqn} \sigma \mathtt{f}_1 + (1-\sigma)\mathtt{f}_2 \succ
\mathtt{g} \,\, \mathrm{if}\,\, \sigma < \rho.\ee
It follows that no number except possibly $\rho$ can satisfy the
equivalence which the theorem requires.  Suppose for contradiction that $\rho$ does not in fact
satisfy it: without loss of generality assume 
\be\label{stepeqn2}\rho \mathtt{f}_1 +(1-\rho)\mathtt{f}_2 \succ \mathtt{g}.\ee  
We can show this is contradictory via QS5a, which entails the existence
of some $\lambda>0$ such that \be\lambda\mathtt{f}_1+ 
(1-\lambda)(\rho \mathtt{f}_1 +[1-\rho]\mathtt{f}_2 ) \succ
\mathtt{g},\ee
but this is equivalent to 
\be (\rho+\lambda[1- \rho])\mathtt{f}_1 + (1-\lambda)(1-\rho)\mathtt{f}_2
\succ \mathtt{g},\ee
in contradiction with (\ref{stepeqn}).

Alternatively, note that an arbitrarily small perturbation of the state 
being measured will change $\rho$ to some $\rho'$ with $\rho'> \rho$.  
By QS5b, there will be some sufficiently small perturbation of this 
sort which preserves the preference (\ref{stepeqn2}), but this is 
in contradiction with (\ref{stepeqn}).

\begin{savstep}
Steps \ref{step522} and \ref{step523} establish, respectively, the von
Neumann-Morgenstern axioms VNM2 and VNM3; VNM0 and VNM1 are immediate consequences
of S0 and S1.  As such, the 
von Neumann-Morgenstern theory of utility can be applied, showing that
to each consequence can be associated a unique numerical value such that
for all games $\mc{G}_i$ with finitely many distinct consequences, 
$\mc{G}_1 \succ \mc{G}_2$ iff $EU(\mc{G}_1) > EU(\mc{G}_2)$.
\end{savstep}
This result goes through unchanged in the quantum case: we have
therefore established the expected-utility rule in that case also.

\subsection{Infinite games}\label{infinitegames}

We have now proved the validity of the expected-utility rule (and with
it the identification of probabilities with weights) for all games with
finitely many distinct consequences --- and, in particular, for all
games played in a finite-dimensional Hilbert space.  This is as far as
Deutsch went in his proof, and possibly as far as we need go: the
Bekenstein bound offers powerful reasons why the Hilbert space of the
Universe must be finite, and more prosaically the state space
 of a finite-volume system (such as a measurement device, or indeed a human
brain) whose energy is bounded above must also be finite.

Nonetheless,  we can fairly straightforwardly extend the 
expected-utility rule from finite to infinite games.  In Savage's
decision theory, this is done in two steps: firstly S7 is used to show
that the value function \mc{V} is bounded, and then this is in turn used
(again via S7) to extend the rule.  We can take this approach directly
over to the quantum case by adding S7 to the quantum axioms (as QS6).

However, it is interesting to note that if we adopt the stability axiom
QS5b in place of QS5a, we can prove the extension to infinite games
without any need to add S7 to the axioms; in the rest of this
section we will prove this.  Infinite games are only possible if it is
possible to measure some operator \op{X} with infinitely many distinct
eigenstates; we will assume henceforth that some such measurement is contained within
$\mc{M}_Q$.

\begin{lemma''}
The set $\mc{V}(\mc{C})$ is bounded.
\end{lemma''}
Proof: we will concentrate on proving that $\mc{V}(\mc{C})$ is bounded
above; proving it to be bounded below proceeds in an essentially
identical way.

Suppose for contradiction that $\mc{V}(\mc{C})$ is unbounded above, and 
let $c_1, c_2, \ldots$ be a sequence of consequences such that $c_i
\succ c_j$ for $i>j$ and $\lim_{i \rightarrow \infty}\mc{V}(c_i)=+
\infty$.  Let $\ket{\lambda_1}, \ket{\lambda_2} \ldots$ be a sequence of
eigenstates of \op{X} with distinct eigenvalues $x_1, x_2, \ldots$ and
set $\mc{P}(x_i)=c_i$.

Since $\langle \ket{\lambda_2}, \op{X},\mc{P}\rangle \succ 
\langle \ket{\lambda_1}, \op{X},\mc{P}\rangle$, it follows from QS5b that
there exists some $\epsilon>0$ such that 
$\langle \ket{\lambda_2}, \op{X},\mc{P}\rangle \succ 
\langle \ket{\psi}, \op{X},\mc{P}\rangle$
whenever $|\bk{\lambda_1}{\psi}|>1-\epsilon$.

In particular, if $\ket{\psi}=\sqrt{1-
\epsilon}\ket{\lambda_1}+\sqrt{\epsilon}\ket{\lambda_i}$, for any $i$,
then this condition is satisfied.  By the expected-utility principle for
finite-outcome games, it follows that
\be (1-\epsilon)\mc{V}(c_1)+\epsilon \mc{V}(c_i)< \mc{V}(c_2)\ee
for fixed $\epsilon$ and all $i$, which contradicts the unboundedness of the set of
consequences.

\begin{theorem}
The expected-utility rule holds for all games, finite and infinite.
\end{theorem}
Proof: Using measurement neutrality, any game \mc{G} can be mapped onto 
some game of the form $\langle \ket{\psi},\op{X},\mc{P}\rangle$, where
\begin{itemize}
\item \op{X} is some fixed non-degenerate observable, and \mc{P} a fixed payoff scheme for that
observable.
\item The eigenstates of \op{X} consist of the doubly infinite sequence
\be\ldots, \ket{\lambda_{-2}}, \ket{\lambda_{-1}},\ket{\lambda_0}, 
\ket{\lambda_1}, \ket{\lambda_2}, \ldots\ee with $\ket{\lambda_i}$ having eigenvalue $x_i$, 
and two additional eigenstates $\ket{\lambda_+}$ and
$\ket{\lambda_-}$, with eigenvalues $x_+$ and 
$x_-$ respectively.
\item \mc{P} satisfies $\mc{P}(x_i)\succeq \mc{P}(x_j)$ whenever 
$i \geq j$.
\item $\mc{V}[\mc{P}(x_+)]$ equals the upper bound of \mc{V},
and $\mc{V}[\mc{P}(x_-)]$ equals the lower bound.
\item $\ket{\psi}=\sum_{i=-\infty}^{+\infty}\alpha_i \ket{\lambda_{i}}$.
(Note that the two `extra' eigenstates $\ket{\lambda_\pm}$ are not
included in this sum.)
\end{itemize}
As such, any game can be referred to just by its state: $\langle
\ket{\psi}\rangle \equiv \langle \ket{\psi},\op{X},\mc{P}\rangle.$

For any game \mc{G}, define $EU(\mc{G})$ as the expected utility
of \mc{G} (the boundedness of \mc{V} guarantees that this is 
well-defined even for infinite games).  In particular, $EU(\ket{\psi})$
denotes the expected utility of $\langle \ket{\psi} \rangle$.

Now for each $n$, define states $\ket{\psi^+_n}$ and $\ket{\psi^-_n}$ by
\be \ket{\psi^\pm_n}=\sum_{|i|<n}\alpha_i \ket{\lambda_i}
+ \left( \sqrt{\sum_{|i|\geq n}|\alpha_i|^2}\right) \ket{\lambda_\pm};\ee
note that $\lim_{n \rightarrow \infty} EU(\ket{\psi^\pm_n})=EU(\ket{\psi}).$

By QS3 (Dominance) we have, for all $n$,
\be \langle \ket{\psi^+_n}\rangle \succeq \langle \ket{\psi} \rangle
\succeq \langle \ket{\psi^-_n}\rangle.\ee
But the games $\langle \ket{\psi^\pm_n}\rangle$ have
finitely many distinct consequences, and so the expected-utility rule
applies to them.  It follows that $\langle \ket{\psi} \rangle \succ
\mc{G}$ whenever \mc{G} is some finite game with
$EU(\ket{\psi^-_n})>EU(\mc{G})$ for any $n$, and hence that 
$\langle \ket{\psi} \rangle \succ \mc{G}$ whenever $EU(\mc{G})<EU(\ket{\psi})$; similarly, 
$\mc{G} \succ \langle \ket{\psi}\rangle$ for any finite game
\mc{G} with $EU(\mc{G})>EU(\ket{\psi}).$

Now, by choosing appropriate weights we can always construct some finite game $\mc{G}$ with
$EU(\ket{\psi})=EU(\mc{G})$; if we can prove that 
$\langle \ket{\psi} \rangle \simeq \langle \mc{G}\rangle$ then the proof will be
complete.

Assume for
contradiction that  $\mc{G} \simeq \crossout\, \langle \ket{\psi}
\rangle$ ---
 without loss of generality we may assume 
$\langle \ket{\psi}\rangle \succ \mc{G}.$  
Then, by the stability postulate QS5b, there exists some $\epsilon > 0$
such that $\langle\ket{\psi'}\rangle \succ \mc{G}$ whenever
$|\bk{\psi}{\psi'}|>1-\epsilon$.  There will always be some
$\ket{\psi^-_n}$ satisfying this condition, so we have
$\langle\ket{\psi^-_n}\rangle \succ \mc{G}$.  But
$EU(\ket{\psi^-_n})<EU(\mc{G})$, in contradiction to the expected-utility
rule.  Since this rule is known to hold for all finite games, the
contradiction is established.

\section{Gleason's Theorem}\label{gleason}

Before going on to discuss the implications of the proofs of sections 
\ref{deutschproof}--\ref{QSproof}, we should consider a possible alternative to
the whole approach.  It has been argued (notably by
Barnum \textit{et al} \citeyear{BCFFS} in their recent criticism of Deutsch) 
that Gleason's Theorem in
any case gives us all we could want in the way of a derivation of the
quantum probability rule.  In this section I will attempt to sketch
a proof of the probability rule using Gleason's theorem, and then argue 
that it is in a number of ways less satisfactory than
the approaches of sections \ref{deutschproof}--\ref{QSproof}.

\subsection{Deducing probabilities from Gleason's Theorem}

Gleason's theorem itself, a piece of pure mathematics, states the
following:
\begin{quote}
Let \mc{H} be a Hilbert space of dimension $>2$.  Let $f$ be any map
from the projectors on \mc{H} to $[0,1]$ with the property that, if
$\{\op{P}_i\}$ is any set of projectors satisfying $\sum_i\op{P}_i =
\id$, then $\sum_i f(\op{P}_i)=1.$ Then there exists a unique density
operator \denop\ on \mc{H} such that $f(\op{P})=\tr(\op{P}\denop)$ for
all projectors \op{P}.[\citeNP{gleason}]
\end{quote}

The theorem is remarkable in its generality: notice in particular that
no restriction of continuity has been placed on $f$ (although of course
the conclusion of the theorem entails that $f$ must in fact be
continuous).

If we are trying to construct some sort of instrumentalist
interpretation of quantum mechanics, Gleason's Theorem is little short
of a panacea.  Such an interpretation needs only to presume
\begin{enumerate}
\item that physically possible measurements on a system are represented
by the algebra of self-adjoint operators on some Hilbert space (and, in
particular, measurements giving YES-NO answers are represented by
projections); 
\item that the theory should assign probabilities to each possible
measurement outcome; and
\item that if $\op{A}$ and $\op{B}$ are commuting self-adjoint operators, 
then the probability of getting a given result upon measuring \op{A}
does not depend upon whether \op{B} is measured simultaneously.  (This
assumption is called \textit{non-contextuality}, and guarantees that
probabilities need be assigned only to individual projectors, rather than 
jointly to sets of them).
\end{enumerate}
Then Gleason's Theorem implies that the only possible probability
choices are those represented by a (pure or mixed) quantum state.  This
even allows the instrumentalist to deny all reality to the quantum
state: quantum states can simply be ways of codifying an agent's beliefs
about the outcomes of possible measurements.

\subsection{Applying Gleason's Theorem to the Everett interpretation}

However, we are not engaged in the project of constructing an
instrumentalist interpretation of QM.  The Everett interpretation is
robustly realist about the quantum state right from the beginning, 
and so applying Gleason's Theorem to it is not in any way needed to justify the
existence of the state.

Nonetheless, might it be possible to use the Theorem to justify the
quantum probabilities?   Only if it can be established somehow that the concept of 
probability is a necessary consequence of decision-making in situations
involving quantum uncertainty, for Gleason's theorem cannot be applied
at all until we can justify the need to assign probabilities to branches.  

We have already seen that this justification can be given if we assume
QS0--2 and QS4 --- but we have also seen that this leads almost immediately
not just to the existence of probabilities, but to the quantum
probability rule itself.  It follows that use of Gleason's Theorem would
only be of use if it can somehow enable us to weaken QS0--2 or QS4.

The only plausible way in which this could work, as far as I can see,
would be via a modification of QS0.  That axiom requires the agent to
have preferences defined over a rather large class of games and
Gleason's theorem suggests that that class could be modified somewhat.

A possible axiom scheme that could implement such an idea might be:
\begin{description}
\item[G0: Act availability] The set of available acts includes \emph{all possible} measurements of
self-adjoint operators on the Hilbert space \mc{H} of some fixed quantum system 
with $\mathrm{Dim}(\mc{H})>2$, as well as some set of bets on the
outcomes of these measurements.
\item[G1: Transitive preferences] The agent has some preference order
$\succ$ on pairs of acts, which is a weak ordering of the set of acts
(defining, as usual, a weak ordering of the set of consequences) 
and satisfies measurement neutrality. (\iec, as for QS1).
\item[G2: Non-triviality] (As for QS2).
\item[G3: Sure-Thing Principle] (As for QS3).
\item[G4: Dominance] (As for (QS4).
\item[G5: Probability ordering]  Any two events are comparable in
probability (in the sense of section \ref{QSprob}).
\item[G6: Structure axiom] (As for QS5, \ie either QS5a or QS5b).
\end{description}

These axioms are quite similar to QS1--5; note, however, that
\begin{itemize}
\item The set of games considered is rather different: now we need consider only 
one physical system, but we need to be able to realise all measurements
on it (or, which is equivalent, perform all unitary transformations).
This means that measurement neutrality leads to rather different results
(in particular, the proof of the Grand Equivalence Theorem now fails).
\item As a consequence of losing the Grand Equivalence Theorem, it is
necessary to postulate that all events are comparable in respect of
probability (\textit{a la} S5) rather than deriving it.
\end{itemize}

The proof of the probability theorem would now go as follows: Firstly,
we know that all events are comparable in respect of probability.  This tells us
that probability defines a weak ordering on the set of all projectors on
\mc{H}.  With the aid of Dominance, we can also prove that the projector
onto the physical state \ket{\psi} of the system is not less probable
than any other projector, as follows: let \ket{\psi} be an eigenstate of
some operator \op{X} with eigenvalue $x_a$; let \mc{P} be a payoff scheme
for \op{X} allocating payoff $x$ to $x_a$ and $y$ to all other
eigenvalues, with $x \succ y$.  (Hence $\langle \ket{\psi},\op{X},\mc{P}\rangle$ 
realises an $x,y$ bet on $x_a$.) Then since payoffs are not
counterfactually defined, and payoff $x$ is guaranteed to occur, we can
replace \mc{P} by a payoff scheme which always gives $x$.  By Dominance,
the resultant game is at least as probable as any other $x,y$ bet, and
the result follows by the definition of the `more probable than'
relation.

Suppose it were the case that the qualitative probability ordering could be
\emph{represented}: that is, that there is some function $f$ from the projectors to
$[0,1]$ which restricts to a probability measure on the spectrum of each
observable operator and which satisfies $f(\op{P}_1)>f(\op{P}_2)$ iff 
$\op{P}_1 \succ \op{P}_2$.  Then Gleason's theorem would imply 
the existence of some density operator \denop\ such that 
$f(\op{P})=\tr (\op{P}\denop).$  

(It may be helpful, here, to observe that measurement neutrality
implies that any representation of the preferences is noncontextual.
This can be seen most directly from the Grand Equivalence Theorem:
let $\op{P}_1$ be a projector, and suppose we consider a bet in which we measure a
state with respect to some commuting set of projectors including $\op{P}_1$, 
getting a reward if and only if the outcome corresponding to $\op{P}_1$ occurs. 
By the Grand Equivalence Theorem, such a bet is specified completely (for 
decision-theoretic purposes) by the weights $\matel{\psi}{\op{P}_1}{\psi}$ and
$1-\matel{\psi}{\op{P}_1}{\psi}$ --- the details of what is measured
concurrently with $\op{P}_1$ are decision-theoretically irrelevant.)

We would not yet have shown that the quantum state generating $f$ was
the physical quantum state, \ket{\psi}; this, however, would be easy.
For the considerations above tell us that the probability of obtaining
$x_a$ upon measuring the observable \op{X} must be no lower than the
probability of any other event --- and hence must equal 1.  The only
quantum state \denop\ satisfying $\tr(\denop\proj{\psi}{\psi})\equiv
f(\proj{\psi}{\psi})=1$ is of course $\proj{\psi}{\psi}$ itself.

This having been established, the remainder of a Gleason's 
Theorem-motivated derivation of the EUT would be able to proceed much as in
section \ref{QMexputil}.

All this relies, however, on our ability to show that any qualitative
probability possesses a representation, and I am not aware of any proof
of that fact.  Effectively, Gleason's theorem provides a uniqueness
proof for probability but not an existence proof, and for all I know it
is necessary to introduce additional axioms of a structural character to
ensure this.

One way of so doing would be to use the additivity strategy to define
the value function on consequences in advance of considering
probability: then, recall, bets can be used to define probabilities.  A
set of axioms implementing this strategy would be virtually identical to
D0$'$--D4$'$, differing only in the replacement of the act-availability
axiom D0$'$ with G0.

\subsection{Disadvantages of the Gleason approach}

How does the derivation of the probability rule from G0 and G2--G4 compare with
those given in sections \ref{deutschproof}--\ref{QSproof}?  The most
obvious  disadvantage of the derivation via Gleason's Theorem 
is of course that it is incomplete, with the existence of \emph{any}
numerical representations of the probability ordering being unproven.
(Granted, adding the assumption of Additivity solves this problem, but
we have already seen that that assumption is implausibly strong).

Let us assume, though, that the existence of a representation can in fact 
be proved from the G axioms without adding any over-strong structure
axioms.   In that case, the assumptions made in this section's
derivation of the probability rule seem pretty much on a par with the
Savage-inspired derivation of section \ref{QSproof}.
Granted, in this section we needed to assume explicitly (G5) that events 
are comparable in respect of probability, whereas this could be deduced
explicitly in section \ref{QSproof}; but G5  is
a highly reasonable-sounding axiom of pure rationality, transferred
\textit{mutatis mutandis} from the classical Savage axioms.  The other
main difference is the set of acts considered: section \ref{QSproof}
requires systems of all dimension to be considered, whereas the present
section required only a fixed dimension but also required that all
unitary transformations be considered performable.  Neither set of acts
seems obviously more reasonable than the other, though: both are clearly
idealisations, but fairly reasonable ones (especially since we only
require agents to have well-defined preferences between such acts and
not necessarily to be able to perform them; I have a definite preference
between exile on Saturn and being appointed World President, though it
is certainly not in my power to bring either consequence about!)

If there is reason to prefer a proof modelled on classical decision
theory to one based around Gleason's theorem, it probably lies more in
the proofs themselves than in the axioms required.  Here it seems that
the former proof has a major advantage in brevity and simplicity: the 
proof of the probability theorem presented in section \ref{QSprob} takes barely
a page, whereas even the proof of Gleason's Theorem is long and
complicated, and we have seen that this would be only one component of a
proof based on the axioms of this section.

Furthermore, conceptually speaking it is important to remember that
Gleason's Theorem is here being used not as an \emph{alternative} to a
decision-theoretic approach, but as a \emph{component} of it. It is not
the case that simple invocation of Gleason's Theorem, unburdened with
considerations of a decision-theoretic character, is sufficient to
resolve the Everettian probability problem.

\section{Quantum probabilities: a discussion}\label{quantprobdiscussion}

In this section I will consider the implications of the proofs just given.   
I will show that they guarantee that quantum weights satisfy the requirements 
(discussed in section \ref{objectivechance}) placed by Papineau  on an
adequate analysis of chance, and strongly support (though they do not
quite entail) a quantum-mechanical version of Lewis's Principal
Principle.  I will also give a brief discussion of some implications of
this new Principle.

\subsection{The decision-theoretical link}\label{papdec}

One of Papineau's two links between objective chances and non-probabilistic facts
was that it is rational to use objective chances as a guide to our
actions.  It might seem that we have justified this already: if I am
offered a bet where the payoff depends on the outcome of a measurement
on a known state, the whole point of our discussion of decision theory
was to prove that the correct strategy was to calculate the expected
utility of the bet using the quantum weights as probabilities.

It is not quite so straightforward as  that, though.  For suppose that
the measurement has \emph{already happened} --- in a sealed box on the
other side of the lab, say.  Even though the box is sealed, decoherence
will still have led to branching, and there will be (at least) one
branch for each possible outcome of the measurement.  My uncertainty is
now simple ignorance: determinately I am in one branch rather than
another (for all that there are copies of me in the other branches) but
I do not know which.

It is surely part of the quantum probability rule that the right
strategy to adopt here is exactly the same as if the measurement had not
yet occurred: namely, maximise expected utility with respect to the
weights of the various measurement outcomes.  This can, in fact, be
deduced from our results so far, provided we accept a further principle
of rationality, called the \emph{reflection principle} by
\citeN{vanfraassen}.
\begin{description}
\item[The reflection principle:] If, at time $t$, I decide rationally to pursue a
certain strategy at a later time $t'$, and if I gain no new information relevant to
that strategy between times $t$ and $t'$, then it is rational not to
change my choice of strategy at $t'$.
\end{description}

Like all  axioms of pure rationality, this seems very obvious (and admittedly, 
like virtually all of them. it has been challenged; see, \egc, \cite{elga}).
We can use it to deduce the post-measurement strategy very easily, as
follows: if you had decided which strategy to adopt \emph{before} the
measurement, you would have opted for maximising expected utility with
respect to quantum weights.  The reflection principle says that you
should adopt the same strategy even after the measurement, as \textit{ex
hypothesi} you have gained no information at all about the results of
the measurement.

With this additional result in place, it seems that we have satisfied Papineau's
requirement that a theory of probability explain probabilistic decision-making.

\subsection{The inferential link}

Papineau's other requirement was that probabilities could be estimated
via frequencies, and we will show the validity of this in the context of
quantum games.  For simplicity, consider only a two-state system,
prepared in some state $\ket{\psi}=\alpha\ket{0}+\beta \ket{1}$ and measured in the
$\ket{0},\ket{1}$ basis.  We have a choice of two bets: either get a
reward of value $x$ if the measurement yields ``0'', and of value $y$ if
it gives ``1''; or get a reward of value 0 with certainty.

Obviously (given our derivation of the EUT in the quantum case) the
correct strategy is to take the first bet if and only if $|\alpha|^2 x +
|\beta|^2 y >0$; but what if we don't have a clue about the values of $\alpha$ 
and $\beta$?
Then it seems that (absent further information) we cannot choose which
strategy to follow.

However, the situation changes substantially if we have access to a
large number of copies of \ket{\psi} and can measure each copy in the
$\ket{0},\ket{1}$ basis.  The classical strategy would then be something
like: measure all $N$ copies; if we get $M$ results of ``0'', then take
the bet provided that $ (M/N) x + ([N-M]/N) y > \epsilon$, for some small
$\epsilon$.  (This is equivalent to using the relative frequencies as
estimates of the probabilities, with $\epsilon$ set non-equal to zero to
hedge against the fluctuations possible in the estimate.)

How do we fare if we adopt this strategy in quantum mechanics?  The
combined weight of the branches with $M$ results of ``0'' will be
\be w (M) = \frac{N!}{M!(N-M)!}p^M q^{N-M}\ee
where for convenience we have defined $p=|\alpha|^2$ and $q=|\beta|^2+$.
The strategy adopted is to bet whenever $M/N\geq p_0$, where $p_0 x + (1-p_0)y =\epsilon$, 
and the expected utility of the strategy is then 
\be EU = \sum_{M\geq p_0 N} \frac{N!}{M!(N-M)!}p^M q^{N-M}(p x+ q y)
\simeq (px+qy)\frac{1}{\sqrt{\pi}}\int^\infty_{x_0} \dr{x} \exp(-x^2) \ee
where $x_0 = \sqrt{N/2pq}(p_0-p)$.

Now, obviously if $px+qy <0$ then this strategy has negative payoff no
matter what the values of $\epsilon$ and $N$ --- but observe that in
such cases, the weight of those branches in which the bet is accepted
will be extremely small.  Thus, although the payoff will be negative it
will be only very slightly negative, and thus only trivially worse than
not taking the bet; on the other hand if the bet is objectively worth
taking then (for large enough $N$) the weight of those branches in which
it is taken will be close to unity.  The strategy, then, seems fairly
well-supported.

\subsection{A quantum Principal Principle}

Lewis's Principal Principle, briefly discussed in section
\ref{objectivechance}, is claimed (with some justice) to ``capture all
we know about chance'' \cite[p.\,86]{lewisprobability}, and it is of some
interest to know whether there is an Everett version of it (we have already
seen three  scenarios --- bets before a measurement, bets
afterwards but in ignorance of the result, and inference to the state 
--- each of which makes a somewhat different use of quantum probability,
so there is certainly reason to desire unification).

To state Lewis's Principle (and our version) it will be useful to follow Lewis
to a more relaxed notion of subjective probability.  Lewis's subjective
probabilities --- which he calls \emph{credences} --- are still in principle 
agent-relative, and still gain their empirical significance by being the
best analysis of an agent's dispositions to act.  But we shall simply
take it as read that they are defined over a very wide class of
situations, rather than carefully deriving this from decision-theoretic
axioms as hitherto.

To be specific, we will require an agent to have a credence function
which ranges over all propositions about the world (understood, if
desired, in Lewis's sense as sets of possible worlds, but this is not
required).  Thus to any propositions $A$, $B$, there will exist
numbers $C(A)$, $C(B)$ giving the agent's credence in the truth of those 
proposition, as well as a conditional credence $C(A|B)=C(A \& B)/C(B)$
giving the credence in $A$ given the known truth of $B$.

Actually, even this is not a sufficiently wide class over which to
define credence.  I wake up, sure of where I am sleeping but unsure of
the time.  I am not ignorant of any facts about my world, only of my
(temporal) location within it.  This suggests an extension of the
credence function so that it ranges over properties (including, `the
property of being in a world where proposition $A$ holds') rather than just
propositions; equivalently (in Lewis's analysis) it ranges over centred
possible worlds --- possible worlds with a preferred agent picked out 
--- rather than just over possible worlds \emph{simpliciter}.

Lewis's Principal Principle is then:
\begin{quote}
Let $t$ be any time, and $x$ a real number, and let $X$ be the proposition
that the objective chance (at $t$) of $A$  holding is $x$.  Let $E$ be
any proposition\footnote{or property, \etc} \emph{admissible} at time $t$.   Then
\be C(A|X\& E)=x.\ee
\end{quote}

Lewis does not give an explicit analysis of `admissible', but essentially
the reason for the restriction is to rule out already knowing what the
result of the chance event is, by time-travel, precognition or other
occult methods.  Such events having been ruled out, the Principle
basically says that credences equal chances.  For Lewis, past events
aren't chancy, so the Principle doesn't directly extend to credences
about a past chance event (such as a measurement) when we don't know
anything about the outcome; however, we can analyse these using the
Reflection Principle.

Here, then, is the Everettian version (which I will defend, below).
\begin{quote}
\textbf{Everett Principal Principle (EPP):} Let  $w(E)$ and $w(A\& E)$
be real numbers; let $A$ be any
statement about a branch; let $E$ any conjunction of propositions, and
statements about a branch.  
Let $X$ be the proposition that the weight of all branches in which $E$ holds
is $w(E)$, and that the weight of all the
branches in which $A$ and $E$ both hold is $w(A\& E)$.  Then
\be C(A|X\& E)=w(A\& E)/w(E).\ee
\end{quote}
Let us consider some aspects of the EPP:
\begin{itemize}
\item The `statements about branches' are to be understood along the
lines of the prototype: `in \emph{this} branch, $x$ is happening/ has happened'.  
As such they are inherently indexical, and cannot be replaced by propositions.
On a technical level, each corresponds to some projector onto a 
coarse-graining of the decoherence-preferred basis, or possibly a string
of such projectors (we will denote such a string, for statement $A$, 
by $\op{P}_A$).
\item There is no need for a division of statements into `admissible'
and `inadmissible'.  It is simply impossible for an agent to gain any
inadmissible information in the Everett interpretation: no amount of
time-travelling, precognition or the like can disguise the simple truth
that there is \emph{no fact of the matter} about the future in the
presence of quantum splitting, and so no way to learn that fact.
\item Despite the unity of the EPP, its credences have an inherently
dualistic nature. If $A$ is determinate relative to the agent then
$C(A|X\& E)$ is his credence in the truth-value of some unknown, but
determinately true or false statement; if, however, $A$ is not 
determinate (if, for instance, it refers to the outcome of a quantum measurement
to be performed in his future) then his credence relates to the sort of
subjective indeterminism discussed in section \ref{dectoQM}.
\item It is easy to see that EPP incorporates the Reflection Principle
as used in section \ref{papdec}.
\item The EPP is perhaps more easily understood if we consider a limiting case: 
suppose that $E=X_0 \& B$
where $X_0$ is the proposition that the Heisenberg state
of the universe is \ket{\psi}, and $B$ is a statement about a branch.
Since $X_0$ determines \emph{all} the weights, it subsumes $X$ (where $X$ is as
defined in the EPP).
If the  projectors
corresponding to $A$ and $E$ are $\op{P}_A$ and $\op{P}_E$ respectively, 
then EPP gives
\be C(A|X_0\& B)=
\frac{\matel{\psi}{\op{A}\op{B}}{\psi}}{\matel{\psi}{\op{A}}{\psi}}.\ee
\end{itemize}

The EPP, I think, gives us everything we might want from probabilities
in the Everett interpretation.  Have we yet done enough to prove it?
A proof along the lines of this paper's discussion might involve the
following three steps:
\begin{enumerate}
\item Show that quantum-mechanical branching events are subjectively
best-understood as situations of indeterminism, hence of ignorance about
the post-branching future.
\item Show that this ignorance is correctly analysed by means of the
quantum probability rule.
\item Extend this result to show that the quantum probability rule gives
the correct analysis of probability even when it concerns the
objectively determinate (such as: what \emph{was} the result of that
measurement?).
\end{enumerate}

(1) was argued for in section \ref{dectoQM}, following Saunders'
analysis.  (2) was proved from general axioms of decision theory in
sections \ref{quantgames}--\ref{gleason}; the only loophole would seem
to be whether my quantum games are too stylised to count as models for
general quantum-mechanical branching events.  And (3) can be argued for
using the Reflection Principle and asking how we would have selected our
strategies if we had made the selection before the original branching.
(There is another loophole here if the multiverse had multiple branches even at
the beginning of time, but this seems cosmologically implausible.)

The EPP, then, seems reasonably secure even if not completely proven.
It is satisfying to note that it  at any rate appears to be on substantially more
secure foundations than any classical analysis of objective chance!

\section{Objections; Replies; Conclusions}\label{conclusion}

The main thrust of the argument of my paper, some of the obvious
objections and replies, and its conclusions, are summarised in the
following dialogue.  The `Everettian' obviously speaks for the author,
but at times so does the `Sceptic' --- his objections are not meant to
be trivial and although I think they can be answered there is probably much scope
for further investigation.

\begin{description}
\sk What's all this attention to `rationality' in aid of?
This is physics, not psychology.
\ev Rationality is the only way in which the concept of
`probability' makes contact with the physical world.
\sk But probability is just relative frequency in the
limit.
\ev If by `limit' you mean after an actual infinity of 
experiments, then you're welcome to define it that way if you like.  But no-one can actually
carry out an actual infinity of experiments.  In practice, we predict
probabilities from the 
relative frequencies of outcomes in \emph{finitely many} experiments.  This
isn't automatically correct --- probability theory itself predicts a finite
probability that it will fail --- so we need some account of why it's
rational to do it.  Also, once we've got these ``probabilities'', what
we actually do with them is use them as a guide to our expectations
about the future, and we need to know why that's rational too.
Arguably, if we could do both we'd have a completely satisfactory
analysis of probability.
\sk Okay, so what do your rationality considerations give you
in the Everett interpretation?
\ev Well, first we consider a quantum measurement, made on a known state, and 
show that it's rational for an agent to act as if the quantum
probability rule were true \ldots
\sk `As if' it were true?  Is it true or isn't it?
\ev Okay, let's backtrack a bit.  The philosophy behind all this is that
probability is an operational notion: if some objective feature of the
physical world plugs into rationality considerations in just the way
that probability does, that feature \emph{is} probability.  Effectively
we're using Lewis's Principal Principle (or qualitative variants like Papineau's
Inferential and Decision-Theoretical links) as a criterion for what some 
chance-candidate has to do in order to \emph{be} chance.  (That's how
Lewis himself is using the Principle: he requires that some account can
be given of how his best-systems analysis justifies the Principal
Principle, and attacks rival accounts of objective chance on the grounds
that they couldn't satisfy this requirement.)

This sort of operationalism about the apparently fundamental is very
much in the spirit of the Everett interpretation, incidentally (the
decoherence-based versions at any rate).  See the author's paper, 
\citeNP{wallacestructure}, for more information.

\sk Okay, grant all this for the moment; where's decision theory come
in?
\ev Decision theory is a tool for decision-making under uncertainty.  It
doesn't introduce a primitive concept of (quantitative) probability at all,
incidentally --- it just shows that rational decision-making requires us
to assign probabilities to events, values to consequences, and then use
them together to maximise expected utility.  

\sk If it's about uncertainty, how can you apply it to a deterministic
theory?

\ev You have to buy into Saunders' account of branching as a
subjectively indeterministic process --- then classical decision theory
goes across \textit{mutatis mutandis}. 

\sk That account is surely open to attack?

\ev Well, you have to accept some reasonably strong philosophical
premises: Parfit's criterion for personal identity, a fairly robust
supervenience of the mental on the physical (probably not much short of
out-and-out functionalism), rejection of temporal becoming, and so
forth.  But those are part and parcel of the Everett interpretation ---
reject them and the interpretation comes apart as soon as we look at the
preferred basis problem, let alone at probability.

\sk So assuming we accept Saunders' account, what goes into the
derivation?  

\ev A standard set of decision-theoretic axioms such as Savage's,
minus most of the structure axioms (they are unnecessary because quantum
mechanics itself supplies all the structure), plus measurement
neutrality: the assumption that it doesn't matter \emph{how} a quantity is
measured provided \emph{that} it is measured.

\sk That innocent-sounding assumption looks like a weak link.

\ev Well, yes.  It seems possible to give it a pretty reasonable
justification from the subjective-indeterministic viewpoint --- and of
course it's intuitively pretty obvious --- but it's obviously not
something we can just read off from the Savage axioms.  And it's doing a
huge amount of work in the proof --- that's mostly tacit in Deutsch's
paper, but the speed of the proof from D0$'$--D4$'$ or S0-S5 of the
quantum probability rule shows how powerful it is.

\sk Don't go appealing to intuitive obviousness here, either.  The
reason we trust our measuring devices is that the operational part of
quantum mechanics --- which is thoroughly tested --- predicts that two
different devices have the same probability of a given outcome when they
measure the same observable on the same state.  But you can't help yourself
to this, since you're trying to justify probability.

\ev Certainly we need to be careful about avoiding circularity in our
justification of measurement neutrality.  But the probabilistic part of
the operational theory is not our only reason for accepting it --- the
major reason is that a measurement device is clearly designed purely to
register a given eigenstate, in a deterministic fashion. 
An experimentalist who's just built a new device 
doesn't deem it a $z$-spin-measurer  because on testing  it gets the same results as some 
`canonical' $z$-spin-measurer
(though to be sure, he might use some such device as a check.)  He deems it 
so because it's designed such that \emph{if} a given particle has a 
definite value of $z$-spin, 
\emph{then} the device will reliably tell us that value.

The property
of measuring devices that leads to collapse (and thus, subjective
\emph{indeterminism}) is simply their property of magnifying
superpositions up to the macro level --- and that property is constant
across different devices.

\sk But for all we know some of the `irrelevant' details of that
superposition-magnifying process are just those details which determine
probabilities!

\ev Don't fall into the trap of assuming that we're after some new
physical law which tells us the probabilities.  \textit{Ex hypothesi} we
know all the laws already --- what we're looking for is a rationality
principle.

\sk Let's move on: quite apart from technical worries about the proof,
the result seems conceptually far too weak to save the Everett
interpretation.  It may tell me that it's rational to bet that the
frequencies on the next zillion repetitions of the experiment will be
what QM says, but what I really need to know is why I should
\emph{expect} them to be!

\ev You expect them to be --- that is, you think they're highly
probable.  But in the account of probability that we're using here,
for something to be highly probable just \emph{is} for it to be rational to bet
on it.  If you wanted to insist on some more fundamental understanding of probability,
or of what we should expect to happen, you should have left long ago.

\sk Okay, point taken.  But isn't there something rather future-directed
about all this?  We're saying how we should act in the future, but we
also need to see how the theory is explanatory of the past --- else we
wouldn't have believed it in the first place.

\ev This is the point of the account (in section \ref{abandonOD}) of how
a theory is explanatory of data.  The idea is essentially that an explanatory 
theory makes the observed data highly probable\footnote{Technically, it only makes 
the observed data a member of some `typical set' of data which is highly 
probable; cf.\,footnote on page \pageref{footnote}.} 
(I mean to say, it makes it rational to bet on its occurrence).  Thus
there's an essentially future-looking aspect even to our assessment of
the theory's relevance for past data: we have to imagine ourselves just
before the data were collected, and ask what we should expect.

\sk So far, we've only been skirmishing.  We've been discussing problems
within the author's program, but now let's move on to reasons why no
such program \emph{could} work.  To begin: the theory is realist about
all branches, so it accepts that there are many branches in which the
frequencies are nowhere near those predicted by the quantum algorithm.

\ev Yes, but this isn't one of them.

\sk How can you tell?

\ev I retract slightly: it's rational to assume that this isn't one of
them (see section \ref{quantprobdiscussion} for the proof).

\sk But it might be.  

\ev Sure, and Elvis might have shot JFK, but it would be rational to
assume neither to be true.  In fact it's overwhelmingly more rational to
bet on Elvis shooting JFK than to bet that this is a wildly anomalous
branch.  This really isn't any different from ordinary probability:
there's always some chance of anomalous statistics.

\sk Everettians often trot out that last response.  The difference is,
in classical probability the anomalous branches aren't actualised.

\ev What you mean is, they probably aren't actualised.  And here's where
we can give a unified account of classical and quantum cases: what we
mean by `they probably aren't actualised' is that it's rational to
assume that they won't be actualised.

\sk But in the quantum case --- and not the classical --- the anomalous
branches actually exist.  In these branches, scientific communities come
up with wildly different theories to yours.

\ev These other theories are wrong, though.

\sk How can you have the confidence to say that? And don't say, `it's rational
to assume it.'  They've also been arrived at by a rational process; 
presumably they'd find it rational to reject \emph{your} conclusion.
Doesn't this undermine faith in your own argument?

\ev Not really.  It's not problematic for my theory to predict that
other people will reject it for rational reasons, provided that it also
predicts that I'm not one of them.   There's no neutral place to stand
in this game.  What's more, there's nothing particularly quantum about
this supposed paradox.  Suppose Alf experiences a truly spooky coincidence, 
the sort of fluke that would only have a one-in-a-billion chance of occurring in 
his lifetime --- unless ghosts exist, in which case it's not surprising at all.
Then it would be rational for him to drastically increase his credence in ghosts, unless
he already thought the chance of their existence far lower than one in a billion.
Yet on a planet with six billion inhabitants, there will be several Alfs even if ghosts do
not exist.  I may
be sure that they're wrong (or at least, overwhelmingly likely to be
wrong) about ghosts, but that doesn't force me to conclude that they're irrational ---
and the consistency of the situation is guaranteed by the prediction
that it's irrational for me to expect to be that person.\footnote{Let
$A$ be ``a one-in-a-billion spooky coincidence'', and $B$ be ``ghosts exist''; 
assume that $\Pr(B) \ll 1$ (uncontroversially, I hope!), that
 $\Pr(A|\neg B)=10^{-9}$, and that $\Pr(A|B)$ is  unity.
 Bayes' Theorem tells us that 
\be \Pr (B|A)\simeq \frac{\Pr(B)}{10^{-9}+\Pr(B)}\ee
which is close to unity even if the prior probability of $B$ is one in
a hundred million --- and how many of us are \emph{that} sure of our twenty-first century
worldview?  (I am quite, quite sure that ghosts don't exist --- sure enough to bet my
life on it, far more certain of it than I am of living until tomorrow --- yet I'm
not \emph{that} sure.  $10^8$ is a big number.)
I am grateful to Simon Saunders for the inspiration
behind this example.}

\sk The strongest objection has been saved till last.  If I accept
Parfit's account of what matters in personal identity, then I care about
my future descendants \emph{because} of the structural and causal
relation which I bear to them.  Why, then, should I give a damn about
their relative weights?

\ev Because it would be irrational not to.  

\sk It is unclear who has the burden of proof: my argument is precisely
that it would be irrational \emph{to} care about the weights.

\ev Your argument is based upon the objective-deterministic viewpoint,
which (it has been argued) is not the right view to assess rationality
considerations in Everett.  From the subjective-indeterministic view it
is a theorem that it is rational to weigh options in proportion to their
quantum weights, and your approach violates that.

\sk But my objection cuts deeper than that: if (as I claim) it is obviously 
rational to care equally about both descendants, and if subjective
expectation is supposed somehow to supervene on the degree to which I do
care for them, surely equiprobability has to go into decision theory as
a premise.  If its denial is a theorem, then the system is inconsistent;
so much the worse, then, for Everett.

\ev That would be fine, \emph{if} you could justify the supervenience of
expectation on ``caring for''.  But no such assumption went into
Saunders' argument.  All he uses is the fact that if I have multiple
descendants then I should expect to be exactly one of them; this is enough to
apply decision theory.  

\sk Suppose I accept this: even so, once I accept the Everett
interpretation then I would be at liberty to use the 
objective-deterministic viewpoint.  From its perspective, I could then
argue that the probability rule should be revised. If this contravened
the Savage axioms, so be it: not even axioms of pure rationality are
immune to revision.

\ev Maybe so, though the rest of us needn't join you in that revision if
we're instead willing to give up your equiprobability argument upon
realising that it clashes with the Savage axioms: apparently reasonable
strategies are certainly not immune to revision if it is shown that they
contravene obvious principles of rationality, and the Savage axioms are
surely more obvious than the correct strategy to take in a case of
fission!

I guess there's a transcendental argument why, if you're sure you'd
adopt the equiprobability rule upon becoming an Everettian, you'd be
irrational to become one.  But if I were sure that I found the de
Broglie-Bohm theory so unaesthetic that I'd lapse into depression and
suicide were I to come to believe it, I would be rational to avoid
believing it \ldots but that wouldn't really bear on its truth.

Incidentally, one possible reason why you should reject your equiprobability rule 
on its own terms is that it assumes that the decoherence process
consists of a discrete number of splittings, each into finitely many
branches.  In reality this is implausible: branching isn't as precisely
defined as all that, and there may well be infinitely or even
continuously many branches.

\sk The Bekenstein bound shows that Hilbert space is finite-dimensional.

\ev I'd prefer to leave quantum gravity out of it.  It's likely to change our theories enough
that effects on the dimension of Hilbert space will be the least of our worries --- but
we don't yet have any clear idea how, and the Bekenstein bound is at present just a plausible
bit of speculation.  Come to that, the decision-theoretic approach
doesn't tell us how to interpret the non-unitary evolution occurring in
black hole evaporation, and that's about as secure (or otherwise) as the
Bekenstein bound.

\sk Let us call a truce for now: \emph{if} your account really is valid,
what do you conclude from it?

\ev Most importantly, that we can vindicate Deutsch's claim:
that applying decision theory to quantum mechanics is enough to solve the probability 
problem in the Everett interpretation.  Decision theory provides a
framework in which we can understand what is involved in deducing
quantitative probabilities for quantum branching, and then shows us that this can be done
satisfactorily even when questionable assumptions like additivity are
abandoned.  Furthermore, the relevant links between quantum probability
and non-probabilistic facts can then be satisfactorily established.  

Just as interesting are the implications of quantum theory for decision
theory and the general philosophy of probability.  On the technical
side, it is noteworthy that the structure axioms required throughout
classical decision theory can be very substantially weakened.  To be
sure, this is only because the mathematical structure of the physical
theory (\iec, QM) in which the decision problem is posed is so rich, but
it seems far more satisfactory to have a richly structured physical theory (whose 
structure is clearly required on directly empirical grounds and in any
case is ontologically on a par with any other postulate of physical theorising) 
rather than introduce axioms governing rationality which are not self-evident
and which fairly clearly are introduced purely to guarantee a
representation theorem.

On a more conceptual level, (Everettian) quantum mechanics seems to
provide a novel route by which the concept of objective chance can be
introduced.  Everettian objective chances supervene entirely on 
particular matters of fact (Lewis's Humean requirement for chance), but 
the Everettian account is an improvement on Lewis's in that chances
supervene on local facts only, rather than on the pattern of all facts up to
the time of the chance event.  Furthermore, an account of how
these chances connect with credence is available that is at least as
secure as the frequency-based account --- indeed, though we do not have
a full derivation of the Everett Principal Principle, we have come
close.

\sk What would be a simple way of seeing how all this is possible: how
quantum mechanics can have these consequences for decision theory, and how 
the derivation of the quantum probability rule
was possible in the first place?

\ev It's long been recognised that the most fruitful guides
to allocation of probability have been frequencies and symmetries, but
the latter has always been somewhat suspect, and it is easy to see why:
how are we to choose which symmetries are respected by the chances?
Appeal to the symmetries of the physical laws seems the obvious method,
but obviously this just begs the question if those laws are
probabilistic.  Even for deterministic laws, though, the situation is
problematic: for if the situation is \emph{completely} symmetric between two outcomes, 
how is it that one outcome happens rather than the other?  In classical
mechanics, for instance, knowledge of the \emph{exact} microstate of a flipped coin 
breaks the symmetry of that coin and tells us with certainty which side
it will land.  The symmetry only enters because we assume the coin's
microstate to be distributed randomly with 50\% probability of leading
to each result, but this introduces probability in advance rather than
deriving it from symmetry.\footnote{This is perhaps mildly unfair: in
statistical mechanics we make a \emph{general} postulate that systems
are randomly distributed in the accessible region of phase space
according to the Liouville measure, and then count the symmetry of the
coin as telling us that the regions of phase space leading to the two
outcomes are of equal measure.  Nonetheless the underlying argument
still contains a strong initial admixture of probability.}

In the Everett interpretation, this circle is squared: it is not the
case that one or the other outcome will occur; from a God's-eye view,
both do.  This allows us to apply symmetry-based reasoning to 
equal-amplitude branching --- the core of the proofs in sections
\ref{deutschproof}--\ref{QSproof} --- and deduce that the branches are
equiprobable.

In a sense, then, the Everett interpretation reverses the primacy of
frequency over symmetry: the frequency of outcomes is an excellent guide
to the symmetry of the state being measured, but ultimately it is the
symmetries which dictate which events are equiprobable.

\end{description}

\section*{Acknowledgements}

For valuable discussions, I am indebted to Hannah Barlow, Katherine Brading, Harvey
Brown, Jeremy Butterfield, 
Adam Elga, Chris Fuchs, Hilary Greaves, Adrian Kent, Chris Timpson,
Wojciech Zurek, to all those at the 2002 Oxford-Princeton philosophy of
physics workshop, and especially to Simon Saunders and David Deutsch.  
Jeremy Butterfield and Simon Saunders also made detailed and helpful comments on an earlier
draft of this paper.

\section*{Appendix: act and consequence additivity}

\begin{theoremapp}\label{app1}
Suppose that we have some set \mc{C}, together with a binary addition
operation $+$ that is associative, commutative and has identity
0.  Let $\succ$ be a weak asymmetric order on \mc{C}, satisfying
\begin{enumerate}
\item $y \succ z$ if and only if $y+x \succ z + x$ for some
$x$;
\item Whenever $y \succ z \succ 0$, there exists some integer $n$
such that $n y \succ $ (and conversely for $y \prec z \prec 0$);
\item Whenever $y \prec 0$ and $x \succ 0$, there exists some integer
$n$ such that $y+ n x \succ 0$.
\item For any $c_1,c_2$ where $c_1 \succ c_2$, and for any $c$, there
exist integers $m,n$ such that $n c_1 \succ m c \succ n c_2$.
\end{enumerate}
Then there exists a function $\mc{V}:\mc{C}\rightarrow \Re$, unique up
to a multiplicative constant, such that (a) $\mc{V}(c_1)> \mc{V}(c_2)$ if
and only if $c_1 \succ c_2$ and (b)
$\mc{V}(c_1+c_2)=\mc{V}(c_1)+\mc{V}(c_2).$
\end{theoremapp}

\noindent \textbf{Proof:} We can suppose without loss of generality that there is at least one
element $x \succ 0$ (if not, but there is at least one $x \prec 0$,
define a new ordering $\succ'=\prec$, construct \emph{its} value
function, and take its negative; if for all $x$, $x \simeq 0$, then just
put $\mc{V}(x)=0$.)  

For all other elements $y \succ 0$, define the set $\val(y)=\{m/n \,|\, ny
\succ mx\}.$ 

\begin{lemma}
$\val(y)$ is closed below: if $m'/n' < m/n$ and $m/n \in
\val(y)$, then $m'/n' \in \val(y)$.
\end{lemma}
Now, suppose $m'/n'<m/n$ (so, $m'n<mn'$). 
\[ m/n \in \val(y) \Longrightarrow ny \succ mx\]
\[\Longrightarrow nn'y \succ mn'x\]
\[\Longrightarrow nn'y \succ m'nx\]
\[\Longrightarrow n'y \succ m'x\]
\[\Longrightarrow m'/n' \in \val(y). \Box \]

\begin{lemma}
$\val(y)$ has no greatest element.
\end{lemma}
Suppose $m/n\in \val(y)$.  We have to show that there exist $m', n'$ such that 
$m'/n' \in \val(y)$ and $m'/n' >m/n$.  
We have $ny \succ mx$, so by postulate 4 there exist integers $a,b$ such
that $nay \succ b x \succ max$.  Put $m'=b$, $n'=na$.  Then since
$b>ma$, $m'/n'>m/n$. $\Box$

Further, $\val(y)$ is non-empty, since it contains 0; it does not
contain all rational numbers, by postulate 3.

But any subset of
rational numbers which is non-empty, not equal to the set of all rationals, closed below, 
and without greatest
element is a \emph{Dedekind cut}, and thus can be regarded as a real
number: we thus take $\mc{V}(y)$ to be the real associated with the
Dedekind cut $\val(y)$.

To establish that \mc{V} has the right properties, suppose first that
$\mc{V}(y) >> \mc{V}(z).$  Then there exist $m,n$ such that $ny \succ mx$ but $ny \preceq mx.$
Hence $ny \succ nz$, so $y \succ z$.

The converse requires another use of postulate 4 (to rule out the
possibility that $z$ is infinitesimally less valuable than $y$).  Suppose
$y \succ z$, then there exist $m$ and $n$ such that $ny \succ mx \succ
nz$.  Hence, $m/n$ is in $\val(y)$ but not $\val(z)$, so
$\mc{V}(y)>\mc{V}(z).$

This completes the construction of the value function for all $y \succ
0$.  For $y \prec 0$ we define $\val(y) = \{m/n \, | \, 0 \succ ny +
mx\}$, and prove in an essentially identical way that this is also a
Dedekind cut. $\Box$

\begin{theoremapp}
Let $\mc{S}=\{s_1, \ldots, s_n\}$ be a finite set (of states); let \mc{C} be a set of
consequences on which there exists an addition operation.  Let \mc{A}, the act space, be defined as the set of
functions from \mc{S} to \mc{C} (to be thought of as bets on which state
will occur).  Define addition of acts in the obvious way: 
$(\mc{P}_1+\mc{P}_2)(s_i)=\mc{P}_1(s_i)+\mc{P}_2(s_i).$
Now suppose that there exists a preference order $\succ$ on \mc{A},
satisfying 
\begin{enumerate}
\item \emph{Additivity}: $\succ$ can be represented by an additive value function \mc{V} on
\mc{A} (equivalently, $\succ$ obeys the postulates of Theorem
A.\ref{app1}).  (We write $\mc{V}(x)$ for the value of the constant act
$\mc{P}(s)=x$ for all $s$.)
\item \emph{Dominance}:  if $\mc{V}[\mc{P}(s_i)]\geq\mc{V}[\mc{P}'(s_i)]$ for all $i$,
then $\mc{V}(\mc{P})\geq \mc{V}(\mc{P}').$
\end{enumerate}

Then there exist positive real numbers $p_1, \ldots, p_n$, satisfying
$\sum_i p_i =1$, such that
for any act \mc{P}, $\mc{V}(\mc{P})=\sum_i p_i \mc{V}[\mc{P}(s_i)]$.
\end{theoremapp}

\noindent \textbf{Proof:} For any consequence $x\succ 0$ and any state index $i$, define
$\mc{P}_{x,i}=x \delta_{i,j}.$ The sum over $i$ of all the acts
$\mc{P}_{x,i}$ is just the constant act $\mc{P}(s)=x$, hence
\[\sum_i\mc{V}(\mc{P}_{x,i})=\mc{V}(x).\]
If we define $p_i(x)=\mc{V}(\mc{P}_{x,i})/\mc{V}(x)$, we have $\sum_i
p_i(x)=1$; by the Dominance assumption, each $\mc{V}(\mc{P}_{x,i})\geq 0$,
so each $p_i(x) \geq 0.$  Our remaining task is to prove that the $p_i$ are 
independent of $x$.

Suppose $y \succ x$.  (We shall not give the proof for $x \succ y \succ
0$ or $y \prec 0$, as it is essentially identical.)  Let $\alpha_j =m_j/n_j$ be an
increasing sequence of rational numbers such that $\lim_{j\rightarrow
\infty}\alpha_j =\mc{V}(y)/\mc{V}(x).$  By Dominance and Additivity, it follows that
\[m_j \mc{V}(\mc{P}_{y,i})=\mc{V}(\mc{P}_{m_j y,i})\geq\mc{V}(\mc{P}_{n_j
x,i})=n_j p_i(x) \mc{V}(x),\]
and hence $\mc{V}(\mc{P}_{y,i})\geq \alpha  p_i(x) \mc{V}(y).$  Similar
consideration of a decreasing sequence gives us
$\mc{V}(\mc{P}_{y,i})\leq  p_i(x) \mc{V}(y)$.  Between these two
inequalities we can read off $p_i(y)=p_i(x)$, and the result is proved.

\end{document}